\documentclass[prd,aps,twocolumn,showpacs,superscriptaddress,floatfix]{revtex4-1}
\usepackage{bm}
\usepackage{pdfpages}
\usepackage{amsmath}
\usepackage{txfonts}
\usepackage{graphicx}
\usepackage{dcolumn}
\usepackage{bm}
\usepackage{color}
\def\comment#1{}

\def\slashchar#1{\setbox0=\hbox{$#1$}           
   \dimen0=\wd0                                 
   \setbox1=\hbox{/} \dimen1=\wd1               
   \ifdim\dimen0>\dimen1                        
      \rlap{\hbox to \dimen0{\hfil/\hfil}}      
      #1                                        
   \else                                        
      \rlap{\hbox to \dimen1{\hfil$#1$\hfil}}   
      /                                         
   \fi}                                         %

\def\nablab{{\mbox{\boldmath $\nabla$}}}

\begin{document}

\title{Thermalization and possible quantum relaxation times in ``classical'' fluids: theory and experiment}

\author{Z. Nussinov*}
\affiliation{Department of Physics, Washington University, St. Louis,
MO 63160, U.S.A.}
\affiliation{Nordita, KTH Royal Institute of Technology and Stockholm University,
Roslagstullsbacken 23, SE-10691 Stockholm, Sweden}
\affiliation{Department of Condensed Matter Physics, Weizmann Institute of Science, Rehovot 76100, Israel}
\email{zohar@wuphys.wustl.edu}

\author{F. S. Nogueira$^\dagger$}
\affiliation{Institut f\"ur Theoretische Physik III, Ruhr-Universit\"at Bochum, Universit\"atsstrasse 150, 44801 Bochum, Germany}
\affiliation{Institute for Theoretical Solid State Physics, IFW Dresden, Helmholtzstr. 20, 01069 Dresden, Germany}
\email{flavio.nogueira@fu-berlin.de}

\author{M. Blodgett}
\affiliation{Department of Physics, Washington University, St. Louis,
MO 63160, U.S.A.}

\author{K. F. Kelton}
\affiliation{Department of Physics, Washington University, St. Louis,
MO 63160, U.S.A.}

\date{\today}

\begin{abstract}
Quantum effects in material systems are often pronounced at low energies and become insignificant at high temperatures. 
We find that, perhaps counterintuitively, certain quantum effects may follow the opposite route and become sharp when extrapolated to high temperature within a ``classical'' liquid phase.  In the current work, we suggest basic quantum bounds on relaxation (and thermalization) times, examine kinetic theory by taking into account such possible fundamental quantum time scales, find new general equalities connecting semi-classical dynamics and thermodynamics to Planck's constant, and compute current correlation functions. Our analysis suggests that {\it on average}, the extrapolated high temperature dynamical viscosity of general liquids may tend to a value set by the product of the particle number density ${\sf n}$ and Planck's constant $h$. We compare this theoretical result with experimental measurements of an ensemble of 23 metallic fluids where this seems to indeed be the case. The extrapolated high temperature viscosity of each of these liquids $\eta$ divided (for each respective fluid by its value of ${\sf n} h$) veers towards a Gaussian with an ensemble average value that is close to unity up to an error of size $0.6 \%$. Inspired by the the Eigenstate Thermalization Hypothesis, we suggest a relation between the lowest equilibration temperature to the melting or liquidus temperature and discuss a possible corollary concerning the absence of finite temperature ``ideal glass'' transitions. We suggest a general quantum mechanical derivation for the viscosity of glasses at general temperatures. We invoke similar ideas to discuss other transport properties and demonstrate how simple behaviors including resistivity saturation and linear $T$ resistivity may appear very naturally. Our approach suggests that minimal time lags may be present in fluid dynamics. \end{abstract}


\maketitle

\section{Introduction}

As long known, at atomic and smaller length scales, quantum mechanical effects are typically extremely important. In condensed matter systems, collectively in crystals and elsewhere, these effects may trigger striking (and often potentially useful) novel low temperature behaviors on far larger spatial scales as in, e.g., semiconductors, superconductors, superfluids, the quantum Hall effects, and countless other systems. Common lore asserts that at high temperatures, quantization is largely inconsequential. 
Historically, Planck first introduced his constant ($h$) in the study of black body radiation \cite{ADP}. Specifically, when attempting to improve ``high temperature'' lightbulb filaments, Planck made an ansatz that sparked his celebrated result for the rate of energy emission per unit time and unit area by photons of frequency $\nu$,
\begin{eqnarray}
\label{bb}
I(\nu,T) = \frac{2 h \nu^{3}}{c^{2}} \frac{1}{e^{\frac{h \nu}{k_{B} T}}-1},
\end{eqnarray}
where $c$ is the speed of light and notably, in the context of our current work, $(k_{B} T)$ sets the only energy scale. 
By simple integration, this leads to another celebrated result for
the total intensity of a black body at temperature $T$,
\begin{eqnarray}
I = \sigma_{SB} T^{4},
\end{eqnarray}
with the Stefan-Boltzmann constant $\sigma_{SB} = \frac{2 \pi^{4}}{15} \frac{k_{B}^{4}}{h^{3} c^{2}}$.
The energy emission (Eq. (\ref{bb})) peaks at a frequency proportional to the temperature.
As is well appreciated, these remarkable textbook formulae are critical in understanding  myriad systems (including those 
at extremely high temperatures such as stellar radiation and volcano lava flows). 

It is colloquially argued that ``classical transport'', such as that manifest in the dynamical viscosity of high temperature liquids, has little to do with quantum mechanics (apart setting the scale of the specific atomic interactions that may vary dramatically from
one liquid to another). In the current work, we argue that the opposite may occur in rather general systems when transport functions are extrapolated to high temperatures. Such an effect may have consequences for the understanding of the behavior of high temperature metals and insulators as well as basic minimal time scales in ``classical'' fluid dynamics. 
It is clear that such a high temperature ($T$) occurrence might not be entirely unexpected. At {\it extrapolated} asymptotically high temperatures such that $k_{B} T$ is larger than any interaction energy scale (or energy associated with a particle having 
a wavelength determined by the inverse volume density, etc.), the dominant energy scale is set by $k_{B} T$. Thus, the only frequency scale is set by $(k_{B} T)/h$ with $h$ being Planck's constant. A similarly natural dominant frequency scale appears in black body radiation (Eq. (\ref{bb})). Thus, it is clear by simple dimensional analysis arguments, that in the very ``classical'' {\it extreme high  temperature limit}, Planck's constant may come in and clear signatures of quantum mechanics may emerge. In what will follow in this work, we will sharpen this intuition and illustrate that at high temperatures, in quantum {\it unentangled} semi-classical systems the relaxation time will, in many cases, be {\it exactly} given by $h/(k_{B} T)$ with no additional numerical prefactors. At room temperatures ($ T \sim 300$ K), this is a very small time scale: $(h/(k_{B} T)) \sim 160$ femtoseconds (fs). A consequence of the current work is that thermalization in equilibrated systems is typically bounded by such times. Although seemingly small, such time scales may, in principle, be probed for and observed in luminescence measurements of both equilibrium and out of equilibrium electronic systems, e.g., \cite{photo-exp,nadav}. As we will further motivate in the current work, electronic, mechanical, or any other measurable thermalization times in equilibrated systems at room temperature might be bounded by this number (and, in the ``worst case'' by $(h/(4 \pi^{2} k_{B} T)) \sim  4$ fs at room temperature). 

Dimensional analysis and more sophisticated constructs are of course prevalent and may suggest similar time scales in disparate regimes (including (entangled) quantum critical and other systems in the diametrically opposite limit of low temperatures, black holes, and others) \cite{Sackur,qcp,entangle_qcp}. 
For instance, in the condensed matter arena, a ``Planckian'' time scale of order ${\cal{O}}(\hbar/(k_{B} T))$ was posited to appear in 
the cuprate superconductors \cite{jan}; such a time scale is of commanding importance in  quantum critical phenomena \cite{qcp}. Amongst others, Damle and Sachdev discussed this time scale in detail 
when analyzing transport in such systems \cite{DS}. More recently, metallic transport properties were studied with this possible 
time scale in mind \cite{bruin,hartnoll}. To avoid confusion with these and other works, we reiterate that apart from refraining from unknown numerical
prefactors, our analysis does not propose scales by examining phenomenology and working back to see if certain constructs might  be useful nor do we focus on quantum criticality or employ conformal symmetries (although the latter may indeed effectively appear in the limit in which the thermal energy scale is far larger than rest mass energy). 

To further couch our work in a broader context, we briefly discuss dynamical viscosity in disparate quantum systems. 
Nearly ten years ago, string theory,  which is a fundamentally quantum theory, made its debut in the field of physical kinetics by 
introducing the concept of {\it perfect fluid} \cite{Kovtun-2005}. Perfect fluidity would be obtained by saturating the following  
lower bound for 
the ratio between the viscosity and the entropy density,
\begin{equation}
\label{Eq:visc-bound}
 \frac{\eta}{s}\geq \frac{\hbar}{4\pi k_B},
\end{equation}
as conjectured by Kovtun and co-workers \cite{Kovtun-2005,Note-1}. 
A perfect fluid as established by the saturation of above inequality (that is, when Eq. (\ref{Eq:visc-bound}) becomes an equality) is inherently 
a quantum state of matter. 
We note that the light velocity does not arise in the bound above, so it is in principle possible to conclude that 
it applies to non-relativistic systems as well, although it has been originally derived for a relativistic theory. 
Experimental evidence for perfect fluidity appears in ultracold gases 
at unitarity  and in the quark-gluon plasma \cite{Schaefer-2006}. 
It has also been argued that graphene might be an almost perfect fluid  \cite{Schmalian-2009}. 
On the other hand, in classical physics perfect fluidity 
is associated to an ideal gas having $\eta=0$. However, even classically, a weakly interacting system typically has 
a large dynamical viscosity, with a small viscosity being associated with a strongly interacting system. This supports the view that 
non-interacting particles have an infinite mean free path, leading in this way to infinite viscosity. Nevertheless, 
if the viscosity of an ideal classical or quantum gas is calculated via correlation functions, a vanishing result is 
obtained (this point will be reviewed in Section  \ref{Sect:correlation}), in agreement with the expectations of classical fluid 
dynamics. 

In general, non-relativistic quantum liquids exhibit a 
viscosity that diverges as $T\to 0$, a behavior contrasting with dilute gases, where the viscosity vanishes as 
$T\to 0$.  In dilute gases at high temperatures 
the viscosity is uniquely determined by two fundamental lengths governing 
the collisional dynamics, namely, for such fluids composed of particles of mass $m$,
the non-relativistic thermal de Broglie wavelength
\begin{eqnarray}
\label{lnr}
\lambda^{nr}_T=\sqrt{2\pi\hbar^2/(mk_BT)},
\end{eqnarray}
and the scattering length, $a_{sc}$, with the diluteness condition 
$\lambda^{nr}_T {\sf n}^{1/3}\ll 1$ holding for a three-dimensional system, where ${\sf n}$ is the particle number per unit volume. 
 As the viscosity has the dimensions of momentum divided by area, we can use the de Broglie relation with the 
 thermal wavelength to write, 
 \begin{equation}
 \label{Eq:classical-dilute}
 \eta\sim \frac{h}{\lambda^{nr}_Ta_{sc}^2}\sim\frac{\sqrt{2\pi mk_BT}}{a_{sc}^2},
 \end{equation}
 and there is no dependence on the Planck constant or the density in this case. At low temperatures, quantum effects 
 in a gas become relevant and a third length, the coherence length, plays also a role. For instance, for a dilute Bose gas 
 in three dimensions and low temperatures such that ${\sf n}a_{sc}(\lambda^{nr}_{T})^2\ll 1$, an explicit dependence on the superfluid 
 density arises \cite{Kirkpatrick-Dorfman} and contains a factor $1/a_{sc}^2$, similarly to Eq. (\ref{Eq:classical-dilute}). 
 By contrast, at sufficiently low temperatures such that ${\sf n}a_{sc}(\lambda^{nr}_T)^2\gg 1$, the viscosity 
 depends on the total density and exhibits a 
 behaviour $\eta\sim T^{-5}$. Interestingly, in this low temperature regime the viscosity is 
 also proportional to $a_{sc}^5$ \cite{Kirkpatrick-Dorfman},  
 indicating a vanishing viscosity in the non-interacting limit.   
Calculations of the viscosity for $^4$He \cite{Khalatnikov} 
yield at 
low temperature, $\eta\sim (a_{sc}/T)^{5}$, in agreement with the previous result, 
while $\eta\sim e^{\Delta/(k_BT)}$ for high temperatures \cite{remark1}, where $\Delta$ represents  
the energy height of the roton minimum of the superfluid spectrum. 
A result that decreases with the temperature is also obtained for a Fermi liquid, where we have 
$\eta\sim 1/T^2$ \cite{Abrikosov-FLT}. For completeness and comparison, we remark that the  viscosity of the unitary Fermi gas has been investigated
in \cite{Cao} where at low temperature the viscosity scale was found to scale with $({\sf n} \hbar)$
while at high temperature, the viscosity scale was set by $\hbar/(\lambda_{T}^{nr})^{3}$.

The high temperature regime of a relativistic system can be also analyzed similarly to our treatment of the 
high-temperature dilute gas. The main difference in the argument is the relativistic thermal 
de Broglie wavelength ($\lambda^{r}_{T}$), which in this case is 
dependent on the light velocity in vacuum $c$ and the spatial dimensionality $D$, 
\begin{equation}
\label{Eq:lambda-rel-1}
\lambda^{r}_T=2\sqrt{\pi}\left[\frac{\Gamma(D/2)}{2\Gamma(D)}\right]^{1/D} \frac{\hbar c}{k_BT}.
\end{equation}
The above thermal de Broglie wavelength holds for a massless relativistic particle. 
The massive case can only be calculated analytically for $D=2$, in which case we obtain,
\begin{equation}
\label{Eq:lambda-rel-2}
 \lambda^{r}_T=\frac{\hbar c}{k_BT}\sqrt{\frac{2\pi}{1+\frac{mc^2}{k_BT}}}.
\end{equation}
Eq. (\ref{Eq:lambda-rel-1}) serves our purpose also  
for massive particles, provided 
the high-temperature regime where $k_BT\gg mc^2$ holds. In particular, in the high temperature limit 
Eq. (\ref{Eq:lambda-rel-2}) reduces to (\ref{Eq:lambda-rel-1}) in $D=2$. It is also interesting to note that 
for $c\to\infty$ we obtain from (\ref{Eq:lambda-rel-2}) the non-relativistic thermal de Broglie wavelength. 

Taking a $\phi^4$ scalar theory of dimensionless interaction strength $u$ in $D=3$  
(in units where $\hbar=c=1$) as an example, it is seen that 
the thermal de Broglie wavelength is the only length scale at high temperature. Therefore, it is easy to obtain  
the high-temperature behavior, 
\begin{equation}
\label{Eq:eta-rel}
 \eta\sim\frac{2(k_BT)^3}{\pi\hbar^2c^3u^2}.
\end{equation}
This form attests to the subtle (and unexpected seemingly {\it intertwined}) ``classical''  non-quantum (i.e., $\hbar \to 0$)
and non-relativistic ($c \to \infty$) limits. An expression having precisely the above behavior has been indeed derived long time ago using Feynman 
diagrams within a non-equilibrium correlation function formalism \cite{Hosoya}. Although the high-temperature relativistic 
result (\ref{Eq:eta-rel}) seems to diverge in the non-interacting limit, it vanishes in the non-relativistic limit, independently 
from the interaction strength. Of course, this result cannot be entirely trusted, as we have already seen in the example 
of the dilute non-relativistic Bose gas at low temperatures. Indeed, there we  would have 
concluded from the moderately low temperature regime,  
corresponding to ${\sf n} a (\lambda^r_T)^2\ll 1$, that the viscosity diverges in the non-interacting limit. However, 
in the even lower temperature regime, the behavior with the scattering length points out to a vanishing result instead.  

Our discussion would not be complete without noting the famous Sackur-Tetrode equation \cite{Sackur,Sackur',tetrode,experiment} for the entropy of a classical three-dimensional ideal gas
of $N$ identical particles,
\begin{eqnarray}
\label{ST}
S = Nk_{B} \Big[ \ln \Big( \frac{1}{{\sf n} \lambda_{nr}^{3}} \Big) + \frac{5}{2} \Big],
\end{eqnarray}
that relates, via the non-relativistic thermal de Broglie wavelength of Eq. (\ref{lnr}), thermodynamics to dynamics. At high temperatures, 
many systems emulate ideal gases. Historically, Eq. (\ref{ST}) enabled estimate of Planck's constant from
thermodynamic measurements (in particular, those of monatomic mercury vapor) \cite{experiment}. 

As a weaker and more generic consideration regarding the specter of quantum effects sharply 
manifesting at high temperatures, we next invoke the energy-time uncertainty relations 
\begin{eqnarray}
\label{ET}
(\Delta E)( \Delta t) \ge \frac{\hbar}{2}.
\end{eqnarray}
If the broadening $\Delta E$ is set by the only scale in the system, 
the thermal energy $(k_{B} T)$, then this will indeed similarly motivate the 
appearance of a minimal uncertainty in time that is of the order of $h/(k_{B} T)$.
The relative phases between energy eigenstates that are superposed in an initial quantum state and ensuing 
generic decoherence time are set by the spread in the energy eigenvalues.
If this spread is of the order of ${\cal{O}}(k_{B} T)$ then a decoherence time of order of $h/(k_{B} T)$ will result. 
More specific and detailed than simple uncertainty relations alone, a plethora of related well-known arguments may be applied to the high temperature limit in which the classical thermal de Broglie wavelength $\lambda_{T}^{nr}$ formally becomes asymptotically small at high temperatures.
In the context of transport measurements, that form the principal motivation 
of this work, the Ioffe-Regel criterion \cite{Ioffe} and other considerations like it applied to
this small wavelength limit may lead to conclusions akin to those adduced from the above dimensional analysis arguments. Physically, for transport to occur the mean-free path $\ell$ must be greater than or equal to the de-Broglie wavelength. For smaller $\ell$, the notion of a particle (or quasi-particle) would be ill-defined. In such cases, constructive interference may lead to localization (similar to that found in electronic systems \cite{Anderson}). Thus, the formal vanishing of the de Broglie wavelength is impossible. Rather, in the liquid phase when quasi-particles are still defined, the de-Broglie wavelength can only be as small as the mean-free path (which in turn can be no smaller than the inter-particle separation). Thus, the uncertainty relations or related de Broglie wavelength scale mandate that at high temperature saturation occurs so long as no phase transitions appear as would be further anticipated for analytic high temperature extrapolations of transport functions within a liquid phase.  
Now here is an important conceptual point of our work. All of the above considerations may indicate that at all temperatures $T$,
\begin{eqnarray}
\eta (T) \ge \eta_{\min}^{\sf quantum}.
\label{limexplain}
\end{eqnarray}
That is, the viscosity cannot be lower that a minimal value $\eta_{\min}^{\sf quantum}$ set by quantum mechanics.
On the other hand, in fluids $\eta(T)$ is generally a monotonically decreasing function of the temperature $T$.
(This behavior may be contrasted with that of a dilute gas in which the viscosity increases with temperature. It is because 
of this different monotonic trend that a minimum of the viscosity and other associated ratios is to be expected, such as that appearing 
in Eq. (\ref{Eq:visc-bound}) \cite{Kovtun-2005}).  In our more general context, this monotonicity, together with Eq. (\ref{limexplain}), suggests that if there is no reason for the lower bound of Eq. (\ref{limexplain}) to not be saturated then {\it an equality} may appear
\begin{eqnarray}
\lim_{T \to \infty} \eta(T) = \eta_{\min}^{\sf quantum}.
\label{vequality}
\end{eqnarray}
Thus, we propose that inequalities akin to those Eq. (\ref{Eq:visc-bound}) might in some cases be replaced by equalities. 
That is, at {\it extrapolated high temperature} values, the viscosity may veer towards points close to a sharply defined quantized value.
The latter qualifier is important as we will consider specific functional forms for the viscosity and examine how these 
may be extrapolated. Furthermore, as we will stress repeatedly throughout this work, in the extrapolations that we will consider 
we will hold the number of particles, etc., fixed and not assume phase transitions in which one functional form 
for the viscosity gives way to another (due to localization or other effects).  Amongst other things, 
especially in the context of many studies concerning the application of string theory motivated AdS-CFT type bounds, 
as temperature increases, the basic pertinent particles (molecules, atoms, quarks, etc.,\cite{Kovtun-2005,Shuryak})
may trivially change as temperature is increased. This change in the number of particles is why {\it bounds} of the form of the ratio in Eq. (\ref{Eq:visc-bound})
may more naturally hold in general phases \cite{Kovtun-2005} as in these the density of the relevant particles (whatever they are)
drops out in such ratios. Along another line of work, high temperature quantum limits on information and entropy (and associated heat flow) were advanced and extended 
in \cite{info1,info2}. The basic ingredient in all of these works starting from Planck's original work \cite{ADP} ultimately relates to the number of available states/channels 
that may occupied. Classical counterintuitive high temperature effects in general systems were advanced in \cite{expandT}.  Apart from the viscosity that we largely focus on here, 
following the considerations outlined in the current work, similar proposals may be advanced for disparate dynamical time scales and a plethora of response functions. 

All of the arguments and caveats above notwithstanding, none of them illustrates that inequalities of the form of Eq. (\ref{limexplain}) may indeed be established with some rigor and that these inequities might, in some cases, become sharp equalities at high temperatures (such that the limiting value of $\lim_{T \to \infty} \eta(T)$ is not larger than an appropriately defined 
$\eta_{\min}^{\sf quantum}$). Towards that end, we will embark on specific calculations that may replace the above inequalities by precise relations of the form of Eq. (\ref{vequality}) with, for semi-classical systems, $\eta_{\min}^{\sf quantum} = {\sf n} h$. To set the scale even for nearly ideal dilute gases, empirically, the viscosity of noble gases such as Argon at atmospheric pressure and room temperature \cite{CRC_hand} ($\sim 0.229 ~cm^{-1} ~gr~sec^{-1}$) well surpasses ${\sf n} h$ 
(which is approximately $1.63 \times 10^{-7} cm^{-1} ~gr~ sec^{-1}$ in this example).
Aside from discussing semi-classical systems, we will further motivate inequalities for broader theories.

We will furthermore suggest that quantum mechanics (in particular, the Eigenstate Thermalization Hypothesis) may shed light on transitions from the high temperature liquid to
a low temperature glass formed by rapid cooling (``supercooling'') below the melting temperature. Notably, we illustrate how simple forms for the relaxation rates in supercooled fluids
may naturally appear.

\section{Outline}
In the sections that follow, we will explicitly study how quantization constraints on the extrapolated high temperature transport functions may arise. The outline of the remainder of this article is as follows. In section \ref{sec:WKB}, we demonstrate how, by virtue of the WKB relations, in the extreme high temperature ``classical limit'', Planck's constant must make an appearance when computing sums over all quantum states. We then proceed to find another relation between the {\it semi-classical dynamics and thermodynamics} appears- i.e., an equality connecting the {\it total time} for periodic motion (summed over different ergodic components) in general bounded many-body systems {\it with thermodynamic entropy}. Armed with this approach, in section \ref{sec:tr}, we apply the WKB borne result and further invoke considerations common in transition state theory applied to deformations of potentials to illustrate that the equilibration time of semi-classical systems may, quite universally, be bounded by $h/(k_{B} T)$. We suggest that this bound is saturated in the high temperature limit. In this case, there are no additional prefactors ``of order unity''. That is, in semi-classical systems, $h/(k_{B} T)$ {\it is the exact extrapolated equilibration time}. We discuss how this time may be found from empirical analysis of the data. In Section \ref{generalt}, we propose general bounds on equilibration times in linear response functions in general (semi-classical or other) systems. Knowing the equilibration time, we apply the Boltzmann equation and find in section \ref{Boltzmann-sec} that at asymptotically high temperatures, the viscosity will decrease and saturate to a lower bound set by ${\sf n} h$ with ${\sf n}$ the number
density of particles. Taking into account multiple relaxation time processes will lead to the use of Gibbs free energies and suggest that, near and at temperatures high enough that the liquid is equilibrated, the leading order approximate form of the viscosity of a semi-classical liquid ``$a$''  will, on average, be given by $ \eta_{a} = A_{a}{\sf n} h \exp(\beta \Delta {\sf H})$, with $A_a \approx 1$ and $\Delta {\sf H}_{a}$ an effective activation barrier. We comment on the relation between our rather general semi-classical analysis vis a vis the special situation in (typically) entangled states that characterize quantum critical points and motivate specific bounds in general quantum systems. In section \ref{Sect:correlation}, we compute, via current correlation functions, the viscosity without the use of transition state frequency considerations to arrive a similar non-identical result. In Section \ref{lower_b}, we propose bounds on the extrapolated forms of the viscosity of semi-classical fluids. In Section \ref{strict_b}, we derive analogous relations to the viscosity to entropy bounds suggested by \cite{Kovtun-2005} without using holography but rather by simply building on Section \ref{generalt}. 
We then point out, in Section \ref{s:Eyring}, that Eyring's form for the viscosity must have no undetermined prefactor (contrary to what Eyring originally suggested) and that it is a different way of couching the considerations of Section \ref{Boltzmann-sec}. In section \ref{normal}, we demonstrate that the prefactors $A_{a}$ experimentally adduced for an ensemble of 23 metallic fluids follow a Gaussian distribution with an average value ${\mathbb{E}}(\{A_{a}\})  =1- 0.00647$ and a standard deviation associated with the average that is $0.09$. In Section \ref{ETH_section}, we discuss the lowest equilibration temperature of ``classical'' fluids and suggest that it may be strongly correlated with the melting or liquidus temperature. As we explain, such a correlation may be naturally expected from the Eigenstate Thermalization Hypothesis. We compare these simple notions with data for metallic liquids. 
These ideas further suggest the absence of finite temperature ``ideal glass'' transitions at which the relaxation times diverge. An extension of these concepts leads to possible dependence of the viscosity of supercooled liquids at general temperatures. As we further discuss, the minimal time scale $h/(k_{B} T)$, for bare minimal moves in local semi-classical fluid dynamics may play a role in fluid analysis as we briefly comment on in section \ref{ns}. We briefly discuss the extension of our analysis of the viscosity to quantum critical critical systems (Section \ref{vis:qc}). In sections \ref{ec} and \ref{tc}, we outline the simplest application of our considerations to other transport functions (the electrical and thermal conductivities). The far too simple analysis provided therein may rationalize the saturation of the resistivity (and linear increase of the resistivity) found in many materials. We further notably provide possible stringent upper bounds on the resistivity of bad metals; these bounds seem to be satisfied by empirical data.  We conclude with a brief synopsis of our main results. In the Appendix, we further elaborate on the experiments performed
and the data analysis carried out in Ref. \cite{metallic-glass}. These {\it experimental results} underlie the empirically found form for the viscosity at high temperatures and were invoked in our analysis of the ensemble of 23 metallic liquids in Sections \ref{normal}, \ref{ETH_section}.

\section{Oscillation frequencies and phase space integrals via WKB}
\label{sec:WKB}
We begin our analysis with a simple and new derivation of how Planck's constant enters some calculations in classical statistical mechanics.   
As long appreciated, classical statistical physics is incomplete unless a prescient constant $h$ is introduced to render phase space integrals dimensionless. 
With such a(n initially seemingly arbitrary) constant $h$ at hand, the classical canonical partition function for a system of 
$N$ (non-identical) particles in $D$ spatial 
dimensions with a Hamiltonian $H$ is set by
e.g.,\cite{huang}, $Z = h^{-DN} \int d^{DN}x ~ d^{DN} p~  \exp(-\beta H)$, with $x$ and $p$ the generalized spatial and momentum coordinates of phase space and 
$\beta = 1/(k_{B} T)$ is the inverse temperature. In a similar related vein of 
textbook statistical mechanics, phase space volumes within the micro-canonical correspond have to be divided by $h^{DN}$ in order to count the number of micro-states. In certain 
simple quantum problems (e.g., a particle in box, harmonic oscillators, etc.), 
a comparison can be made between exactly solvable classical and quantum partition functions ($Z= Tr  ~\exp(-\beta H)$) and this factor $h$ turns to be exactly 
equal to Planck's fundamental 
constant of quantum mechanics. That basic blocks of classical phase 
space of volume $h^{DN}$ cannot correspond to numerous states may be motivated by the uncertainty principle. In most non-relativistic problems of practical 
interest where the total particle number 
is conserved, the actual value of $h$ in the classical partition function 
is irrelevant as it is merely an innocuous arbitrary constant and
cancels out when computing nearly all probabilities (and all observables computed with these probabilities or alternatively calculated by direct differentiation of the 
thermodynamic free energies where $h$ simply appears 
as an immaterial additive constant to the free energies). This cancellation is somewhat reminiscent to phase factors in electrodynamics and other 
gauge theories that may cancel in all physically meaningful final (gauge invariant) results. An earlier notable exception to this ``rule" concerning the unimportance 
of the numerical value of $h$ is afforded by 
reaction rates \cite{kramers,langer,hanggi,eyring-tst,wigner,miller,wolynes,WSW} in chemical, nuclear, and other systems where the number and nature of the degrees of freedom 
may vary at transitions. An even earlier example is furnished by the well-known Sackur Tetrode relation of Eq. (\ref{ST}) for the entropy of an ideal gas 
that can be experimentally verified \cite{Sackur,Sackur',tetrode,experiment}.

In what follows, we quickly derive the appearance of Planck's constant in the semi-classical limit of quantum systems with a single degree of freedom ($x$) that, in some cases, may serve as a 
pertinent generalized coordinate in 
a many body system. Given a general time independent Hamiltonian, {\it all such one dimensional systems} are trivially integrable in their classical limit. To make the link with the semi-classical description of the quantum system, 
we first recall a generic outcome of the lowest order terms in the
WKB approximation as applied to {\it bounded} classical phase space trajectories
yielding a celebrated Bohr-Sommerfeld type relation
\begin{eqnarray}
\label{bohr}
J_n \equiv \oint_{n-th ~ state} p ~ dx = h(n+C),
\end{eqnarray}
with $p$ the canonical momentum, $n$ an integer labeling the state, and $C$ a constant that is set by the character (and number) of the turning points. In the standard cases, $C$  vanishes for steep ``hard boundaries'', assumes a value of $C=1/2$ for soft potentials, and $C=3/4$ for one dimensional potentials with one hard and one soft boundaries. The leading order WKB 
result of Eq. (\ref{bohr}) becomes {\it progressively more precise as the classical limit is approached}. 
With an eye towards things to come,
we remark that in systems such as fluids (that form the focus of our interest in the current work), atomic motion is largely {\it bounded}. In Eq. (\ref{bohr}), the momentum $p_{n}$ is that associated with the $n$-th state and the integral is
performed over a closed orbit (i.e., a complete periodic one dimensional semi-classical trajectory).
The integral in Eq. (\ref{bohr}) is the micro canonical classical phase space area associated with all states of energy less than $E_{n}$ 
in a system with a single remnant degree of freedom. The same holds for theories in which the $x$ coordinate is decoupled from all others. 
From Eq. (\ref{bohr}), 
the area of the phase space annulus between two consecutive values of $n$ is none other than Planck's constant $h$, 
thus establishing the appearance of this exact fundamental constant in the micro-canonical one-dimensional one particle ``ensemble" 
and all ensembles derived from it. 

It is useful to recall the relation between $J_n$ and the density of states of the quantum system \cite{Berry-Mount}.   
The action-angle variable describing the classical trajectory
\begin{equation}
 J_n=J(E_n),
\end{equation}
where, 
$E_{n}$ are the energy eigenvalues and
\begin{equation}
 J(E)=\oint dx \sqrt{2m[E-V(x)]},
 \label{JE}
\end{equation}
In Eq. (\ref{JE}), we denote the arc length along the path by $dx$. 
Putting all of the pieces together, it follows that
%
\begin{equation}
\label{Eq:DOS}
 \rho(E)=\frac{\tau(E)}{\hbar}\sum_{n} \delta\left(\frac{J(E)}{\hbar}-2\pi(n+C)\right),
\end{equation}
yields precisely the density of states provided Eq. (\ref{bohr}) holds. The period 
of the cyclic motion \cite{LL}
%
\begin{equation}
\label{Eq:tau}
 \tau(E)=\frac{\partial J(E)}{\partial E}=\sqrt{\frac{m}{2}} \oint \frac{dx}{\sqrt{E - V(x)}}.
\end{equation}
We briefly remind the readers where the very standard, yet in some disciplines not often used, expression of Eq. (\ref{Eq:tau}) for the period 
$\tau(E)$ comes from. The speed of a classical particle of energy $E$ in a potential energy field $V(x)$ is $v= \sqrt{2(E-V(x))/m}$ and the time increment required to traverse a distance $|dx|$ is $|dx|/v$. This illustrates that the period $\oint |dx|/v$ is given by Eq. (\ref{Eq:tau}). 
From Eqs. (\ref{Eq:DOS}) and (\ref{Eq:tau}) we obtain the well-known result that the classical frequency \cite{Berry-Mount},
%
\begin{equation}
 \label{Eq:nu-cl}
 \nu(E)=\frac{h}{\tau(E)}=\frac{1}{\rho_{\rm cl}(E)},
\end{equation}
where $\rho_{\rm cl}(E)$ is the classical density of states. 
Therefore, an immediate {\it corollary of the WKB 
type relations of} Eq. (\ref{bohr}) and 
the fact
that $J_{n}$ is, geometrically, the phase space volume of
states bounded by an energy $E_{n}$, is that sums over energy eigenstates $n$ can be replaced by
\begin{eqnarray}
\label{sums}
\sum_{n} \to \frac{1}{h} \int dJ = \frac{1}{h} \int dx \int dp,
\end{eqnarray}
since $J(E)$ is related to the number of states, $N(E)$, bounded by $E$ through $J(E)=hN(E)$. 
Albeit exceedingly simple, we are not aware of an earlier derivation of Eq. (\ref{sums}) with $h$
being Planck's constant that instantly follows from the WKB type
relation of Eq. (\ref{bohr}). 
Similar expressions follow relating quantum systems to their classical counterparts. 
For instance, replacing the partial derivative
in Eq. (\ref{Eq:tau}) by finite differences, $\nu_{n} \sim (E_{n+1}- E_{n-1})/(J_{n+1} - J_{n-1})$
and invoking the quantization condition of Eq. (\ref{bohr}) for the $n$-th and the $(n+1)$-th levels, we obtain
the semi-classical frequency of the $n$-th state,
\begin{eqnarray}
\label{nun}
\nu_{n} \sim \frac{E_{n+1} - E_{n-1}}{2h}.
\end{eqnarray}
The constant $C$ in Eq. (\ref{bohr}) drops out in the subtraction between $J_{n+1}$ and $J_{n-1}$. 
Eq. (\ref{nun}) applies to general potentials. To establish trivial intuition, we briefly regress to a harmonic
oscillator of resonant frequency $\overline{\omega}$, with energies $E_{n} = (n+ 1/2) \hbar \overline{\omega}$
and for which (as indeed apparent in Eq. (\ref{nun}) as it must be) 
the angular frequency of oscillations of all semi-classical levels is $\overline{\omega}$.  
Akin to Eq. (\ref{Eq:nu-cl}),
\begin{eqnarray}
\label{te}
 \tau~  \Delta E = h.
\end{eqnarray}
Here, $\Delta E$ is the average of the energy differences between the consecutive quantum levels that involve $E_{n}$ (namely, ($E_{n+1}- E_{n}$) and $(E_{n}- E_{n-1})$) 
and $\tau$ the period with the semiclassical $n-$th level. Note that this looks like the time-energy uncertainty relation of Eq. (\ref{ET})
 yet now an equality appears instead of a lower bound inequality. The above relations, exact to lowest non-trivial order in WKB, and many other similar variants that follow from the applications of these ideas (some of which will be perused in subsection \ref{new_thermo}) are exceedingly simple yet nevertheless seem to be largely new.  
 Assumed relations invoked since the beginning of quantum mechanics were of a very different nature. A century ago, Sackur \cite{Sackur,experiment} posited an equality of the form of Eq. (\ref{te}) instead of (and before the advent of) the uncertainty inequality bounds. Here, we see how precise relations may appear quite rigorously in the leading order semiclassical WKB limit. 
 We wish to emphasize that the quantization that we study in this work is no different from that of topological or other initially surprising quantum effects. Most quantization effects ranging from the Bohr atom to those attributed to topological effects can be simply understood from the quantization of the mechanical action. For instance, in the Aharonov-Bohm effect for a charge encircling a magnetic flux solenoid, the accumulated phase is the contribution to the classical action stemming from the minimal coupling of the current to the electromagnetic vector potential: $\frac{e^{*}}{\hbar c} \oint A_{\mu} dx^{\mu}$
 (where $e^{*}$ is the effective charge, $A_{\mu}$ the electromagnetic four vector potential). in the standard time independent Aharonov-Bohm effect \cite{AB}, the latter integral is performed over a closed path encircling a flux $\Phi = \oint {\bf{A}} \cdot {\bf{dr}}$; this phase is trivial only for quantized flux. This origin of the quantization may, of course, be rationalized as a generalized coherence condition within the Feynman path integral formulation of quantum mechanics (an integer winding number of the relative phase). 

To summarize, results such as  Eq. (\ref{sums}) follow as a precise equalities with $h$ being exactly equal to Planck's constant by noting that consecutive semi-classical trajectories $n$ and $(n+1)$ have, by Eq. (\ref{bohr}), a variation in their action of size $(J_{n+1}-J_{n}) = h$. This difference in the action associated with consecutive levels immediately gives rise to the fraction of $1/h$ multiplying the integrals on the righthand side of Eq. (\ref{sums}). We may similarly link the semi-classical frequencies to the energy eigenvalues of the quantum problem (Eq. (\ref{nun})). 

It is in the semi-classical high temperature (or high energy) limit with a divergent number of states $n$ that the sum over the many viable quantum states $n$ can be replaced by the continuous integral of Eq. (\ref{sums}). For a sum containing $M(=(n_{2}-n_{1})+1)$ discrete $n$ values in the sum of Eq. (\ref{sums}), there are corrections of typical relative strength ${\cal{O}}(1/M)$ augmenting the integral on the righthand side as is seen from the well known Euler-MacLaurin formula, 
\begin{eqnarray}
\sum_{n=n_{1}}^{n_{2}} f(n) \sim \int_{n_{1}}^{n_{2}} f(x) ~dx ~+~ \frac{f(n_{1}) + f(n_{2})}{2}  \nonumber
\\ + \sum_{n'=1}^{\infty} \frac{B_{2n'}}{(2n')!} (f^{(2n'-1)}(n_{1}) - f^{(2n'-1)}(n_{2})),
\end{eqnarray}
with $\{B_{2n'}\}$ the Bernoulli numbers. 
Thus, from the Euler-MacLaurin formula, in the semi-classical limit, not only is Eq. (\ref{bohr}) asymptotically exact but also the ensuing conversion from a discrete sum to a continuous integral in Eq. (\ref{sums}) {\it may emerge as an exact relation} with vanishing corrections. Albeit trivial, it is important to stress that use of the replacement of Eq. (\ref{sums}) {\it does not }mandate that
the system is in highly excited states. All that matters is that one may replace the sum over a set of states $\{|n \rangle\}$ 
by a continuous integral; the contributions from these states are smooth functions (even if the occupancy of this set of states is small).

\section{Thermalization and recurrence rates in semi-classical systems}
\label{sec:tr}
In this section, we largely focus on relaxation rates in semi-classical systems. We start by discussing relations between dynamics and thermodynamics.
As we will explain, in these equalities, classical recurrence type times (and action) appear in unison with Planck's constant and the entropy (subsection \ref{new_thermo}). We then turn to thermalization rates (subsection \ref{transition_wkb}) in general semi-classical systems. 
As we will explain in subsection \ref{transition_wkb}, the found semi-classical thermalization rates are bounded
by the recurrence rates of subsection \ref{new_thermo}. In the subsequent subsections, we explain how entropic and enthalpy differences may be exactly ascertained from 
measurements of the rates as a function of the temperatures (subsection \ref{measurements}) and discuss possible aspects of prethermalization (subsection \ref{pret}).
The relations in subsection \ref{measurements}  
will reappear when we turn to the viscosity in later sections. We conclude with speculations concerning quintessential non semi-classical systems- theories involving quantum critical points (subsection \ref{qcp:sec}). In subsequent sections we will suggest that when the semi-classical analysis no longer holds, more general inequalities hold instead of more constrained semi-classical equalities. 

\subsection{New relations between classical dynamics and thermodynamics}
\label{new_thermo}

The expression for the density of states (\ref{Eq:DOS}) can be used to compute the 
partition function of a quantum mechanical system. We divide the system into different
sectors that are linked to each other via ``one dimensional'' semi-classical dynamics along an arc as
the system evolves in time. Given initial velocities and coordinates, the system evolution is {\it uniquely determined}. 
As earlier and similar to \cite{bender}, we may parameterize the unique configuration space trajectory by the one dimensional coordinate (e.g., $x$)
and transform our high dimensional many body system (with particles $i$ of generally different masses $m_{i}$, etc.) onto one-particle systems in one dimension. Towards that end, we may rescale the coordinates ${\bf{r}}_i$ of any such particle $i$ by $\sqrt{m_i}$ (that is, ${\bf{r}}^{\sf~ rescaled}_{i} =  {\bf{r}}_{i} \sqrt{\frac{m_{i}}{m}} $ with $m$ an arbitrary mass) such that the kinetic energy of any such particle is $\frac{1}{2} m \Big( \frac{d {\bf{r}}^{\sf ~rescaled}_i }{dt} \Big)^2$. With all of the above in tow, we can express the many-particle potential energy $V (\{{\bf r}_{i}\})$ in the rescaled coordinates (denoted below by $V_{\sf ~rescaled})$. This transforms the classical Hamiltonian of the many body (MB) system, 
\begin{eqnarray}
H_{MB} = \sum_{i} \frac{{\bf p}_{i}^{2}}{2m_{i}} +  V (\{{\bf r}_{i}\}),
\end{eqnarray}
with $V$ containing external potentials as well as two- and higher-body interactions,
into an eftective one 
 along its one-dimensional arc trajectory in configuration space. The resultant effective Hamiltonian 
 \begin{eqnarray}
 \label{MBarc}
 H_{MB}^{\sf arc} &&= \frac{1}{2} m \sum_{i=1}^{N} \Big( \frac{d {\bf{r}}^{\sf ~rescaled}_{i}}{dt} \Big)^{2} + V_{\sf ~rescaled}\Big( \{{\bf{r}}_{i}^{\sf~ rescaled} \}_{i=1}^{N} \Big) \nonumber
 \\ &&= \frac{1}{2} m \Big( \frac{dx}{dt} \Big)^{2} + V_{\sf arc}(x)  \equiv \frac{p^{2}}{2m} +  V_{\sf arc}(x),
 \end{eqnarray}
 with $p$ the momentum in the dual one-dimensional system. 
 In Eq. (\ref{MBarc}), the potential energy
 $V_{\sf ~arc}$ is the function $V_{\sf ~rescaled}$ when it is parameterized along the one-dimensional arc coordinate $x$. 
 As the value of the integral of Eq. (\ref{bohr}) is quantized along the one-dimensional arc coordinate so is the integral
 \begin{eqnarray}
 \label{config:}
J_{n} = \oint_{n-th ~state} \sum_{i=1}^{N} {\bf{p}}_{i} \cdot {\bf{dr}}_{i} 
\end{eqnarray} 
in high dimensional configuration space. We very briefly discuss high dimensional extension of the standard one-dimensional quantization condition of Eq. (\ref{bohr}) in \cite{explain_highd}. We may denote the density of states within each ergodic phase space component by $\rho_{\sf ec}$ and the associated period of motion along any one-dimensional path by $\tau_{\sf ec}(E)$. 
In the classical system, $J$ is a continuous function of the energy $E$ and initial direction $\hat{n}$ (or, equivalently, initial momenta). Thus, as $E$ is increased (as well as when $\hat{n}$ varies) continuously, quantization occurs when the integral of Eq. (\ref{config:}) assumes values with continuous consecutive integers $n$. For fixed $\hat{n}$, it is trivial to see that 
$J$ is monotonic in the energy $E$ and Eq. (\ref{Eq:tau}) holds in any number of dimensions \cite{explain_J}.


Armed with Eq. (\ref{MBarc}), in what follows, we relate semi-classical dynamics and time scales therein to
thermodynamic quantities. We may write the partition function $Z={\rm tr}(e^{-\beta H})$
as a sum over all different ergodic components (${\sf ec}$) with given initial velocities (or momenta) and spatial coordinates (i.e., $2DN$ dimensional phase space coordinates) that are not connected to each other by the temporal evolution 
of the system,
\begin{eqnarray}
\label{Zsumec}
Z = \sum_{\sf ec} Z_{\sf ec}.
\end{eqnarray}
As is well known, the uniqueness of solution of Hamilton's equations of motion ensures that ($2DN$ dimensional) phase space trajectories cannot self intersect (and thus that phase space behaves as an incompressible fluid). That is, for periodic bounded motion, the phase space evolution is that along non-intersecting closed loops. 
The union of all such non-intersecting loops (the ergodic components) fills up all of phase space. 
By contrast, in $DN$ dimensional configuration space, different configuration space evolutions can overlap. 
Associated with each ergodic component in phase space, similar to Eq. (\ref{MBarc}), we may describe the spatial (-only) coordinates of the many body system by an arc coordinate $x$ in $(DN)$ dimensional configuration space as it evolves in time.  With these, for each such ergodic sector of phase space, we will have an associated partition function
\begin{eqnarray}
 Z_{\sf ec}&=&\int_0^\infty dE~ \rho_{\sf ec}(E)e^{-\beta E}
 \nonumber\\
 &=&\frac{1}{\hbar}\int_0^\infty dE~\tau_{\sf ec} (E)e^{-\beta E}\nonumber
 \\ &&\times \sum_{n=-\infty}^\infty\int_{-\infty}^\infty
 \frac{d\lambda}{2\pi}
 e^{i\lambda[J_{\sf ec} (E)/\hbar-2\pi(n+C)]},
\end{eqnarray}
where we have used the integral representation of the delta function. 
The ``single ergodic component'' period and density of states are what we have designated the simple classical density of states $\rho_{cl}$ and period $\tau(E)$ in Section \ref{sec:WKB} for a single particle in one dimension. In this section we will employ this more precise notation 
to highlight and derive relations that appear in many body systems in a high number of spatial dimensions.
Using the Poisson summation formula, the above equation can be rewritten as,
\begin{eqnarray}
 Z_{\sf ec}&=&\frac{1}{h}\int_0^\infty dE\tau_{\sf ec} (E)e^{-\beta E}
 \nonumber\\
 &+&\frac{2}{h}\sum_{n=1}^\infty\int_0^\infty dE\tau_{\sf ec} (E)e^{-\beta E} \nonumber
 \\ &&\times \cos\left\{n\left[\frac{J_{\sf ec}(E)}{\hbar}
 -2\pi C\right]\right\}.
\end{eqnarray}
Since at high temperatures the zero mode yields the largest contribution, the first term of 
the equation above yields a good approximation in this regime, namely
\begin{equation}
\label{Zperiod}
 Z_{\sf ec}\approx \frac{1}{h}\int_0^\infty ~dE~\tau_{\sf ec} (E)~e^{-\beta E}.
\end{equation}
This relation is new and may be quite illuminating in general problems. 
An equivalent derivation of Eq. (\ref{Zperiod}) is given by the following change of
variable along with the use of the equation $\nu = \partial E/ \partial J$, i.e., semi-classically, 
\begin{eqnarray} 
Z_{\sf ec}=\frac{1}{h} \int dJ_{\sf ec} ~ e^{-\beta E} \nonumber
\\ = \frac{1}{h} \int \frac{dJ_{\sf ec}}{\partial J_{\sf ec}/ \partial E} ~ \frac{\partial J_{\sf ec}}{\partial E}
 e^{-\beta E} \nonumber
 \\ = \frac{1}{h}
\int dE ~\tau_{\sf ec}(E) ~ e^{-\beta E}.
\label{Zperiod*}
\end{eqnarray}
If we define the ``total'' period $\tau_{tot}(E)$ by
$\tau_{tot}(E) \equiv \sum_{\sf ec} \tau_{\sf ec}(E)$
and note that total density of states is given by the sum of the density of states over all disjoint ergodic components, $\rho_{tot}(E) \equiv \sum_{\sf ec} \rho_{\sf ec}(E)$ (or one single-dimensional particle density of states $\rho_{cl}$) then we will see that the semi-classical period is given by the density of states, 
\begin{eqnarray}
\tau_{tot} (E) \approx \frac{\rho_{tot}(E)}{h}.
\label{ES}
\end{eqnarray} 
That is, with the above definitions, by Eqs. (\ref{Zsumec},\ref{Zperiod*}), 
\begin{eqnarray}
Z = \frac{1}{h} \int dE ~\tau(E)~ e^{-\beta E} = \int dE~ \rho_{tot}(E)~ e^{-\beta E}.
\end{eqnarray}
Applying an inverse Laplace transform on the above yields Eq. (\ref{ES}).
This equation constitutes a generalization of Eq.(\ref{Eq:nu-cl}) to large macroscopic systems.
The density of states is set by thermodynamic entropy $S(E)$. That is the entropy is defined by $\rho_{tot}(E) \equiv e^{S(E)/k_{B}}$, with $\rho_{tot}(E)$ the total number
of microstates with energies on the interval $[E,E + \delta E]$ where the arbitrary interval $\delta E$ is fixed. Thus, we have that
\begin{eqnarray}
\label{entropy!}
\boxed{(\delta E)( \overline{\tau}_{tot}(E)) = h e^{\frac{S(E)}{k_{B}}}}
\end{eqnarray}
with $\overline{\tau}_{tot}(E)$ the average (over energies) of periods, or recurrence times, of all orbits with energy in the interval $[E,E+\delta E]$ summed over all ergodic components. Eq. (\ref{entropy!}) {\it universally relates semi-classical dynamics (namely, the accumulated period} $\tau_{tot}(E)$ {\it summed over all ergodic components) to thermodynamics (the entropy) and explicitly illustrates how Planck's constant relates the two}. 
Regardless of the complexity (or simplicity) of the system dynamics, even if there are multiple particles and the physical system resides in many spatial dimensions, if the orbits are bounded and simple periodic, Eqs. (\ref{Zperiod},\ref{ES},\ref{entropy!}) must hold. 
These, as far as are aware, simple new relations connect the period of orbits in semi-classical dynamics to the partition function sums and associated entropies. When ergodic dynamics are present, the system evolves through all states in a given ergodic component and the periods $\tau_{\sf ec}(E)$ may be larger than those in solvable systems with simple cyclic evolution. In integrable systems, conserved quantities restrict the number of points 
(or volume size of space) that may be related to each other by a dynamical evolution. An important requisite of the above derivation is that the period $\tau(E)$ along each path starting from a given initial configuration space point and velocity direction $\hat{n}$ is a continuous function of the energy $E$. That is, the partial derivative of Eq. (\ref{Eq:tau}) is indeed equal to the period $\tau$ in any number of spatial dimensions (see, e.g., \cite{explain_J}). If the left and right derivatives of $J(E)$ (given a fixed initial coordinates and velocity direction) as a function of $E$ are ill defined or do not match
with each other then Eq. (\ref{Zperiod*}) need not hold. Thus, Eq. (\ref{entropy!}) is valid for general non chaotic systems (or, more generally, over regimes in which they are non chaotic). 
As our interest in this work is not in turbulent fluid dynamics, this restriction and similar ones like it will not be of pertinence.

Let us now consider the average period rate $r_P=\langle \tau^{-1}(E)\rangle$ associated to the period $\tau(E)$. 
Thus, 
\begin{eqnarray}
 r_P&=&\frac{e^{\beta F}}{h}\int_0^\infty dEe^{-\beta E}
 \nonumber\\
 &+&\frac{2e^{\beta F}}{h}\sum_{n=1}^\infty\int_0^\infty dE e^{-\beta E}\cos\left\{n\left[\frac{J(E)}{\hbar}
 -2\pi C\right]\right\}
 \nonumber\\
 &=&\frac{k_BT}{h}e^{\beta F}
 \nonumber\\
 &+&\frac{2e^{\beta F}}{h}\sum_{n=1}^\infty\int_0^\infty dE e^{-\beta E}\cos\left\{n\left[\frac{J(E)}{\hbar}
 -2\pi C\right]\right\},
\end{eqnarray}
where $F$ is the free energy of the system. Once more, in the high-temperature regime the zero mode dominates and we obtain, 
\begin{equation}
 r_P\approx\frac{k_BT}{\int_0^\infty dE\tau_{tot}(E)e^{-\beta E}}.
\end{equation}
The above result has been obtained assuming an equilibrium situation. When shear forces are present such that the original 
potential is modified like in Fig. \ref{Potential}, non-equilibrium transition 
processes from one potential well to the other are induced. In such a case,  
unitarity, which is a fundamental quantum mechanical 
property, relates different transition rates in a way independent of the temperature. Indeed, as discussed by 
Weinberg \cite{Weinberg-Book}, unitarity implies many of the most fundamental results of statistical mechanics, including 
a derivation of Boltzmann's H-theorem without assuming time-reversal invariance. This follows from the fact that by assuming 
unitarity time-dependent probability distributions obey a very general master equation which depends on the transition rates 
between initial and final states. For example, if we denote by $N(E,t)$ the non-equilibrium particle distribution 
for a given potential in the presence of shearing, the transition rate for increasing the number of particles on one well, $r_i$ is 
related to the transition rate for decreasing the number of particles on the other well, $r_d$, by $r_d=e^{\beta E}r_i$ \cite{Kobes}. 
Therefore, the deviation from equilibrium particle distribution at time $t$ is given by,  
\begin{equation}
\label{decay'}
 \delta N(E,t)=\delta N(E) e^{-r(E)t},
\end{equation}
where $r(E)=r_d-r_i=(1-e^{-\beta E})r_d$ and $\delta N(E)$ is the particle fluctuation at $t=0$. Evidently, at high temperature 
we have, 
\begin{equation}
 r(E)\approx\frac{E}{k_BT}r_d.
\end{equation}
Since $r_d$ yields the probability per unit time for decreasing the number of particles from the initial state, it should 
be proportional to the Boltzmann factor $e^{-\beta(E_{\rm escape}-E_{\rm initial})}$, with $E_{\rm escape}$ and 
$E_{\rm initial}$ as shown schematically in Fig. \ref{Potential}. Thus, simple dimensional considerations 
imply naturally that $r_d\sim k_BTe^{-\beta(E_{\rm escape}-E_{\rm initial})}/h$. In the following we will substantiate 
further the unitarity arguments given here. The main message we want to convey is that unitarity is the main mechanism 
determining particle flow for non-equilibrium systems. For liquids this additionally leads to transport 
coefficients which depend explicitly on the particle density in the case of non-relativistic systems. 

Although Eq. (\ref{entropy!}) applies to non chaotic systems, we can trivially write down other relations that link entropy to dynamics in all instances
(including theories with chaotic dynamics). Towards this end, we may invoke earlier equalities \cite{Fl,badiali1} connecting dynamics to system entropy in the Feynman path integral representation, e.g., a simple relation for the entropy,
\begin{eqnarray}
\label{sfeyn}
S = k_{B} \ln \Big[  \frac{1}{N!} \int dx(0) \int Dx(t) e^{- \frac{1}{h} \int_{0}^{\beta \hbar} dt  [H -U]} \Big],
\end{eqnarray}
for a system of $N$ identical particles,
where the classical Hamiltonian $H$ and $U=U[x(t)]$ is the potential energy evaluated for closed classical trajectories of period $\beta \hbar$. In other words, $x(0) = x(\beta \hbar)$. 
To cast this correspondence in our form, we observe \cite{explain_J} that for closed periodic orbits the integral $ \int_{0}^{\tau(E)} dt  [H -U]= \frac{J(E)}{2}$ is none other than 
exactly half the {\it classical action} associated with each closed path. Thus we may turn Eq. (\ref{sfeyn}) around and link the weighted sum of all exponentiated actions $\exp(-J_{\alpha}/h)$ over all paths $\alpha$ of recurrence time $\tau$
to the entropy at a temperature $T = \frac{\hbar}{k_{B} \tau(E)}$, 
\begin{eqnarray}
\label{exact!}
S(T=\frac{\hbar}{k_{B} \tau(E)}) = k_{B} \ln \Big[  \frac{1}{N!} \int dx(0) \int Dx(t) e^{- \frac{J(\tau(E))}{h}} \Big].
\end{eqnarray}

\subsection{Thermal transition rates via WKB}
\label{transition_wkb}

\begin{figure}[htp] 
\centering
\includegraphics[width=7cm]{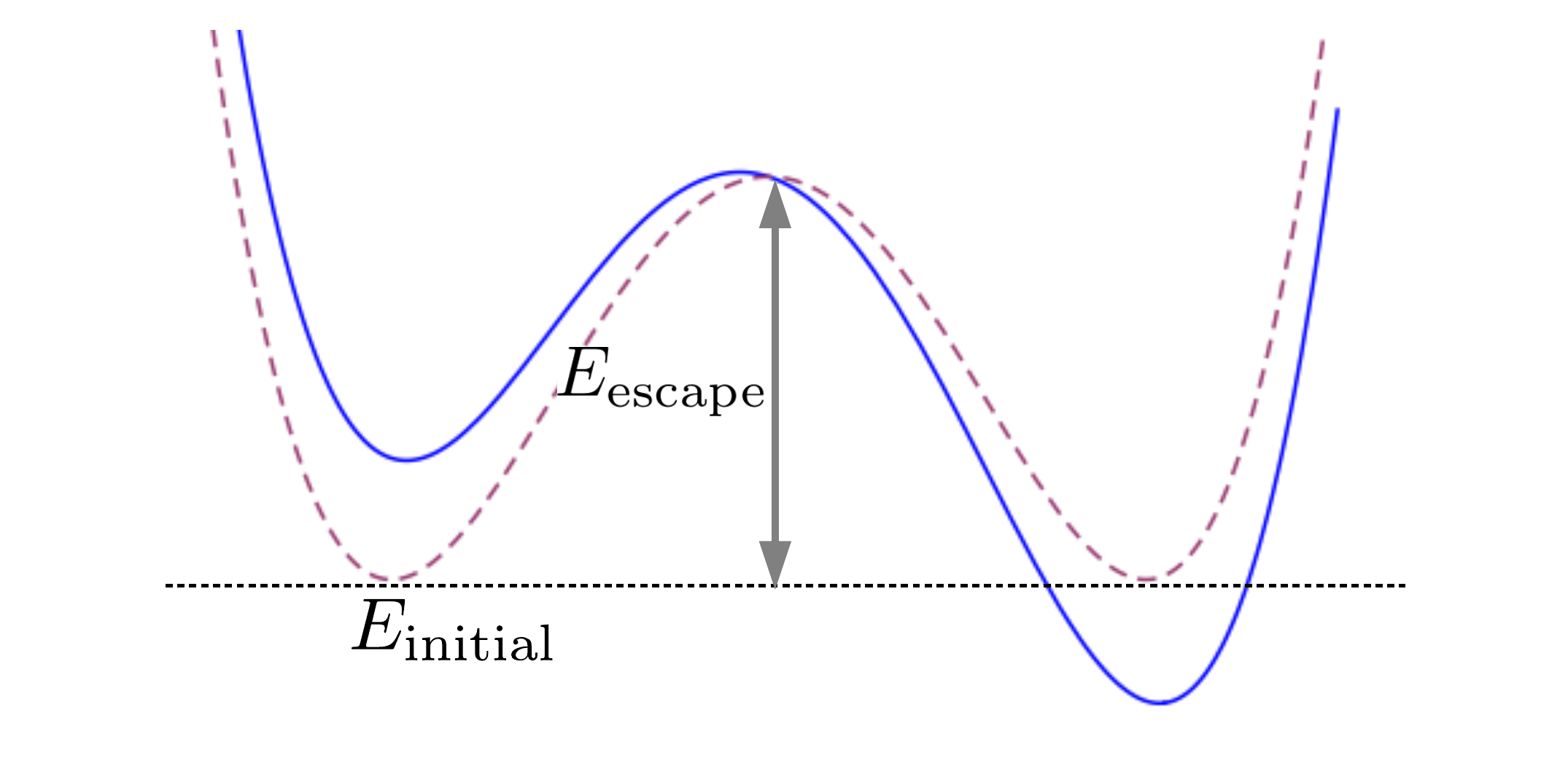}
\caption{A cartoon of the transition process from an $initial$ state lying anywhere within the basin on the left 
to a $final$ state (this state may have a lower or a higher energy as compared to the $initial$ state) on the right. In this cartoon, for the transition to occur, a threshold escape energy
needs to be exceeded.}
\label{Potential}
\end{figure}

In this subsection, we will revisit an old problem- that of transition rates that has been looked by numerous researchers, e.g., \cite{kramers,langer,hanggi,eyring-tst,wigner,miller,wolynes,WSW,badi2}.
Our aim to motivate possible broad rigorous results and to understand their content in a manner that is free from saddle point jargon concerning molecules with specific interactions and transition rates as often
employed in physical chemistry.  In later sections, we will employ the below results to suggest that these might hold for complex systems such as multi-component metallic fluids where nearly exact quantization might appear. In this subsection, we will widely discuss thermalization times in semiclassical systems (for which well defined particles or quasi-particles are present). By a general deformation of the many body Hamiltonian, we will rederive a well known result (Eq. (\ref{r(T)})) for the particular aforementioned physical problem of fluid dynamics that we wish to address. The result will suggest, for many body systems in general dimensions, a possible limiting functional form for the transition rates when these are extrapolated to high temperatures (Eq. (\ref{rTL})). 
Towards that end, we will employ the one-dimensional Hamiltonian of Eq. (\ref{MBarc})
describing the unique evolution of the system once an initial spatial coordinates and velocities are prescribed. 
As we described earlier, 
Eqs. (\ref{Eq:tau}) 
and
(\ref{sums})
provide the means to compute the oscillation frequency 
and quantities relying on it within
the semi-classical limit. We now combine this relation with 
an expression for acceptance rate for a transition from a complete set of $initial$ states to those in $final$ states
(that may be of higher or lower energy). The well known problem is sketched in Fig. \ref{Potential}.
Although our formulation is somewhat different, the above logic emulates the considerations laid out by Eyring \cite{eyring-tst} and many subsequent works in transition state theory \cite{hanggi,miller}. 

 In the current derivation, we assume, in a semi-classical spirit, that once the system has a sufficiently large energy $E>E_{escape}$ then
the system may overcome the energy barrier and go to a $final$ lower energy state, see Fig. \ref{Potential}.
As sketched in this figure, at low temperatures the $initial$ energy $E_{initial}$ of the system may, trivially, be close to the bottom of the potential well. 
We will assume that initial the semi-classical system is in thermal equilibrium (and is described 
by a coordinate $x_{\rm initial}$ in configuration space). We can express the initial quantum mechanical state
via a superposition of the Hamiltonian eigenstates ($|n \rangle$). The system (and each of these eigenstates) evolves simply in time
until at at a later ``$final$'' time, the system is equilibrated anew. At the ``$final$'' location $x_{\rm final}>x_{\rm initial}$,
individual original eigenstate superposition is changed via contact with the heat bath. The distance $L=|x_{\rm final} - x_{\rm initial}|$
is the thermalization length scale. In a free system, $L$ will be set by the thermal de Broglie wavelength $\lambda^{nr}_{T}$ 
or mean free path in a dilute gas beyond which momentum eigenstates are changed due to collisions. The thermalization length determining the density of photons and other scatters may generally depend on the temperature, i.e., $L=L(T)$. As the reader can appreciate from our discussion, this length is not an exact constant and one may anticipate a distribution of possible $L$ values. However, as we will see in the calculation below, for a fixed such length, the value of $L$ will cancel. It is important to highlight that
$L$ is bounded from above in any system. In liquids and elsewhere, when potential energy effects may be important, as in the caricature of Fig. (\ref{Potential}), and the system gets trapped in energy wells, the $final$ location at
which thermalization may occur anew may be associated with the bottom of the wells at which the system can spend much time before
ultimately veering elsewhere. Semi-classically, the effective one-dimensional system can veer towards the lower energy $final$ 
state if the momentum $p>0$. Although in Fig. (\ref{Potential}) we sketch one particular path, the simple calculation that we will detail
is for any trajectory that leads to thermalization (requiring a threshold energy $E_{escape}$). In an unbiased equilibrated system, the rate 
of transitions from left to right is equal to that from right to left. We arbitrarily focus on one of the two directed paths. 

As emphasized in subsection \ref{new_thermo}, unlike the phase space trajectories that cannot self-cross or overlap, when momenta are no longer coordinates (as indeed
in configuration space), there may be several configuration space trajectories that overlap. In particular, especially at low energies in which the potential energy is significant and 
the system need not evolve along a fixed momentum direction, there may be different configuration space motions, of different momenta, link $x_{initial}$ to $x_{final}$. In what follows we will need to sum over all
such paths (all of the aforementioned positive momenta 
$p>0$ along the arc coordinate ($x$) and of energies $E>E_{\rm escape}$) that link the initial and final states. We ignore any tunneling between the $initial$ and $final$ states via intermediate states of energy lower than $E_{escape}$ and similarly disregard quantum borne reflections occurring at energies larger than $E_{escape}$.  
 Within any state $n$, the particle spends, on average, a time given by $\nu_{n}^{-1}/2$ before transitioning into the lower energy $final$ state.
This is so as the motion between $x_{\rm initial}$ and $x_{\rm final}$ constitutes half of a full period
if infinite potential barriers existed at $x=x_{\rm initial}$ and $x_{\rm final}$.
For computing this time, we invoke Eq. (\ref{Eq:tau}) for a system in which the original potential $V$ was {\it modified} only for~$x<x_{\rm initial}$ and $x>x_{\rm final}$ by the insertion of an infinite potential barrier. 
Thus, we introduce a hard wall deformed Hamiltonian,
\begin{eqnarray}
  H^{transition} \equiv \begin{cases}
    H, & \text{if $x_{initial} \le x \le x_{final}$}.
    \\ \infty, & \text{otherwise}.
  \end{cases}
  \label{Ht}
\end{eqnarray}
We may always place such reflecting hard walls irrespective of how complicated (or chaotic) the classical one-dimensional arc trajectory is. 
Such a modification of the Hamiltonian will not alter the classical ``time of flight'' between $x_{\rm initial}$ and $x_{\rm final}$ 
with the true potential (which does not diverge at $x=x_{\rm final}$). That this is true is evident by, e.g., 
examining half of Eq. (\ref{Eq:tau}). The integral 
of Eq. (\ref{Eq:tau}) will clearly not be affected by change of the potential
$V(x)$ only at point $x=x_{\rm final}$. Putting a hard wall boundary conditions corresponds to a fixed value of $C$ in Eq. (\ref{bohr}) for all levels and just as for the undeformed system,  the measure $dJ$
of Eq. (\ref{sums}) is unchanged. Similarly, when invoking Eq. (\ref{sums}), to rewrite quantities in terms of semi-classical integrals over restricted phase space domains, the classical energy $H(x,p)$ will be unchanged for all $x_{\rm initial} < x< x_{\rm final}$  and $p>0$. To avoid confusion, we should state that even though the hard wall replacement of Eq. (\ref{Ht}) clearly does not change the leading order semi-classical trajectories, it will alter higher order quantum corrections. Our objective here is, however, to obtain the leading order finite temperature behavior in equilibrated liquid motion. 

In a quantum mechanical setting, the frequency associated with moving from $x_{initial}$ to $x_{final}$ is given by the trace 
\begin{eqnarray}
\label{TRP}
Tr \{P ~\hat{\rho}~ (v/L)~ P\},
\end{eqnarray} 
with $v$ the velocity operator, $\hat{\rho}$ the density matrix to be described below, and $P=P_{1} P_{2}$ is the product of two projection operators over space and momentum direction. Specficallly, $P_{1}$ is the projection onto the real space states corresponding to $x$ in the interval, namely,
\begin{eqnarray}
P_{1} = \sum_{x_{initial} \le x \le x_{final}} |x\rangle \langle x |,
\end{eqnarray}
and $P_{2}$ is the projection onto states with positive momenta,
\begin{eqnarray}
P_{2} = \sum_{p>0} |p \rangle \langle p |.
\end{eqnarray}
We will effectively work in the basis of eigenstates of the projected Hamiltonian $P H P$ in the interval $[x_{initial}, x_{final}]$ (emulating $H^{transition}$) and sum over those states $n$ of high enough energies that enable motion from the left to right end of the interval. 
At length scales beyond $L$, thermalization scatters the states amongst themselves and the simple calculation below assuming that the states $|n \rangle$ do not change as the system evolves between and $initial$ and $final$ location will become incorrect. We thus consider a nearly equilibrated system, for which on the left of Fig. \ref{Potential} (at $x_{initial}$), there are numerous impinging positive momentum states
 of energy $E_{n}$ that have an energy $E_{n}>E_{escape}$ and for which $v_{n}$ is such that the semi-classical frequency 
 $\nu_{n} = (\partial E/\partial J)_{E=E_{n}}$. The velocity $v_{n}$ is related to the frequency $\nu_{n}$ via $v_{n} = \nu_{n} L$.
 The result of this slightly different physical case (that of already an equilibrated system) largely mirrors our earlier analysis.
 If the system defined by Eq. (\ref{Ht}) may, be described by a normalized wavefunction inside the box $x_{initial} \le x \le x_{final}$ (or normalized unit trace density matrix over such 
 wave functions), one may view it as having one ``generalized configuration space particle'' in this region initially. 
 An initial state is in contact with a heat bath at a temperature $T$. The thermal occupancy $\langle {\sf{n}}^{\sf configuration~space}_{n} \rangle$ of each of the configuration space states $n$ is set by $(\hat{\rho})_{nn}=\exp(-\beta E_{n})/{\cal{Z}}$ with ${\cal{Z}}$ the partition function (with all off-diagonal elements of the density matrix $\hat{\rho}$ vanishing in the 
 $\{|n \rangle\}$ eigenbasis). The occupancy can be related to very standard forms for the ``single particle'' distribution 
 \begin{eqnarray}
 \label{unit}
 &&\langle {\sf{n}}^{class.}_{n} \rangle = e^{-\beta(E_{n} - \tilde{\mu}_{\sf config.})}, 
 \\ &&\langle {\sf{n}}^{F-D}_{n} \rangle = \frac{1}{1+ e^{\beta(E_{n} - \tilde{\mu}_{\sf config.})}}, \nonumber
 \end{eqnarray} 
 a semi-classical Boltzmann distribution relevant to our analysis  and a Fermi-Dirac  ($F-D$) distribution occupancy (added here only for conceptual understanding and comparison to known transport problems). The distributions of Eq. (\ref{unit}) are normalized and adhere to unitary dynamics  \cite{explain-TST}.
 In the current case of a ``generalized particle'' describing the state of the system in high-dimensional configuration space, the fugacity 
 \begin{eqnarray}
 \label{mueq}
 z= \exp(\beta \tilde{\mu}_{\sf config.}) = 1/{\cal{Z}},
 \end{eqnarray}
 ensures normalization of the Boltzmann distribution, $\sum_{n} \langle {\sf{n}}_{n} \rangle =1$. To avoid cumbersome notation, we will denote the effective configuration space 
 $\tilde{\mu}_{\sf config.}$ by $\mu$. 

 Putting all of the pieces together, the semi-classical transition rate $r$ from the $initial$ to $final$ state
(or, more precisely, between sets of such states containing $x_{initial}$ and $x_{final}$) is
\begin{eqnarray}
\label{r(T)}
r(T) = \sum_{E_n>E_{\rm escape}} (2 \nu_{n})~ \langle {\sf n}_{n}(T) \rangle~ \Theta( \langle p_{n} \rangle).
\end{eqnarray}
We remark that simple intuitive notions such as those of caging are trivially accounted for in Eq. (\ref{r(T)}) \cite{caging}.

The system will not reach $x_{final}$ unless it is excited at some time step into any one of the high energy states $n$ with $E_{n}>E_{escape}$. 
The Heaviside function $\Theta(\langle p_{n} \rangle)$ ensures that the momentum $p$ along the trajectory arc is positive. 
In Eq. (\ref{r(T)}), the factor of two multiplies $\nu_{n}$ as the time of flight from $x_{initial}$ to $x_{final}$
is, as stated above, half that of an entire period, and thus the frequency of the motion from $x_{initial}$ to $x_{final}$ is
double the frequency associated with a full oscillation between these two turning points of $H^{transition}$.
We next invoke the simple correspondence of Eq. (\ref{sums}) for half of the states $n$ 
(viz., those with positive
momentum $p$ in Eq. (\ref{sums})), and replace the frequency $\nu_{n}$ 
by $\partial H^{transition}/\partial J$ evaluated at an energy $E=E_{n}$ (Eq. (\ref{Eq:tau})) implying a uniform sign of this quantity) applied to the deformed classical potential with hard wall boundaries. 
The shape of the high dimensional trajectory in configuration space varies with energy. We will first focus on a one-dimensional problem with a fixed arc direction and then explain how our results may change and lead to Gibbs free energy differences when the shape of the classical configuration space trajectory varies.
The total rate  $r(T)$ (the rate is the total ``particle'' current impinging from left to right divided by distance $L$ between the initial and final locations,  $r(T)=\langle j \rangle/L = \sum_{E_n>E_{\rm escape}} \langle j_{n} \rangle/L$), 
 we very explicitly have
 \begin{eqnarray}
 \label{particle_derivation}
 r(T) &=& \sum_{E_n>E_{\rm escape}} (2 \nu_{n})~ \langle {\sf{n}}^{class.}_{n} \rangle~ \Theta( \langle p_{n} \rangle) \nonumber
 \\ &=&  \frac{1}{h} \int dJ \frac{\partial H^{\rm transition}}{\partial J} 
e^{-\beta(H^{\rm transition} - 
\mu)}\nonumber 
\\ &=& \frac{1}{h} \int_{E_{\rm escape}}^{\infty} dH^{\rm transition} e^{-\beta(H^{\rm transition} -  \mu)}
\nonumber
\\ &=& \frac{k_{B} T}{h} e^{\beta(\mu - E_{\rm escape})}.
\end{eqnarray}
We reiterate that prior to its conversion to an integral via Eq. (\ref{sums}), in the original sum of Eq. (\ref{r(T)}) (the top line above), a factor of $(1/2)$ (that, in the aftermath, cancels the prefactor of two in $(2 \nu_{n})$) arises from the $ \Theta( \langle p_{n} \rangle)$ term; this factor of a half is reminiscent to that present in standard textbook calculation of the pressure of an ideal gas on 
wall containers (with only a half of the particles moving with a positive velocity component so as to impinge on a given wall). 
In the integration above, we  effectively set the upper cutoff on the energy to be infinite. In reality, of course, an ultra violet cutoff would bound the highest possible rate 
and set it equal, at infinite temperatures, to $(E_{\rm highest}  - E_{\rm escape})/h$. We will, however, assume that we are the liquid state and the highest possible energy 
$E_{\rm highest}$ far exceeds any measured temperature scale and the escape energy. For completeness, we remark that if the Fermi-Dirac function in the second line of Eq. (\ref{unit}) is used then $r(T)= \frac{k_{B} T}{h} \ln (1+ e^{-\beta (E_{escape}- \mu)})$; when  $k_{B} T \ll (E_{escape}- \mu)$, this latter form coincides with that of Eq. (\ref{particle_derivation}). For any given momentum direction, there is a minimal energy $E_{escape}$ required for the system to transition (including those paths for which the momentum direction changes due to later scattering with the potential walls and a ``recrossing'' occurs).


 When $(E_{\rm escape}- \mu(T))$ becomes comparable to $(k_{B} T)$, the relative weight $q \equiv \sum_{E_{n} \ge E_{\rm escape}} \langle {\sf{n}}^{class.}(T) \rangle = {\cal{O}}(1)$ and the particle is no longer confined in the original system (not that defined by 
Eq. (\ref{Ht}) which we used for computational convenience). In such a case most of the system weight resides
in nearly continuum free particle states. Therefore within the liquid phase, $q$ is bounded by a small finite number less than unity. Within the liquid phase, $(E_{escape} - \mu(T))$ is bounded. An extrapolation from the liquid phase to high temperature may thus keep $(E_{escape}- \mu(T))$ bounded by a temperature independent value. 
One may conjecture that the barrier $E_{\rm escape} = {\cal{O}}(k_{B} T_{\sf evaporation})$ \cite{qv} where
$T_{\sf evaporation}$ is the transition temperature of the liquid into a gas. This may be so as in a gas, particle motion is no longer confined. 
Within such an interpretation, those energetic atoms with $E>E_{\rm escape}$ are analogous to those of a gas; correspondingly, the length $L = |x_{\sf initial} - x_{\sf final}|$  
plays the role of a mean free path in a gas \cite{evaporation_footnote}. 

For the applications that we consider in later sections concerning the viscosity or conductivities associated with the response of the system
along a specific direction, e.g., the application of elastic shear or electric field along one direction, the pertinent dynamics will largely be one-dimensional
(albeit with a possible distribution of escape energies $E_{escape}$). In all of these cases, the system will tend to evolve along the direction of externally applied field or stress and the one-dimensional calculation of Eq. (\ref{particle_derivation}) will be directly relevant.

We now turn to dynamics in higher dimensions. The rate $\nu$ can be associated with the motion of a single particle with the coordinates (and velocities of all other particles held fixed) or it may similarly be defined for the rate of motion along the many-body initial and final configurations in the high dimensional configuration space. In both cases, if we define as $x$ as the coordinate along the respective arc traced by (i) a {\it single particle} or (ii) the {\it many particle system in which $x$ denotes an arc coordinate along a trajectory in a high dimensional configuration space} (as in Eq. (\ref{MBarc})) within a given ergodic component, Eq. (\ref{Eq:tau}) will hold. In case (ii), in the notation of Eq. (\ref{MBarc}), $V(= V_{\sf arc}$) will represent the rescaled many body potential $V(\{ {\bf{r}}_{i} \}_{i=1}^{N})$ of the $N$ particles as the system evolves in time along the arc parameterized by $x$ in $(DN)$ dimensional configuration space. In all instances, once the direction $\hat{n}$ of the single-particle velocity (or, equivalently, the set of all velocities of the $N-$ particle system in $DN$ dimensional configuration space)
is given along with the energy $E$ and initial spatial coordinate(s), the semi-classical one-dimensional trajectory of the system is uniquely defined. 
In what follows, we explicitly consider what occurs when the system may transition along not a particular one-dimensional path from, e.g., right to left, as above but rather along all possible different directions. 
We now consider motions in a general $d (=DN)>1$ dimensional space.
The product of the complete set of eigenstates $\{|m_{\hat{n}_{a}} \rangle\}$ corresponding to one-dimensional problems
along $d$ orthogonal Cartesian directions $\{{\hat{n}}_{a}\}_{a=1}^{d}$  trivially form a complete eigenbasis for any wave function.
That is, 
\begin{eqnarray}
\label{standard}
\langle x_{1}  x_{2} \cdots x_{d} | \psi \rangle =  \sum_{m_{1} \cdots m_{d}} {\cal{A}}_{m_{1} \cdots m_{d}} \Big(  \langle x_{1}  |m_{1} \rangle \langle x_{2} |m_{2} \rangle  \nonumber
\\ \cdots
\langle x_{d}  |m_{d} \rangle \Big),
\end{eqnarray}
with $\{ {\cal{A}}_{m_{1} \cdots m_{d}} \}$ normalized amplitudes, and ensuing density matrices well approximated in the basis spanned by such states. Entangled states (such as those that we will further discuss in subsection \ref{qcp:sec}) may generally require the appearance of numerous non-vanishing amplitudes $\{ {\cal{A}}_{m_{1} \cdots m_{d}} \}$. 
 If  $d'>d$ non direction-orthogonal directions $\{\hat{n}_{a}\}_{a=1}^{d'}$
are chosen then the basis spanned by the tensor products of the form $\{|m_{1} \rangle \otimes |m_{2} \rangle \otimes \cdots  \otimes |m_{d'} \rangle\}$ 
will be over complete; a unique decomposition of the type in Eq. (\ref{standard}) will not be possible.
Contrary to entangled states, as in transition state theory, in semi-classical systems, effective excited modes (associated with the semi-classical paths) that cross the barrier may, to a good (saddle point type) approximation,
be expressed as a sum of states evolving along different directions $\hat{n}$ in configuration space,
\begin{eqnarray}
\label{semi-classical-state}
| \psi \rangle \approx a_{1} | \psi_{\hat{n}_{1}} \rangle + a_{2} | \psi_{\hat{n}_{2}} \rangle + ... + a_{l} | \psi_{\hat{n}_{l}} \rangle,
\end{eqnarray}
with $l$ a finite integer. Semi-classically, each wave function $|\psi_{\hat{n}} \rangle$ is a caricature of the system initially evolving in along the direction $\hat{n}$ (whose later evolution is set by the classical equations of motion). Excited semi-classical states may thus be expressed in the ``overcomplete'' basis as $|\psi \rangle = \sum_{\hat{n}} a_{\hat{n}}~  |\psi_{\hat{n}} \rangle$, with amplitudes $a_{\hat{n}}$ (or measure $a(\hat{n})$ in the continuum limit) that trivially adheres to normalization. If the semi-classical states $| \psi_{\hat{n}} \rangle$ are localized along ray ${\hat{n}}$ 
and thus $\langle \psi_{\hat{n}} | \psi_{\hat{n}'} \rangle =0$ is ${\hat{n}} \neq  \pm {\hat{n}}'$, the normalization condition becomes $\sum_{\hat{n}} |a_{\hat{n}}|^{2} =1$, or in the continuum limit
of numerous semi-classical paths,
\begin{eqnarray}
\label{normalize}
1= \int d \hat{n} ~ |a(\hat{n})|^{2}.
\end{eqnarray} 
The calculation proceeds identically for each direction $\hat{n}$ \cite{identical_n}.

As $E$ increases for a fixed momentum direction $\hat{n}$ in configuration space, the trajectories will
define a curvilinear {\it plane}. For each semi-classical path along a ray, the
action must be an integer multiple of Planck's constant $h$, Eq.
(\ref{config:}) as seen from the WKB solution to the effective
one-dimensional problem in the arc coordinates for a specific energy $E$ and
direction $\hat{n}$ defining self-consistently the potential $V$ (and the WKB
approximation to the eigenvalue problem with the energy $E$) or
from the requirement that the paths add coherently in the high dimensional
formulation of \cite{explain_highd}. If there are many possible rays having their own
final locations $x_{final}$, this rate will have identical form along each such ray (viz., the rate
of motion from an initial point to a set of final points attained by
evolution along different initial momentum directions will always have the
same form of Eq. (\ref{r(T)})). As along each initial ray direction $\hat{n}$, one-dimensional results are obtained, in the sum we will have their weighted superposition.
In thermal equilibrium, we may (ensuring unitarity and normalization) average over all different rays, to obtain the relaxation rate,
\begin{eqnarray}
r(T) = \frac{k_{B} T}{h} \frac{\sum_{\hat{n}} e^{-\beta (E_{escape~ \hat{n}} - \mu_{\hat{n}})} e^{-\beta \mu_{\hat{n}}}}{\sum_{\hat{n}} e^{-\beta \mu_{\hat{n}}}} \nonumber
\\ = \frac{k_{B} T}{h} \frac{\sum_{\hat{n}} e^{-\beta E_{escape~\hat{n}}}}{\sum_{\hat{n}} e^{-\beta \mu_{\hat{n}}}},
\end{eqnarray}
where we used the fact that the $e^{-\beta \mu_{\hat{n}}}$ is the sum of Boltzmann weights associated with ray direction $\hat{n}$.
(For continuous directions $\hat{n}$, the sum, naturally, becomes an integral.)
The total partition function of the system is the sum over all rays (in our assumption for the form of the wave functions which are linearly independent along all directions),
\begin{eqnarray}
Z_{transition} \equiv \sum_{\hat{n}} e^{-\beta \mu_{\hat{n}}}.
\end{eqnarray}
This is the total partition function as we are summing over all rays and energies and thus are summing over all states. 
Similarly, we can regard the sum
\begin{eqnarray}
Z_{escape} \equiv \sum_{\hat{n}} e^{-\beta E_{escape~\hat{n}}} 
\end{eqnarray}
as a fictive partition function associated with escape energies along different rays.
The average rate summed over all paths (or rays) then becomes
\begin{eqnarray}
\label{ratio}
r(T) = \frac{k_{B}T}{h} \Big( \frac{Z_{escape}}{Z_{transition}} \Big) \nonumber
\\ \equiv \frac{k_{B} T}{h} e^{-\beta \Delta G}.
\end{eqnarray} 
Here, 
\begin{eqnarray}
\label{dg}
\Delta G = G_{escape} - G_{transition}
\end{eqnarray}
is the difference between the Gibbs free energy associated with the entire system and that associated with the fictive system defined by the escape energies.
This energy difference corresponds to the thermalization process. 


In situations in which the transition would have been between two different materials with varying chemical potentials
(for which the $initial$ and $final$ states would lie in regions with different local chemical potentials), the escape energy
$(E_{\rm escape}- E_{\sf initial})$ could be associated with a ``work function'' reflecting a difference in local chemical potentials.
Such a local shift in the chemical potentials is akin to a variation in the Gibbs free energies per ``generalized configuration space particles"
between the left- and right-hand sides of Fig. (\ref{Potential}). This may also emulate problems involving chemical reactions in which
the $initial$ and $final$ states may have very different chemical potentials.  
In our above calculation however, we are considering the very different situation of a regular system- the chemical potential is largely 
uniform in space. The application of an external field, stress (which we will return to in later sections), or doping may lead to an effective change in the local chemical potential as it does in semiconductors and countless other systems, e.g., \cite{AM}.  

Albeit not potentially realizing the general derivation of Eq. (\ref{particle_derivation}) for {\it arbitrary Hamiltonians} using the deformation by newly inserting hard wall boundaries, invoking the classical relation of Eq. (\ref{Eq:tau}), the leading order WKB borne substitution of Eq. (\ref{sums}), and simple normalization, the final result of Eq. (\ref{particle_derivation}) that we obtained above was found long ago by Eyring \cite{eyring-tst} and discussed in numerous textbooks for chemical reactions, e.g., \cite{Atkins,Laidler},
in which the initial and final products have different chemical potentials.  We reiterate that Eq. (\ref{particle_derivation}) is completely
general for {\it arbitrary Hamiltonians} and is {\it not confined} to that of free particles in a box having translational degrees of freedom (as first considered in \cite{eyring-tst}) or of harmonic oscillators, etc., on which nearly all of the work has been done to date. In Kramers' original work \cite{kramers} and Planck's constant never appeared nor was temperature directly invoked with the Boltzmann distribution (instead, temperature was discussed ad hoc via noise or effective friction). To make contact with standard cases, we remark that, e.g., the standard Arrhenius form $r = \nu_{0} \exp(-\beta \Delta E)$ 
with a fixed gap $\Delta E$ and $\nu_{0}$ a constant frequency follows from Eq. (\ref{ratio})
for harmonic potential \cite{harmonic}. One may, of course, extend such trivial calculations to free particles in a box \cite{particle_in_a_box} to emulate caging effects. There are Kramers (and Langer) corrections to such forms \cite{kramers,langer}. In our case, there is a nearly uniform many body system. 
Recently, a possible illuminating link between Kramers formalism and activated dynamics of a form similar to Eq. (\ref{particle_derivation}) have been suggested \cite{badi2} based on the use of a harmonic heart bath and the relation between Fokker Planck equations and quantum dynamics (see, e.g., \cite{t2t}, for a recent extension and application of this relation to general dynamical correlations). In \cite{badi2}, an additional prefactors appear that result in a form similar yet a bit different from Eq. (\ref{particle_derivation}). We remark that, In principle, the steps that we invoked in going from Eq. (\ref{particle_derivation}) may be reproduced nearly verbatim on the reaction rate form of \cite{badi2} to arrive at an analog form of Eq. (\ref{ratio}) in which additional non-universal prefactors appear. We conclude this subsection by noting that our above approach may apply to both liquids as well as dense gases (for which the notion of ``holes'' or ``vacancies'' such as those historically considered by Eyring \cite{eyring-viscosity,tabor} is somewhat ill-defined and thus the ensuing derivations in these classical works not transparent insofar as their physical assumptions). Disparate relaxation rates might be given by Eq. (\ref{ratio}) yet with largely varying free energy barriers. The rate associated with events requiring low escape energies may, naturally, be far larger than that associated with processes entailing more substantial energy barriers.

To close our circle of ideas and make further contact with subsection \ref{new_thermo}, we discuss and emphasize the extreme limits that our results imply. In Eq. (\ref{entropy!}), the time $\overline{\tau}_{tot}$ is the sum of the system period over all ergodic components averaged over the energy interval $[E,E+\delta E]$. As such, by virtue of being a {\it sum} over different components, it serves as an upper bound on all energy averaged Poincare recurrence times for such orbits. Each Poincare recurrence time for motion not cut short by thermalizing events is, in turn,
larger than the time $\tau = 1/r(T)$ for a closed orbit associated with the Hamiltonian $H^{transition}$ of Eq. (\ref{Ht}) in which the system suffered a thermalizing event at $x_{final}$
that disrupted its full recurrence path sans such a perturbation. In the micro-canonical ensemble, thermalization events correspond to a redistribution of the energy density. In the this ensemble, once a system returns back to its original state after a complete recurrence cycle,
its evolution is repeated once more and so on ad infimum. Thus, if no thermalization events occurred in a recurrence cycle then none will ever occur. The Poincare orbits correspond to extreme events (ones in which no thermal agitations occur) and, consequently, 
$\overline{\tau}_{tot} = \frac{h}{\delta E} \exp(S/k_{B}) >  \tau = \frac{h}{k_{B} T} \exp(\beta \Delta G)$.

\subsection{Measurable relations between classical thermodynamics and dynamics involving a quantum time scale for thermalization}
\label{measurements}

We now briefly return to the objective of subsection \ref{new_thermo} and suggest how Planck's constant may be measured to high accuracy 
by combining both thermodynamic and dynamic quantities. As we mentioned earlier, in fluids with quasi-one dimensional dynamics,
the thermodynamic functions (e.g., the free energy or effective chemical potential $\mu(T)$) and the relaxation rates $r(T)$ are related via
Eq. (\ref{particle_derivation}). Thus by experimental thermodynamic and dynamic measurements of classical fluids, it is in principle 
possible to adduce, or provide estimates for, the value of Planck's constant.

Before turning to details we remark that Eq. (\ref{ratio}) demonstrates that in an appropriate $\beta \to 0$ limit (one in which the pertinent $\Delta G$ is, effectively, held bounded as we
will explain next), the {\it extrapolated} rate 
\begin{eqnarray}
\label{rTL}
r(T) \sim \frac{k_{B} T}{h}.
\end{eqnarray}  
The generality of Eq. (\ref{rTL}) may relate a result 
concerning time in ``typical non-equilibrium states'' \cite{random}
being of order $h/(k_{B}T)$ as well as certain explicit calculations for specific systems, e.g., \cite{Nussinov3}. At high temperatures, all states are equally likely
and may be randomly chosen. A ``Planckian time scale'' {\it of order} ${\cal{O}}(\hbar/(k_{B} T))$ was proposed in several investigations, e.g., \cite{jan,qcp,DS,bruin,hartnoll}. 
In the current work, we will suggest that not only the order is fixed but rather that {\it exact asymptotic equalities} will appear in semi-classical systems. 

In Eq. (\ref{particle_derivation}), we suppressed the temperature dependence of the various quantities. As the temperature $T$ is varied, the atomic positions vary at at typical snapshot and consequently the effective potential $V$ along each ray ${\hat{n}}$ varies. Consequently, the eigenstates $\{|n \rangle_{\hat{n}}\}$ along different rays ${\hat{n}}$ change, the ensuing chemical potential $\mu_{\hat{n}}(T)$ is modified (as it even does for a system with fixed (temperature independent) effective $V$ such as the one that we examined in subsection \ref{transition_wkb}) as do the barrier heights $E_{\sf escape}$ and the effective probabilities $|a_{\hat{n}}(\beta)|^{2}$ that the particle may assume different trajectories in configuration space are altered as the temperature changes. 
In order for a barrier to meaningfully exist along any ray $\hat{n}$, in Eqs. (\ref{ratio}, \ref{dg}), the difference $\Delta G > 0$.
In general, for any fixed potential $V$, the chemical potential $\mu(T)$ is monotonically decreasing in temperature.
Thus, for semi-classical systems for which our considerations apply, the relation 
\begin{eqnarray}
\label{rmax}
r(T) \le \frac{k_{B} T}{h}
\end{eqnarray}
follows as {\it a strict inequality at all temperatures}. Clearly, if the system may return to the original state albeit having enough energy, the rate
would be even smaller (and thus the inequality of Eq. (\ref{rmax}) more stringent). Consequently, there clearly is a {\it rigorous minimal equilibration time} 
 \begin{eqnarray}
 \label{tmin}
  \boxed{ \tau_{\min} \equiv \frac{h}{k_{B} T}}
 \end{eqnarray}
 for general {\it semi-classical} systems at all temperatures $T$ (that is, the thermalization decay times $\tau \ge \tau_{\min}$). Supplanting the earlier arguments in the Introduction, we now obtained $\tau_{\min}$ as a bona lower bound on the relaxation time. As we will elucidate in Section \ref{Boltzmann-sec} and sections thereafter, the relaxation time scales and viscosity are very generally linked. Eq. (\ref{tmin}) is thus related to Eq. (\ref{limexplain}) and constitutes an inherent limit. Similar to Eq. (\ref{vequality}), this inequality may be saturated and lead to a well defined quantized value for the minimal relaxation time as we will briefly explain. 
 
Practically, for the analysis of experimental data, we wish to motivate a simple approach. Towards this end, we observe that trivially from Eq. (\ref{ratio}), the derivative 
\begin{eqnarray}
\label{rG}
- \frac{\partial \ln (h \beta r)}{\partial \beta} = \frac{\partial \Delta G}{\partial \beta} = \Delta {\sf H},
\end{eqnarray} 
where $\Delta {\sf H} = {\sf H}_{escape}- {\sf H}_{system}$ is the change in the enthalpy between the fictive system set by the escape energies and the physical system at hand
in the transition region. (In our case with the single generalized particle representing the state of the entire system, we may as well replace think of the enthalpy as representing the change in the internal energy.) By the simple thermodynamic relation 
\begin{eqnarray}
\Delta G = \Delta {\sf H} - T \Delta S,
\label{HG}
\end{eqnarray}
 with $\Delta S = S_{escape} - S_{transition}$ and even simpler and more direct relation of Eq. (\ref{ratio}) yielding the effective Gibbs free energy allows the construction of all thermodynamic functions given the values of the rate $r$ as a function of the temperature $T$. 
 For instance, when the system is in equilibrium and the calculation of Eqs. (\ref{particle_derivation},\ref{ratio}) applies, simple thermodynamic relations relate, e.g., the entropy change to that rate $r(T)$,
\begin{eqnarray}
\label{dst}
\Delta S(T) = \Delta S(T_{A}) + \int_{T_{A}}^{T} (\frac{\partial {\sf \Delta H}}{\partial T})  \frac{dT}{T}  \nonumber
\\  = \Delta S(T_{A}) +  k_{B} \int_{T_{A}}^{T} \frac{\partial}{\partial T} \Big(T^{2} \frac{\partial}{\partial T} \ln (h \beta r) \Big) \frac{dT}{T},
\end{eqnarray}
with $T_{A}$ an arbitrary temperature.  
  We now seek to directly obtain non-trivial quantities from the seemingly all too banal
 prevalent operation of fitting relaxation rates to Arrhenius forms. By analogy to standard calorimetric approaches, by the third law of thermodynamics, in the low temperature (viz., $T_{A}$) limit $\Delta S_{A} \to 0$ and a measurement of the rate $r(T)$ could in principle enable a computation of $\Delta S(T)$ from which $\Delta G(T)$ may be calculated and ultimately Planck's constant may be extracted via Eq. (\ref{ratio}). Liquids, and in particular, the metallic glass formers that we study in this work are in equilibrium only at sufficiently high temperature. Thus, an integration similar to that in Eq. (\ref{dst}) can only be done from a finite non-zero temperature upwards. As we will aim to motivate in later sections, at 
 the lowest temperature at which the system is still in equilibrium, the effective entropy difference at it appears in the transition region $\Delta S_{A} \to 0$ (in a manner superficially reminiscent to the effective vanishing
 of an assumed entropy form in the so-called ``Kauzmann temperature'' in glasses \cite{kauz}). To relate to usual Arrhenius forms, we follow a more pragmatic approach. Suppose that at a certain temperature $T_{A}$, a tangent to the plot of $ \ln (h \beta r)$ as a function of the inverse temperature  $\beta$ is given by the line
 \begin{eqnarray}
 \label{rH}
 \ln (h \beta r) = \ln (h \beta_{A} r_{A}) -(\beta - \beta_{A}) \Delta H_{A}.
 \end{eqnarray}
 In Eq. (\ref{rH}), the local slope of the tangent is the enthalpy change $\Delta H_{A}$ evaluated at temperature $T_{A}$ (as follows from Eq. (\ref{rG})).
 As $\ln (h \beta_{A} r_{A})  = - \beta \Delta G_{A}$, by Eqs. (\ref{ratio}, \ref{HG}), the local tangent at $T_{A}$ is associated with the Arrhenius form
 \begin{eqnarray}
 \label{rA}
 r = \Big( \frac{\beta^{-1}}{h} e^{\Delta S_{A}/k_{B}} \Big)
e^{-\beta \Delta {\sf H}_{A}}.
 \end{eqnarray}
Quite naturally, $\Delta S_{A} \le 0$ (i.e., $S_{transition} > S_{escape}$) \cite{micro_obvious} and thus, as in Eqs. (\ref{rmax},\ref{tmin}), we find that 
$r_{A} \le \frac{k_{B} T_{A}}{h}$.  All thermodynamic functions may be determined from derivatives of the Gibbs free energy $\Delta G$ (set by $\ln r$).
Thus, in principle, the entropy difference $\Delta S_{A}$ at any temperature $T_{A}$, such as that appearing in Eq. (\ref{rA}) may be computed from the rate.  
If $\Delta S_A \le 0$ at all temperatures then, as a consequence of the third law of thermodynamics (in the low temperature ($T_{A}$) limit, $\Delta S_{A} \to 0$), 
Planck's constant $h$ is given by 
an infimum over all temperature $T_{A}$ of the reciprocal of the prefactor in Eq. (\ref{rA}) multiplied by $(k_{B} T_{A})$, viz.,
\begin{eqnarray}
\label{hb}
h = \inf_{\beta}  \{ \frac{1}{\beta r} e^{-\beta (\frac{\partial \ln r}{\partial \beta})} \}.
\end{eqnarray}
This bound is saturated in the low temperature (or large inverse temperature $\beta$) limit. 
To make explicit that $T_{A}$ is an arbitrary temperature $T$ at which the system is in equilibrium, in Eq. (\ref{hb}), we replaced $T_{A}$ by $1/(k_{B} \beta)$.
The local tangent extrapolation $\Delta S_{A} \to 0$ employed for the local Arrhenius form for the rate $r(T)$ on a log-linear scale at inverse temperature $\beta_A$ may yield $r \sim (\beta_A^{-1}/h) e^{-\beta \Delta {\sf H}_{A}}$. 
Thus, if Arrhenius forms for the relaxation rates are locally performed at disparate temperatures $T_A$ {\it at which the system is ergodic and in equilibrium} then the rate will be given by Eq. (\ref{rA}). 
In the liquid there are no equilibrium states at temperatures $T<T_{A}^{\min}$ where $T_{A}^{\min}$ is defined as the lowest temperature at which the liquid remains in equilibrium. 
Thus, with such a defined temperature, thermodynamics becomes ill defined at $T<T_{A}$. For instance, the microcanonical entropy is not defined at energies lower than the energy associated with $T_{A}^{\min}$.  Lower energy states may be inaccessible in a rapidly cooled liquid liquid as it falls out of equilibrium and thus 
$S_{transition}(T \le T_A^{\min}) = 0$. The existence of a temperature $T_{A}^{\min}$ far above the glass transition temperature at which a supercooled liquid falls out of equilibrium 
and, e.g., no longer satisfies the Stokes-Einstein relation has been numerically seen \cite{numerics}.  According to the above, a tangent to the log of the rate as a function of inverse temperature at this temperature, will yield $r= \beta_{A}^{-1}/h \exp(-\beta \Delta)$ with $\Delta \equiv \Delta {\sf H}_{A}$. At higher temperatures, the rate is given by Eq. (\ref{rA})
where the prefactor $\exp(\Delta S_{A}/k_{B}) < 1$.

\subsection{Prethermalization and local relaxation times}
\label{pret}

As made clear in the discussion above and will be reiterated throughout this work, Eqs. (\ref{particle_derivation},\ref{ratio}) for the thermalization rate
(and Eq. (\ref{etax}) for the viscosity which we will derive later on by employing these relations and other equilibrium properties) are only valid for systems at or near thermal equilibrium. 
That is, these equalities are only true at sufficiently high temperatures when the liquids are equilibrated. At lower temperatures or in the presence of perturbations that remove the system from equilibrium, 
these relations need no longer hold. We may consider, however, what occurs if the system is in equilibrium on sufficiently local spatial scales. In such a case, we may analytically continue $\Delta G$ to temperatures in which equilibrium no longer holds but an effective relaxation of the form of Eq. (\ref{ratio}) persists. In supercooled liquids that fall out of equilibrium it is tempting to ask whether these forms may describe local shorter time so-called $\beta$ or Johari-Goldstein \cite{JG} relaxations for which an Arrhenius behavior (with a nearly constant $\Delta G$) for the relaxation persists. One may posit that this effective $\Delta G$ does not vary much relative to the lowest temperature $T_A$ at which the liquid is still in global thermal equilibrium. 
One may then apply the relations introduced in Section \ref{measurements} with $T_{A}$ changed to a temperature of local equilibrium with relaxation rates are 
those of local equilibrium processes. This Arrhenius type behavior may be contrasted with the $\alpha$ relaxation \cite{tr} in which relaxation increases with $(1/T)$ in a manner that is faster than Arrhenius (``super-Arrehnius'')-  the effective $(\Delta G)$ varies with temperature (especially more dramatically in ``fragile'' glass formers that we will turn to in Section \ref{ETH_section}). A broad distribution of local relaxation processes may trigger 
fragile dynamics \cite{banerjee}. 

Another, possibly related, viable extension of Eq. (\ref{ratio}) to systems out of equilibrium that may equilibrate on a short time scale locally is associated with 
``prethermalization'' \cite{prethermal}. That is, in certain quenched systems, relaxation occurs in steps- first to a  metastable``pre thermal'' state and only
after a sufficiently long time to true thermal equilibrium. It is conceivable that a few such systems may be described by a local thermal state
(for which the derivation of Eqs. (\ref{particle_derivation},\ref{ratio}) may apply with effective $\Delta G$ and temperature).

\subsection{Remarks on quantum critical systems}
\label{qcp:sec}

Thus far, we largely focused on semi-classical systems. We now briefly turn to very different systems exhibiting quantum critical points.  
In the quantum critical regime, $k_BT$ is the only energy scale available for 
temperatures above the quantum critical point (QCP). Thus, in this case it can be deduced from general 
arguments that for all $T>0$ a thermal energy gap of the form $\Delta=a_\Delta k_BT$ arises, where $a_\Delta$ is 
a universal number \cite{qcp}. Furthermore, the pertinent free energy difference
$a_+k_BT$, where again $a_+$ is universal. The above thus suggests that
\begin{equation}
\label{rt:qcp}
 r(T)|_{\rm QCP}=\frac{k_BT}{h}e^{-a_+}.
\end{equation}
This equation provides an example where the coefficient of $k_BT/h$ is {\it not} unity at high 
temperatures. That is, a richer array of asymptotic high temperature forms may be possible at quantum critical points. 
In fact, at a quantum critical point there is no such a thing as high and low temperature 
regime, as scale invariance holds in this case and there is no other energy scale to which $k_BT$ may be 
compared with. 

We further remark on the limitations of our general derivation in subsection \ref{transition_wkb} when applied to quantum critical systems.
Unlike the typical situation of Fig. (\ref{Potential}), quantum critical systems are {\it gapless} and the standard caricature of Fig. \ref{Potential} no longer applies. 
That is, at any intermediate finite temperatures $T$, there might not be a natural energy scale $E_{escape} \gg k_{B} T$. Rather, for certain distributions of gapless modes, the latter may indeed yield a numerical prefactor $e^{-a_{+}}$ by comparison to 
Eq. (\ref{particle_derivation}) or more general forms. 
Naturally with their gapless character, {\it long range quantum entanglement} is a typical hallmark of quantum critical phenomena, e.g., \cite{entangle_qcp}. 
By contrast, as our treatment in subsection \ref{transition_wkb} (including Eq. (\ref{semi-classical-state})) may have made clear, in 
the semi-classical picture the $initial$, $final$, and intermediate states are trivial unentangled ``states''.   At high energy end of the spectrum when the potential becomes irrelevant, in the derivation of subsection \ref{transition_wkb}, each particle (or high dimensional  many body configuration space) state has a well-defined momentum (momenta). The system is unentangled
in this limit. This can be contrasted with the entangled quantum critical
system in which the dynamics are far more complicated and constrained at
all temperatures above the quantum critical point. Collective (non single particle) descriptors underlie quantum criticality.

\section{General bounds on thermalization rates and chaos} 
\label{generalt}
In this brief section, extending the results above and those in our earlier work \cite{qv}, we will introduce and motivate a general bound of the form of Eq. (\ref{tmin}) (yet of
a weaker variety) that may hold for {\it all} systems
(semi-classical, quantum critical, or others). Here, we will investigate relaxation times in general systems at a fixed temperature $T$.
Our suggestion for the below bound on the relaxation time is stimulated by recent bounds \cite{juan} concerning a bound on the growth 
of chaos in quantum thermal systems. Specifically, one may examine the finite temperature dynamics of general systems
with time independent Hamiltonians by endowing observables with quantum dynamics, $W(t) = \exp(i H t/ \hbar)  W  \exp(- i H t/\hbar)$
and computing correlation functions such as \cite{juan}
\begin{eqnarray}
F(t) \equiv Tr[yVyW(t)yVyW(t)]
\label{fte}
\end{eqnarray}
between $W$ and a general perturbation $V$ with $y^4= \exp(-\beta H)/Z$ with $Z$ the partition function at inverse temperature $\beta$. 
That is, in Eq. (\ref{fte}) the Boltmzann factor of $y^{4}$ has been split into four identical factors. 
The authors of Ref. \cite{juan} compared the expectation value of Eq. (\ref{fte}) to the product of the individual thermal averages,
\begin{eqnarray}
F_{d} \equiv Tr [y^2 V y^2 V] ~~ Tr[y^2 W(t) y^2 W(t)],
\label{fd}
\end{eqnarray}
and based on a simple requirement for analytically and boundedness in a complex time strip (of width set by $\beta/2$ in the imaginary direction)
proposed that for positive times relative to the appearance of thermalization,
\begin{eqnarray}
|F_{d}- F(t))| \le K  \exp(\lambda_{L} t),
\label{fdft}
\end{eqnarray}
 with $K$ a large positive constant and $\lambda_{L}$ a Lyapunov exponent that is bounded
from above, 
$\lambda_{L} \le \frac{2 \pi k_{B} T}{\hbar}$.
Specifically,  this bound \cite{juan} may hold at intermediate times far larger than the dissipation time yet shorter than the so-called ``scrambling time'' \cite{suskind} beyond which $ \langle [W(t), V(0)]^{2} \rangle$ becomes significant. We remark that $(2 \pi  k_{B}T)/\hbar$ is nothing but the reciprocal of the minimal difference between (consecutive) Matsubara frequencies whether bosonic ($\omega_{n} = \frac{2 \pi n}{\beta \hbar}$ with integer $n$) or fermonic ($\omega_{n} = \frac{(2n+1) \pi}{\beta \hbar}$). Imaginary time calculations lead to decay set by the Matsubara frequencies, e.g., \cite{Nussinov3}. 

In what follows, we wish to turn around the argument of \cite{juan} and argue that if the above holds then it may strictly follow that the thermalization time in all quantum systems (i.e., not necessarily semi-classical ones) is trivially bounded,
\begin{eqnarray}
\tau_{quantum}(T) \ge \frac{h}{4 \pi^{2} k_{B} T}.
\label{tmq}
\end{eqnarray}
The logic behind this suggestion of ours is exceedingly simple. Eq. (\ref{fd}) is the expectation value anticipated at a long time thermal equilibrium. While \cite{juan} studied the 
deviations from the decoupled product $F_{d}$ when evolving forwards in time towards a maximally chaotic state, we can similarly examine the {\it backwards in time} evolution
(with the same Hamiltonian $H$) from a given thermal state to a maximally chaotic one. The deviation between equilibrium correlators and those at finite times cannot be 
arbitrarily large for a finite time $t$ and is in fact bounded by Eq. (\ref{fdft}). By setting $t \to (-t)$ and evolving backwards in time towards a thermal state with
factorized correlations (Eq. (\ref{fd})) we see that in the appropriate regime
 it is impossible to converge to equilibrium exponentially with a time constant that is smaller than that of Eq. (\ref{tmq}). The derivation of the bound of Eq. (\ref{fdft}) in \cite{juan} can be reproduced
 {\it mutatis mutandis} for the negative $t<0$ half-strip of width $\beta/2$ in the complex plane. It is precisely the decay time in correlators such as Eq. (\ref{fte}) that is of pertinence to the response functions (such as viscosity) that are of interest to us in this work \cite{explain_fd}.  
 
 Relying on limits on the transfer of quantum information in a time $\tau$, invoking the definition of Shannon entropy and the third law of thermodynamics 
 a slightly stronger bound was suggested in \cite{hod} for the decay time, $\tau_{quantum}(T) \gtrsim \frac{h}{2 \pi^{2} k_{B} T}$. We may invoke
 time reversal on any proposed bound on the decay rate to obtain a corresponding bound on Lyapunov exponent. We find that in this case, the latter bound of $\tau_{quantum}$ 
 would suggest a corresponding twice as strong bound on the Lyapunov exponent than that proposed by \cite{juan},
 viz.,
 \begin{eqnarray}
 \lambda_{L} \le \frac{ \pi k_{B}T}{\hbar}.
 \end{eqnarray} 
 
Our bound of Eq. (\ref{tmq}) for the relaxation time in a thermalized system is weaker than those of than we derived earlier for semi-classical systems (Eqs. (\ref{tmin}, \ref{rt:qcp}) \cite{qv}). 
Similarly, from our relations of Eqs. (\ref{tmin}, \ref{rt:qcp}), the Lyapunov of a semiclassical liquid system (for which the considerations of
subsections \ref{transition_wkb} and \ref{measurements} apply) may be bounded,
\begin{eqnarray}
\label{chaosus}
\lambda_{L}^{semi-classical} \le \frac{k_{B} T}{h}.
\end{eqnarray}

\section{Quantum Kinetic theory of the viscosity and its analytic continuation to asymptotically high temperatures}
\label{Boltzmann-sec}

Armed with Eqs. (\ref{particle_derivation},\ref{ratio}), we now turn to the calculation of the viscosity by employing the standard Boltzmann equation. We briefly review elements essential to our analysis, e.g., \cite{Reif}.
In what follows, $f_{1}^{(0)}({\bf{r}}_{1}, {\bf{v}}_{1};t)$ denotes {\it a single particle} equilibrium distribution. That is,
the equilibrium probability of finding a particle at location in a region of size $d^{3}r$ about $\vec{r}$ and velocity
that is within $d^{3}v$ of $\vec{v}$ at time $t$
is $f_{1}^{(0)}({\bf{r}}_{1}, {\bf{v}}_{1},t) ~ d^{3} r_{1} ~ d^{3} v_{1}$. This equilibrium distribution
is set by the Boltzmann weights associated with the Hamiltonian $H = \sum_{i} p_{i} ^{2}/(2m) + V$. 
Let us denote the $N$ body distribution function by $f_{N}(\{\bf{r}_{i}\}_{i=1}^{N}, \{\bf{v}_{i}\}_{i=1}^{N})$.
As is, e.g., evinced by the Stokes-Einstein relation, in equilibrated liquids, $f_{N}^{(0)} (\{ {\bf{r}}_{i}\}_{i=1}^{N}, \{ {\bf{v}}_{i}\}_{i=1}^{N}) = \frac{1}{\cal{Z}}  \exp(-\beta H)$.
It is important to emphasize that again that our extrapolation is from low temperatures (yet high enough that Arrhenius type dynamics appears as we derive via the use of 
Eq. (\ref{particle_derivation}) with the equilibrium distribution function as we do in this section. That is, in Section \ref{normal} in comparing to experimental data we will
extrapolate from the tangent to the log linear curve of $\ln \eta$ vs. $(1/T)$ at the temperature at which Arrhenius behavior onsets. It is this curve and its extrapolated value at high $T$ that we compare to theory. At emphasized in Section \ref{sec:tr},
at high temperatures the chemical potential $\mu$ acquires a temperature dependence that may obscure the simple form of Eq. (\ref{particle_derivation}) with an effectively temperature independent value of $(E_{\rm escape} - \mu)$.  

With all of these preliminaries we turn to our simple derivation. The single particle distribution $f_{1}$ results from an integral over all but one 
of the particle coordinates, 
\begin{eqnarray}
\label{f1:eq}
f_{1}
({\bf{r}}_{1}, {\bf{v}}_{1};t) = \int d^{3} r_{2} \int d^{3} p_{2} \cdots \nonumber
\\ \int d^{3} r_{N} \int d^{3} v_{N} ~~
f_{N}(\{ {\bf{r}}
_{i}\}_{i=1}^{N}, \{ {\bf{v}}_{i}\}_{i=1}^{N};t).
\end{eqnarray}
 Akin to the memory time in linear response functions, e.g., \cite{explain_fd}, 
the probability that
the system will transition in a time increment $dt$ is $(dt/\tau) e^{-t/\tau}$. 
The relaxation time relates naturally to the rate that we computed in the earlier sections (see, e.g., Eq. (\ref{decay'})). To explicitly relate the relaxation rate that we computed earlier to
the probability of staying in an initial lore we briefly review an all too standard derivation. If the rate of transitions is $r(T)$ then the probability of having a transition occurring in an interval $dt$ is set by $r~dt$ and the probability of not undergoing a transition in an interval $dt$ (i.e., staying in the original state at time $(t+dt)$ is $f(1- r~dt) = fe^{-t/\tau})$
where $f$ is the probability of being in the original state at time $t$ and, as in earlier sections, $\tau = 1/r(T)$. Normalization sets the exponential function $f(t)$ to be $(e^{-t/\tau}/\tau)$.
The non-equilibrium distribution function satisfies
\begin{eqnarray}
\label{collide} 
f_N({\bf{r}}_{1}, {\bf{v}}_{1};\{{\bf{r'}}_{i} \}_{i=2}^{N} ,\{ {\bf{v'}}_{i}\}_{i=2}^{N};t) \nonumber
\\ = \int_{0}^{\infty} f_{N}^{(0)} ({\bf{r}}_{10},{\bf{v}}_{10};
\{ {\bf{r'}}_{i0}\}_{j=2}^{N}, \{ {\bf{v'}}_{i0}\}_{j=2}^{N}; t-t') \nonumber
\\ \times e^{-t'/
\tau ({\bf{r}}_{1}, {\bf{v}}_{1})_{\{
{\bf{r}}_{i} 
\}_{i=2}^{N} ,
\{
{\bf{v}}_{i}
\}_{i=2}^{N}}}
\frac{dt'}{\tau({\bf{r}}_{1},  {\bf{v}}_{1})_{\{\bf{r}_{j} \}_{j=2}^{N} ,\{\bf{v}_{j}\}_{j=2}^{N}}},
\end{eqnarray} 
where ${\bf{r}}_{10} \equiv {\bf{r}}_{1}(t_{0})$ and ${\bf{v}}_{10} \equiv {\bf{v}}_{1}(t_{0})$ are the initial coordinates and velocity of particle ``$1$''
and $\{{\bf{r'}}_{j 0}, ~{\bf{v'}}_{j 0}\}$ those of all other $(N-1)$ particles $( j \ge 2)$ such that for a closed $N$ body system governed by Hamilton's equations of motion, the system will obtain coordinates and velocities 
$\{{\bf{r}}_{j}\}_{j=1}^{N}$ and $\{{\bf{v}}_{j}\}_{j=1}^{N}$ at time $t$. The relaxation time $\tilde{\tau}({\bf{r}}_{1},  {\bf{v}}_{1})_{\{\bf{r}_{j} \}_{j=2}^{N} ,\{\bf{v}_{j}\}_{j=2}^{N}}$ is associated with equilibration of the $N-$particle system (as it appears in the linear response functions \cite{explain_fd} and manifest in the calculation for the viscosity as the system linear response to applied shear). We have written it with the location ${\bf{r}}_{1}$ and velocity ${\bf{v}}_{1}$ of an arbitrary particle (particle $1$) highlighted as we wish to highlight that it may be used to
describe all possible relaxation times associated with particle $1$ when all of the remaining $(N-1)$ particles assume any possible positions and velocities. 
Although not cardinal to our argument, we remark that if the system is sheared and the viscosity is directly examined, the relaxation times will be associated with the one-dimensional dynamics along the applied external shear direction. If there is only single dominant relaxation time, then for the one particle distribution function of Eq. (\ref{f1:eq}),
\begin{eqnarray}
\label{collide1} 
f_{1} ({\bf{r}}_{1}, {\bf{v}}_{1};t) = \int \frac{dt'}{\tau} ~ f_{1}^{(0)} 
({\bf{r}}_{10},{\bf{v}}_{10}; t-t') e^{-t/\tau}.
\end{eqnarray} 
A relation of the form of Eq. (\ref{collide1}) is typically invoked in textbook treatments of the kinetic theory of gases. However, in what follows, we allow for the {\it distribution} of
single particle relaxation times for all velocities and locations of particles $2, 3, \cdots, N$. 
The more precise multi-particle linear order relation of Eq. (\ref{collide}) expresses that particles reaching $(\bf{r}, \bf{v})$ at time $t$ could have been in equilibrium for any amount of time $0<t'<\infty$ beforehand and that the probability for staying in equilibrium for such an amount of
time is set by $1/\tilde{\tau} \times \exp[-t'/\tau]$ with, for a single particle, $\tau = 1/r(T)$ of Eqs. (\ref{particle_derivation},\ref{ratio}). That is, we will employ the relaxation time found in earlier sections. To compute the viscosity within the Boltzmann equation formalism, one typically examines the system response of the system
to a $z$ dependent shear force along the $x$ direction that leads to a constant average drift velocity.  The shear has an associated local equilibrium single particle distribution function $f_{1}^{(0)}(v_{1x} - u_{1x}(z), u_{1y}, u_{1z}) \equiv f_{1}^{(0)}(U_{1x}, U_{1y}, U_{1z})$
with similar generalizations to arbitrary order distribution functions. By translational invariance, the shear does not depend on the center of mass coordinate or, equivalently, on the coordinate ${\bf{r}}_1$ of the ``first'' particle, We will forgo repeating standard steps as those may be found in excellent standard texts. {\it Multi-particle} corrections to single particle motion are, e.g., often afforded by a BBGKY type hierarchy that are truncated at a particular order.  In what follows, we start with the general multi-body relation of Eq. (\ref{collide}). The final result, e.g., \cite{Reif}, that we now trivially generalize to include arbitrary order particle correlations, is that the viscosity.
\begin{eqnarray}
\label{etaint}
\eta &=& mN \frac{ \int \prod_{i=1}^{N} d^{3}U_{i} \prod_{i=2}^{N}d^{3} r_{i} \tau(r_{1},U_{1})^{r_{2}, \cdots r_{N}; U_{1}, \cdots, U_{N}} U_{1z}^{2} f_{N}^{(0)}}
{ \int \prod_{i=1}^{N} d^{3}U_{i} \prod_{i=2}^{N} d^{3} r_{i} f_{N}^{(0)}} \nonumber
\\ &&\times \int \prod_{i=1}^{N} d^{3}U_{i} \prod_{i=2}^{N} d^{3} r_{i} f_{N}^{(0)}
 \nonumber
\\ & \equiv & mN \langle \tau U_{1z}^{2} \rangle  \int \prod_{i=1}^{N} d^{3}U_{i} \prod_{i=2}^{N} d^{3} r_{i} f_{N}^{(0)},
\end{eqnarray} 
where the average in the last line of Eq. (\ref{etaint}) is performed with the distribution function $f_{N}^{(0)}$ integrated over all particle velocities and the coordinates except those of a particle number ``$1$''.  The quantum kinetic calculation for the viscosity is carried out with
the relaxation time $\tau({\bf{r}}_{1}, {\bf{U}}_{1})^{{\bf{r}}_{2}, ..., {\bf{r}}_{N}, {\bf{U}}_{2}, ...., {\bf{U}}_{N}} $
of a single particle with its initial coordinates ($r_{1}$) and those of all
other particles and their velocities ${\bf{U}}_{2 \le b \le N}$ fixed. If this
case, we simply insert the values of the coordinates and velocities in the
many body potential to form a one-body Hamiltonian for particle $\#1$ which
can move along any ray
$\hat{U}_{1}$ in three-dimensional space.
The factor of $N$ in the first two lines of Eq. (\ref{etaint}) appears as each of the $N$ particles contributes to the viscosity. The integral  $\int  \prod_{i=1}^{N} d^{3}U_{i} \prod_{i=2}^{N} d^{3} r_{i} f_{N}^{(0)} = n_{1}$, namely the probability of finding particle number $1$ in a unit volume. The product $Nn_{1} = {\sf n}$ is equal to the total particle density. This enables us to write
$\eta = 2  {\sf n} \langle \tau E_{kz~(1)} \rangle$,
with $E_{kz~(1)} \equiv \frac{1}{2} m U_{z1}^{2}$. The product of time multiplied by energy of some sort is 
superficially reminiscent of Eq. (\ref{entropy!}). One might therefore anticipate that
Planck's constant may, be related to the expectation value of product in the original sum over
semi-classical states. 

The sole assumption made in deriving Eq. (\ref{etaint}) is that the external shear is a small perturbation, e.g.,  \cite{Reif}.
Thus, Eq. (\ref{etaint}) is an exact relation within the Boltzmann formalism of Eq. (\ref{collide}) that includes arbitrarily high order correlations in the liquid. 
Clearly,  $\langle  \tau U_{1z}^{2} \rangle = \langle \tau \rangle  \langle U_{1z}^{2} \rangle + \langle (\delta \tau) (\delta (U_{1z}^{2}) )\rangle$,
where $\delta A \equiv A - \langle A \rangle$ (where $A = \overline{\tau}$, $U_{1z}^{2}$).
Thus,
\begin{eqnarray}
\label{etax}
\boxed{\eta = {\sf n} k_{B} T \tau + mN  \langle (\delta \tau) (\delta (U_{1z}^{2}) ) \rangle.}
\end{eqnarray}
In a mean field type approach in which the fluctuations are neglected, $\eta = {\sf n} k_{B} T \tau$.
The fluctuations encode a velocity dependence of the relaxation (or collision time) vis a vis
the general average relaxation time $\tau$.

In Eq. (\ref{etax}), we employed
\begin{eqnarray}
\label{u2n}
\langle U_{1z}^{2} \rangle \equiv  \frac{\int 
\prod_{i=1}^{N} d^{3}U_{i}  \prod_{i=2}^{N} d^{3} r_{i} ~U_{1z}^{2}  f_{N}^{(0)}}{
\int 
\prod_{i=1}^{N} d^{3}U_{i}  \prod_{i=2}^{N} d^{3} r_{i} f_{N}^{(0)}} = \frac{k_{B} T}{m}.
\end{eqnarray}
This is a trivial consequence of the equipartition theorem for general non-relativistic Hamiltonians $H$ with {\it arbitrary interactions}
(i.e., those with with a $p^{2}$ momentum dependence with arbitrary potentials $V$)- not solely those describing free non-interacting gas particles as in most elementary treatments. As highlighted earlier, in Eqs. (\ref{etaint}, \ref{etax}, \ref{u2n}), we accounted for many body correlations via the $N-$ body distribution function $f_{N}^{(0)}$ and  $\tau(r_{1},U_{1})^{r_{2}, \cdots r_{N}; U_{1}, \cdots, U_{N}}$ representing the equilibration time of the entire system. 
If we perform a single particle calculation (that is, one employing only $f_{1}^{(0)}$ after integration of all $(N-1)$ particles $i \ge 2$), 
we will similarly obtain $\eta= {\sf n} k_{B} T \tau_1$ with $\tau_1$ marking explicitly 
the assumed uniform (position independent) average single particle relaxation time in such a calculation. Once again, via equipartition, Eq. (\ref{u2n}) holds true in a general interacting system. In this case, the fluctuation borne corrections in Eq. (\ref{etax}) arise from higher order correlations. In individual non-dilute fluids, the higher order corrections 
to particle correlations (even those beyond second order- the so-called ``Born-Green-Yvon equation'' \cite{bgy,bgy'}) may be notable. Formally, one may redefine the relaxation time
so that only the first term in Eq. (\ref{etax}) remains.  That is, Eq. (\ref{etaint}) may, of course, trivially be cast as
\begin{eqnarray}
\label{etaint'}
\eta &=& mN \frac{ \int \prod_{i=1}^{N} d^{3}U_{i} \prod_{i=2}^{N}d^{3} r_{i} \tau(r_{1},U_{1})^{r_{2}, \cdots r_{N}; U_{1}, \cdots, U_{N}} U_{1z}^{2} f_{N}^{(0)}}
{ \int \prod_{i=1}^{N} d^{3}U_{i} \prod_{i=2}^{N} d^{3} r_{i} U_{z1}^{2} f_{N}^{(0)}} \nonumber
\\ &&\times \frac{ \int \prod_{i=1}^{N} d^{3}U_{i} \prod_{i=2}^{N} d^{3} r_{i}  U_{z1}^{2} f_{N}^{(0)}}{ \int \prod_{i=1}^{N} d^{3}U_{i} \prod_{i=2}^{N} d^{3} r_{i}  f_{N}^{(0)}} \nonumber
\\ &&\times  \int \prod_{i=1}^{N} d^{3}U_{i} \prod_{i=2}^{N} d^{3} r_{i}  U_{z1}^{2} f_{N}^{(0)}
 \nonumber
\\  && = {\sf n} k_{B} T \tau',
\end{eqnarray} 
where 
\begin{eqnarray}
\tau' \equiv  \frac{ \int \prod_{i=1}^{N} d^{3}U_{i} \prod_{i=2}^{N}d^{3} r_{i} \tau U_{1z}^{2} f_{N}^{(0)}}
{ \int \prod_{i=1}^{N} d^{3}U_{i} \prod_{i=2}^{N} d^{3} r_{i} U_{z1}^{2} f_{N}^{(0)}}
\end{eqnarray}
is the relaxation time averaged not with $f_{N}{(0)} = \exp(-\beta H)$ but rather with $U_{z1}^{2} \exp(-\beta H)$. Clearly, at extrapolated asymptotically high temperatures when the velocities are high, the logarithmic in 
$U_{1z}$ dependence in the exponential might lead to little difference relative to the average with $\exp(-\beta  H)$ with quadratic (in $H$) velocity dependence.

We return to Eq. (\ref{etax}) and note that within the liquid phase, the viscosity is, generally, a monotonically decreasing function of the temperature. In {\it the extrapolated high temperature and/or dilute limit} of the liquid system, the first term in Eq. (\ref{etax}) is the dominant contribution. Invoking Eq. (\ref{tmin}), we have a lower bound on liquid viscosity at any temperature,
\begin{eqnarray}
\label{nh}
\eta \ge {\sf n} h.
\end{eqnarray}
It is interesting to note that Eq. (\ref{nh}) also follows from Maxwell's equation $\eta \propto \rho v \ell$ 
for the viscosity of a dilute gas (of mass density $\rho= m {\sf n}$,  the velocity $v$, and mean free path $\ell$) \cite{maxwell} if it is ``quantum limited'' meaning that
the mean free path saturates at high temperature and is set equal to the de Broglie wavelength $h/(mv)$. Although somewhat less well known, not only in dilute gases but also in fluids, there is an analogue of the mean free path termed the liquid ``mixing length'' as conceived by Prandtl \cite{prandtl}- the distance over which an element of the liquid will remain unaltered before mixing with the outside fluid. In the most commonly studied cases, physically, for transport to occur the mean-free path, $\ell$ must be greater than or equal to the de Broglie wavelength; for smaller values of $\ell$, constructive interference may lead to localization, e.g., \cite{Ioffe,Anderson}. 
For ideal gases with no collisions and relaxation processes, the viscosity may become arbitrarily small. In subsection \ref{ideal:sec}, we will indeed discuss how such a vanishing value may be realized the case. By contrast, if intermediate time thermalization events (or gaseous collisions) occur, as they do in general equilibrated non-dilute systems, then the rates will be exactly given by Eq. (\ref{ratio}) with the lower bound of Eq. (\ref{tmin}). We will further investigate the viscosity in such a ``hydrodynamic'' regime in subsection \ref{finite_relax}.
In the high energy (and temperature) limit, the interaction becomes irrelevant, correlations become faint, and the appearance of
known ideal gas type results as an exact equalities in general multi-particle system in the first lines of Eq. (\ref{etaint})
 is to be expected. 
 
So far, our exceedingly simple calculations have been exact within the Boltzmann formalism. Clearly, in different fluids, 
the connected correlation function $\langle (\delta \tau) (\delta (U_{1z}^{2}) ) \rangle$
attains disparate values. There may be no general simple relations that constrains and relates
the deviations $\delta \tau$ and $\delta(U_{1z}^{2})$ to each another. If, over an ensemble of fluids, apart from the universal
leading order result, higher corrections for individual fluids are randomly distributed relative to one another
$\mathbb{E}(\langle (\delta \tau) (\delta (U_{1z}^{2}) ) \rangle) =0
$ where $\mathbb{E}$ denotes an ensemble average (or, equivalently, the contributions to
the viscosity from the higher order correlation functions $f_{i \ge 2}^{(0)}$) then, obviously,
\begin{eqnarray}
\label{EnT}
\mathbb{E}(\frac{\eta}{ {\sf n} k_{B} T \tau}) = 1.
\end{eqnarray}
Although it would obviously constitute a simplification, the assumption of Eq. (\ref{EnT}) is not rigorous, to say the least.

Before concluding this section, we explicitly remark that trivially inserting Eq. (\ref{ratio}) into Eq. (\ref{etax}), 
\begin{eqnarray}
\label{enh}
\eta = {\sf n} h e^{\beta \Delta G} +  mN  \langle (\delta \tau) (\delta (U_{1z}^{2}) ) \rangle.
\end{eqnarray}
The first term in Eq. (\ref{enh}), that remaining if Eq. (\ref{EnT}) holds, is the form that we will use to investigate our experimental results. 
In Section \ref{s:Eyring}, we will review the old derivation of this first term and remark how this result may similarly emerge from an appropriate ensemble average. 
Before we do so, we first derive the viscosity generally anew via a correlator formalism that will highlight the subtle low frequency limit that needs to be taken 
when discussing the hydrodynamic regime.


\section{The Current correlation function and the viscosity}
\label{Sect:correlation}

In earlier sections, we highlighted quantum processes and their associated rates (or time scales) in governing dynamics. 
In this section, we will depart from this microscopic point of view and first analyze the viscosity careful via continuous current correlation functions
in which discrete dynamics are not enforced from the outset. We will then turn to examine what occurs for finite relaxation times. 
The form of the resulting expressions for the viscosity are similar to those of Eqs. (\ref{etaint},\ref{etax}) yet due to the underlying premise are not identical to these. 

While response theory calculations 
of the viscosity in relativistic system must necessarily be cmputed via a stress tensor correlation function, the current 
correlation function can be used in non-relativistic systems. By  
the Kubo formulas \cite{Kadanoff-Martin}, 
\begin{equation}
 \label{Eq:eta-C}
 \eta=m^2\lim_{\omega\to 0}\lim_{q\to 0}\frac{\omega}{q^2}{\rm Im}C_t(\omega,{\bf q}),
\end{equation}
where $C_t(\omega,{\bf q})$ is the transverse component of the retarded current correlation function, 
\begin{equation}
\label{Eq:Cij}
 C_{ij}(t,{\bf r};t',{\bf r}')=-i\theta(t-t')\langle [ j_i(t,{\bf r}),j_j(t',{\bf r}')]\rangle.
\end{equation}
A related formula allows us to also calculate the bulk viscosity, $\zeta$, 
\begin{equation}
 \frac{4}{D}\eta+\zeta=m^2\lim_{\omega\to 0}\lim_{q\to 0}\frac{\omega}{q^2}{\rm Im}C_l(\omega,{\bf q}),
\end{equation}
where $D$ is the dimension and  $C_l(\omega,{\bf q})$ is the longitudinal component of the current correlation function. 
Kubo formulas were employed to derive exact non-perturbative results \cite{mohit}.

\subsection{The ideal semiclassical case}
\label{ideal:sec}

Let us verify explicitly that $\eta=0$ in the case of ideal semi-classical gases. In this case the current correlation function 
is given in imaginary time simply by,
\begin{widetext}
\begin{equation}
\label{Eq:cc-correl}
 C_{ij}(i\omega,{\bf q})=\frac{\hbar g}{4m^2} k_BT\sum_{n}\int\frac{d^Dk}{(2\pi)^D}(2k_i+q_i)(2k_j+q_j)G_0(i\Omega_n,{\bf k})
 G_0(i\omega-i\Omega_n,{\bf k}+{\bf q}),
\end{equation}
where $g$ is the degeneracy, and 
\begin{equation}
 G_0(i\omega,{\bf k})=\frac{\hbar}{i\hbar\omega+\mu-\frac{\hbar^2k^2}{2m}},
\end{equation}
and the Matsubara sum is either fermionic or bosonic. 
After evaluating the Matsubara sum and assuming a classical regime such that 
the equilibrium distribution function is given by the Maxwell-Boltzmann one, we obtain, 
\begin{equation}
 C_{ij}(i\omega,{\bf q})=\frac{\hbar {\sf n}(\lambda_T^{nr})^D}{4m^2}\int\frac{d^Dk}{(2\pi)^D}\exp\left[-\beta\left(\frac{\hbar^2k^2}{2m}-\mu\right)\right]
 \left[\frac{(2k_i+q_i)(2k_j+q_j)}{i\omega-\frac{\hbar}{2m}(q^2+2{\bf k}\cdot{\bf q})}-
 \frac{(2k_i-q_i)(2k_j-q_j)}{i\omega+\frac{\hbar}{2m}(q^2-2{\bf k}\cdot{\bf q})}\right],
\end{equation}
where ${\sf n}$ is the density and $\lambda_T^{nr}= \sqrt{2 \pi \hbar^{2}/(mk_{B} T)}$ is the non-relativistic de Broglie thermal wavelength that we discussed in the Introduction. We remark that $e^{\beta \mu}$ is not the bare standard fugacity 
as evident by our factorization of the $({\sf n}(\lambda_T^{nr})^D)$ prefactor- the value to which the fugacity approaches in the ideal classical gas.  
Analytically continuing to real frequencies, $i\omega\to\omega+i\delta$, with $\delta\to 0+$, allows us to easily 
take the imaginary part of the current correlation function to obtain, 
\begin{equation}
\label{Eq:ImCt}
 {\rm Im}C_{ij}(\omega,{\bf q})=-\frac{\pi \hbar{\sf n}(\lambda_T^{nr})^D}{4m^2}\int\frac{d^Dk}{(2\pi)^D}
 \exp\left[-\beta\left(\frac{\hbar^2k^2}{2m}-\mu\right)\right]
 \sum_{s=\pm}s(2k_i+sq_i)(2k_j+sq_j)\delta((\hbar/m){\bf k}\cdot{\bf q}-F_s(\omega,q)),
\end{equation}
\end{widetext}
where 
\begin{equation}
\label{Eq:Fs}
 F_\pm(\omega,q)=\omega\pm\frac{\hbar q^2}{2m}.
\end{equation}
The integral can now be evaluated exactly and we obtain in the ``classical'' regime (i.e., $\hbar\to 0$), for instance,
\begin{equation}
\label{Eq:ideal}
 {\rm Im}C_t(\omega,q)={\sf n}e^{\beta\mu}\sqrt{\frac{\pi\beta}{2m}}
 \left(\frac{\omega}{q}\right)
 \exp\left\{-\frac{m\beta}{2q^2}\left[\omega^2+\left(\frac{\hbar q^2}{2m}\right)^2\right]\right\},
\end{equation}
where ${\sf n}$ is the density, 
and we see that $\eta$ vanish. As we have already mentioned, this 
differs from the relativistic result where $\eta\to\infty$ in the non-interacting case. Note that $\eta=0$ immediately violates 
the inequality (\ref{Eq:visc-bound}). At this point we recall that Eq. (\ref{Eq:visc-bound}) was derived assuming 
a strongly interacting system. Moreover, it is frequently argued that viscosity in non-interacting systems is an ill-defined 
concept, as in absence of collisions no meaningful hydrodynamic limit can be defined.

\subsection{The case with a finite relaxation time}
\label{finite_relax}

The most difficult task within the correlation function approach is not only to obtain a non-vanishing viscosity, but also finite 
moments of the frequency defined by \cite{Martin-Book,Forster-1},
\begin{equation}
\label{Eq:moments}
 \langle \omega^n({\bf q})\rangle=\frac{1}{C_t(0,{\bf q})}\int_{-\infty}^\infty \frac{d\omega}{\pi}\omega^n\tilde C_t(\omega,{\bf q}),
\end{equation}
where $n$ is a positive integer, and $\tilde C_t(\omega,{\bf q})={\rm Im}C_t(\omega,{\bf q})/\omega$. 
Both quantities depend crucially on the behavior of the relaxation time, which in general may be frequency- 
and momentum-dependent. Eq. (\ref{Eq:ideal}) naturally fulfills the finiteness criterion for the moments of 
the frequency, but it yields a vanishing viscosity as it has a  divergent relaxation time. 

Up to now, we considered examples of characteristic time scales which do not depend neither on frequency nor on 
the wavevector. Within this framework, 
an easy way to account for a finite relaxation time $\tau$ arising from collisions 
between the particles is to replace the delta function in Eq. (\ref{Eq:ImCt}) by 
\begin{equation}
 \Delta_s(\omega,{\bf q})=\int_{-\infty}^\infty\frac{dt}{2\pi}e^{it[(\hbar/m){\bf q}\cdot{\bf k}-F_s(\omega,q)]
 -\frac{t^2}{2\tilde \tau^2}},
\end{equation}
thus providing a phenomenological model for dissipation. Here $\tilde \tau=\tau/(2\pi)$.  
Performing the integrals, we obtain,
\begin{widetext}
\begin{equation}
\label{Eq:cc-int}
 {\rm Im}{\rm C_t}(\omega,q)=\frac{\sqrt{2\pi}{\sf n}\tilde \tau e^{\beta\mu}}{m\beta\hbar\sqrt{1+\frac{q^2\tilde \tau^2}{m\beta}}}
 \sinh\left[\frac{\hbar\omega q^2\tilde \tau^2}{2m\left(1+\frac{q^2\tilde \tau^2}{m\beta}\right)}\right]
 \exp\left\{-\frac{\tilde \tau^2}{2\left(1+\frac{q^2\tilde \tau^2}{m\beta}\right)}\left[\omega^2+\left(\frac{\hbar q^2}{2m}
 \right)^2\right]\right\}.
\end{equation}
Note that when $\tau\to\infty$ and $\hbar\to 0$ we recover Eq. (\ref{Eq:ideal}), as it should. As before in the 
non-interacting case, all moments of the frequency are finite. However, in the case of Eq. (\ref{Eq:ideal}) we 
obtain the standard deviation $\Delta\omega(q)=\sqrt{\langle\omega^2(q)\rangle}=0$, while in the case of 
Eq. (\ref{Eq:cc-int}), we have $\Delta\omega=\lim_{q\to 0}\Delta\omega(q)=1/\tilde \tau=2\pi/\tau$. As we have argued 
in the Introduction, 
$\Delta E=\hbar\Delta\omega\sim k_BT$, so we obtain once more that $\tau\sim \tau_{\min}$. 

It is easy to see that the $q\to 0$ limit in Eq. (\ref{Eq:eta-C}) does not vanish any longer for the 
case of Eq. (\ref{Eq:cc-int}). Thus, let us consider the frequency-dependent viscosity,
\begin{equation}
 \label{Eq:eta(w)}
 \eta(\omega)=m^2\lim_{q\to 0}\frac{\omega}{q^2}{\rm Im}C_t(\omega,q)
 =\sqrt{\frac{\pi}{2}}k_BT{\sf n}\omega^2\tau^3\exp\left(\frac{\mu}{k_BT}-\frac{\omega^2\tau^2}{2}
 \right).
\end{equation}
If $\tau$ is $\omega$-independent, we still have a vanishing viscosity, just like in the free case.  
However, we see that if the relaxation time scales like $\tau\sim\omega^{-2/3}$, a finite 
viscosity is obtained. We will see below, that $\tau\sim\omega^{-2/3}$ is also possible beyond the simple phenomenological model 
above.

Another phenomenological calculation using current correlation functions 
considers the damping in a manner 
reminiscent to that of a canonical quantum theory of screening with a relaxation time \cite{Harrison-book},  
\begin{equation}
\label{Eq:Cij}
C_{ij}(\omega,{\bf q})=-\frac{\hbar g}{4{m}^2}\int\frac{d^Dk}{(2\pi)^D}
 f(k)
\left[\frac{(2k_i+q_i)(2k_j+q_j)}{\omega-\frac{\hbar}{2m^*}(q^2+2{\bf k}\cdot{\bf q})+i\gamma}-
 \frac{(2k_i-q_i)(2k_j-q_j)}{\omega+\frac{\hbar}{2m^*}(q^2-2{\bf k}\cdot{\bf q})+i\gamma}\right],
\end{equation}
where $m^*$ is the effective mass,   
$f(k)$ is the distribution function, and we have introduced an unknown parameter $\gamma$, which as we will emphasize later on may depend on 
both the frequency and momentum. The one particle distribution function $f(k)$ is the analog of that of the exact one-particle 
equilibrium distribution function of Eq. (\ref{f1:eq}).
 
The transverse component of Eq. (\ref{Eq:Cij}) is obtained by contracting it with  
$(D-1)^{-1}(\delta_{ij}-q_iq_j/q^2)$, which yields, 
\begin{equation}
 C_t(\omega,q)=-\frac{\hbar g}{(D-1){m}^2}\int\frac{d^Dk}{(2\pi)^D}
 \frac{(\hbar/m^*)[k^2q^2-({\bf k}\cdot{\bf q})^2]f(k)}{\omega^2-\gamma^2-\left(\frac{\hbar q^2}{2m^*}\right)^2\left[1-\frac{(2{\bf k}\cdot{\bf q})^2}{q^4}\right]
 +2i\gamma\left[\omega-\frac{\hbar}{m^*}({\bf k}\cdot{\bf q})\right]}.
 \end{equation}
 \label{ct} 
\end{widetext}
Since $\eta(\omega)=m^2\lim_{q\to 0}q^{-2}{\rm Im}C_t(\omega,q)$, it is enough to expand the equation 
above up to order $q^2$. Thus, we obtain the long wavelength form, 
\begin{equation}
\label{Eq:effoscill}
 \frac{m^2}{q^2}{\rm Im} C_t(\omega, q)=\frac{\varepsilon_K}{D}\frac{4\omega\gamma}{\left[\omega^2-\omega_0^2(q)\right]^2+4\gamma^2\omega^2},
\end{equation}
where $\varepsilon_K$ is the average effective kinetic energy density (per unit volume), and 
\begin{equation}
 \omega_0(q)=\sqrt{\gamma^2+\left(\frac{\hbar q^2}{2m^*}\right)^2},
\end{equation}
such that $q^{-2}{\rm Im} C_t(\omega, q)$ is formally identical to the imaginary part of the 
susceptibility of a damped oscillator of frequency $\omega_0(q)$, ``mass'' $Dm^2/(2\varepsilon_K)$, and 
friction $2\gamma$. 

From Eq. (\ref{Eq:effoscill}) the frequency-dependent viscosity
\begin{equation}
 \label{Eq:visc}
 \eta(\omega)=\frac{\varepsilon_K}{D}\frac{4\omega^2\gamma}{(\omega^2+\gamma^2)^2}, 
\end{equation}
which, once again, vanishes in the hydrodynamic limit if $\gamma$ is assumed to be uniform. 
The analysis based on Eq. (\ref{Eq:effoscill}) in which, unlike the derivation of Eq. (\ref{etaint}) no interactions 
(potential energy terms) were introduced is indeed {\it valid only in the hydrodynamic regime} where the 
``collision rate'' set by $\gamma$ (or $1/\tau$) is far smaller than $\omega$.  Thus, if the 
square in the denominator of Eq. (\ref{Eq:visc}) is replaced 
by 
\begin{eqnarray}
\label{hw}
(\omega^2+\gamma^2)^2 \to (\omega^{4} + 2 \gamma^{2} \omega^{2}),
\end{eqnarray} 
then 
the extrapolated $\lim_{\omega \to 0} \eta(\omega) = (2 \varepsilon_K)/(D\gamma)$. Performing the integrals to calculate the Gaussian average of the kinetic energy $\varepsilon_{K}$,
viz., trivially invoking equipartition, we find that
$\varepsilon_K = D {\sf n}k_{B} T/2$ and $\gamma = 1/\tau$. When these inserted, {\it we regain the general result} of Eq. (\ref{etaint}).
Albeit possibly pleasing, the replacement of Eq. (\ref{hw}) is not rigorous, to say the least.  The difficulty stems from the requisite dependence of 
the non-constant scattering rate $\gamma$
{\it on both} $q$ and $\omega$ in order to properly amend the non-interacting theory in the hydrodynamic regime, namely, $\gamma = \gamma(q,\omega)$.

We briefly consider some of the simplest possible forms for $\gamma$ and will then turn in Eq. (\ref{Eq:ImCt-1}) to another independent and far more rigorous derivation of the universal outcome of Eq. (\ref{etaint}).
The two simplest possibilities for achieving a finite  $\omega\to 0$ limit in Eq. (\ref{Eq:visc}) are $\gamma\sim \omega^{2/3}$ and 
$\gamma\sim \omega^2$. The former arises in an  Ising-Nematic system \cite{Metlitski-Sachdev}. Note, however, that 
in such a system the dispersion is anisotropic. The latter occurs in a Fermi liquid  
in three dimensions at low temperatures. There 
typically $\gamma\approx  \tilde{A}\hbar\epsilon_F\omega^2/(k_BT)^2$, where $\tilde{A}$ is a numerical 
coefficient. In this case we obtain for the viscosity,
\begin{equation}
 \eta\approx\tilde A\frac{4\hbar}{5}\left(\frac{\epsilon_F}{k_BT}\right)^2,
\end{equation}
where we have specialized to a spin degeneracy $g=2$. The above formula 
corresponds to the well-known $1/T^2$ behavior for the viscosity of a Fermi liquid \cite{Abrikosov-FLT}. 

As the above forms may be valid only in a limited hydrodynamic range of the frequencies, Eq. (\ref{Eq:effoscill})
does not lead to finite moments 
of the frequency for all $n$ in Eq. (\ref{Eq:moments}). Assuming a dependence on the frequency for $\gamma$ is 
an ad-hoc procedure which is not completely consistent- the sum rule \cite{Martin-Book}, 
\begin{equation}
\label{Eq:sum-rule}
 \int_{-\infty}^\infty\frac{d\omega}{\pi}\frac{{\rm Im}C_t(\omega,q)}{\omega}=\frac{{\sf n}}{m},
\end{equation}
is not fulfilled for a simple Ansatz $\gamma=c\omega^2$ with $c$ being a positive constant, since 
the integral diverges in this case. In the case of Fermi liquid theory the situation is different, since there $\gamma$ 
just behaves like $\sim\omega^2$ at low temperatures, while the complete result leads to a finite result for the 
integral appearing in the sum rule above. On the other hand, a convergent integral is obtained if one assumes 
that $\gamma$ depends only on $q$. Indeed, we obtain, 
\begin{equation}
\label{Eq:gamma-q}
 \gamma\approx\sqrt{\frac{k_BT}{m}}q,
\end{equation}
in the long wavelength limit. Unfortunately, in this circumstance we would obtain $\eta(\omega)=0$. 
This indicates that the best Ansatz should assume a $\gamma$ that depends both on $\omega$ and $q$. 
In this case our analysis is reminiscent from the interpolating Ansatz to compute the viscosity of simple fluids 
introduced by Forster {\it et al.} long time ago \cite{Forster-2}.  This approach is, in fact, more general, and 
is based on few principles involving sum rules relative to moments of the frequency, and analyticity arguments. 
In this sense, our simple phenomenological Ansatz (\ref{Eq:Cij}) can be viewed as a motivation to a more rigorous approach, 
where ${\rm Im}C_t(\omega, q)$ can be written quite generally as \cite{Martin-Book,Forster-2},   
\begin{equation}
\label{Eq:ImCt-1}
\frac{{\rm Im}C_t(\omega,q)}{C_t(0,q)}=\frac{\omega^3q^2D_t(\omega,q)}{[\omega^2-c^2(q)q^2]^2+[\omega q^2D_t(\omega,q)]^2},
\end{equation}
where $D_t(\omega,q)$ is  a dynamical diffusion coefficient,  
which can be determined from sum rules and has the form, 
\begin{equation}
\label{Eq:Dt}
 D_t(\omega,q)=[\tilde c^2(q)-c^2(q)]\tau(q){\cal D}[\omega\tau(q)],
\end{equation}
where  possible phenomenological interpolating Ans\"atze for the function ${\cal D}[\omega\tau(q)]$ are \cite{Forster-2}, 
\begin{equation}
\label{Eq:Ansatz}
{\cal D}[\omega\tau(q)]=
\left\{
\begin{aligned}
\frac{1}{1+\omega^2\tau^2(q)}, \\
 e^{-\omega^2\tau^2(q)/\pi},\\
e^{-2|\omega|\tau(q)/\pi},
\end{aligned}
\right.
\end{equation}
with $\tau(q)$ being a $q$-dependent relaxation time, which will actually have a different functional form  
depending on the chosen Ansatz for the function ${\cal D}[\omega\tau(q)]$.  
From the three possibilities in Eq. (\ref{Eq:Ansatz}), the first one is the less accurate, since it fails to yield a 
finite fourth moment of the frequency. This approach has the virtue of being more general and able to potentially yield 
the correct large time limit, which should correspond  to the non-interacting regime (\ref{Eq:ideal}).  

Upon accounting for the sum rule (\ref{Eq:sum-rule}) and evoking 
the Kramers-Kronig relaton, we easily obtain that $C_t(0,q)={\sf n}/m$.  
Thus, by identifying $2\gamma$ with $q^2 D_t(\omega,q)$, we see by comparing with Eq. (\ref{Eq:effoscill}) that  
${\sf n}\omega^2/m$ should match $2q^2\varepsilon_K/(Dm^2)$. This indicates that $\varepsilon_K$ would have to be 
dependent on $\omega/q$, something we have not assumed so far. A dependence on $\omega$ would indeed arise in a 
non-equilibrium situation where the distribution function is also time-dependent. In this case,
\begin{equation}
 \varepsilon_K(\omega/q)=\frac{\hbar^2}{2m}\int\frac{d^Dk}{(2\pi)^D}k^2f(k,\omega/q).
\end{equation}
On the other hand, by insisting on what we have been doing so far, we would be led to the identification of 
${\sf n}m\omega^2/(2q^2)$ and ${\sf n}k_BT/2$, obtaining in this way a frequency spectrum of the same form as 
the damping in Eq. (\ref{Eq:gamma-q}). 

For simple liquids the shear modulus vanishes, which implies in general that 
$c(0)=0$ in Eqs. (\ref{Eq:ImCt-1}) and (\ref{Eq:Dt}), while \cite{Martin-Book} (in particular, Eq. (47) in chapter C therein)
\begin{equation}
\label{MB}
 m\tilde c^2(q)=\frac{2\varepsilon_K}{D{\sf n}}+ 
 {\sf n} \int d^{D}r~ g^{(2)}(|{\bf r}|) \frac{\sin^{2}{{\bf q} \cdot {\bf r}}}{q^{2}}[\nabla^{2} - (\hat{q} \cdot \nablab)^{2}] ~\Phi_{l}(|{\bf{r}|}),
\end{equation}
where $g^{(2)}(|{\bf{r}}|)$ is the pair distribution function, and $\Phi_{l}$ the liquid pair potential.  In this way the hydrodynamic limit yields the 
viscosity in the form $\eta=m{\sf n}D_t(0,0)$. The two terms in Eq. (\ref{MB}) play a role analogous to those in Eq. (\ref{etax}).
In an ideal gas (for which the total kinetic energy is $(D V {\sf n}k_B T)/2$) with no higher order correlations (and consequent connected correlation function for general fluctuations), 
only the first term in Eq. (\ref{MB}) is important. We remark that at low temperatures, the second term in Eq. (\ref{MB}) is notably important.
In fact, as the temperature is elevated from very low values and the system becomes less stiff, 
the speed of sound $\tilde{c}$ initially decreases with the temperature. By the Newton-Laplace equation,
$\tilde{c}(0) \sim \sqrt{G_{\infty}/(m {\sf n})}$ where $G_{\infty}$ is the infinite frequency shear modulus. 
(Although, as stated above, a defining proper of liquids is that they exhibit no resistance to shear at long times, they do behave like solids 
with an effective modulus $G_{\infty}$ at very short times.) The speed of sound in low temperature systems is notably different from the rms 
thermal speed. The relaxation time $\tau$ that we computed in earlier sections was associated with the transitions from
one state to another (not the speed of a uniform acoustic mode in a nearly static solid or liquid- one whose state does not change). 
That is, there is a difference between the speed within a local element of the fluid and that associated
with the motion of a fluid element relative to the external fluid which is linked to mixing.

\section{Lower bounds on the extrapolated high temperature viscosity in semi-classical systems}
\label{lower_b}

In Section \ref{Boltzmann-sec}, employing the monotonicity of the viscosity in temperature within the liquid phase, we derived the bound of Eq. (\ref{nh}) on the viscosity.
We will now elaborate on how this bound follows from the results in Section \ref{Sect:correlation}. Our aim is to reinforce the perhaps this all too simple
lower bound on the viscosity and to, later on, motivate that this bound might be saturated (i.e., becomes an equality) in the extrapolated high temperature
limit of the viscosity in the liquid phase.

 Similar to the discussion following Eq. (\ref{etax}), in {\it the low density limit and/or extrapolated high temperature limit}, the first term in Eq. (\ref{MB}) dominates. 
The result of Eq. (\ref{MB}) may be compared and equated with the weighted average $\overline{\tau}$ of kinetic relaxation times $\tilde{\tau}({\bf{r}}_{1}, {\bf{U}}_{1})_{\bf{r}_{2}, \cdots \bf{r}_{N}, {\bf{U}}_{2}, \cdots {\bf{U}}_{N}}$ in Eq. (\ref{etaint}). 
Therefore, from Eq. (\ref{Eq:Dt}), in $D$ spatial dimensions, and at extrapolated high temperatures where the first term in Eq. (\ref{MB}) dominates,
\begin{equation}
\label{etac}
 \eta(0)=m{\sf n}\tilde c^2(0)\tau \sim \frac{2\varepsilon_K\tau}{D},
\end{equation}
where $\tau=\tau(0)$ and in the argument of $\eta$, we highlight the neglect of the second, higher order, term valid in the low density limit.
As $\varepsilon_K=D{\sf n}k_BT/2$, we obtain once more the result of Section \ref{Boltzmann-sec}, viz., 
\begin{eqnarray}
\label{enk}
\eta(0)={\sf n}k_BT\tau.
\end{eqnarray}
This result is true irrespective of the Ansatz used in 
(\ref{Eq:Ansatz}). Thus, the simple kinetic theory calculations of 
Section \ref{Boltzmann-sec} leading to Eq. (\ref{etaint}) may be generally rederived by employing a potent approach
based on the dynamical diffusion coefficient $D_t(\omega,q)$ when the latter is examined at high temperatures.  
We highlight that Eqs. (\ref{etaint}, \ref{etac}) are more general than Eq. (\ref{nh}) in which the particular limit of Eq. (\ref{rTL}) is invoked. 

Related observations have been made in other works, e.g., \cite{hartkamp} along the following lines. The zero shear viscosity of the fluid is 
the ratio of the hydrostatic pressure to the collision rate , $\eta(0) ={\sf P}/r$. For dilute fluids, the hydrostatic pressure ${\sf P} = {\sf n} k_{B} T$
and thus we have Eq. (\ref{enk}).
For finite densities,  the viscosity may be empirically may be written as an exponentiated polynomial multiplying the low density limit \cite{rowler},
\begin{eqnarray}
\eta({\sf n}) \equiv  \eta(0) W =  \eta(0) e^{\sum_{i=1}^{s} B_{i} {\sf n}^{i}}
\end{eqnarray}
where the coefficients $\{ B_{i} \}$ are not of a fixed sign. 
We remark that, clearly, in those systems in which $\eta({\sf n}) \ge \eta(0)$ then by Eqs. (\ref{tmin},\ref{enk}), we will have the result of Eq. (\ref{nh}) yet again.
A more practical related bound can be generally derived.
Inserting Eq. (\ref{rA}), we have from Eq. (\ref{enk}) for a dilute liquid
\begin{eqnarray}
\label{eta0s}
\eta(0) = {\sf n} he^{- \Delta S_{A}/k_{B}}  e^{\beta \Delta H_{A}},
\end{eqnarray}
for the viscosity of a dilute fluid in the vicinity of an equilibrium temperature $T_{A}$.
Thus if we fit the viscosity curve to an Arrhenius form 
\begin{eqnarray}
\label{Arr}
\eta(0) = A e^{\beta \Delta},
\end{eqnarray}
by examining the {\it local tangent} to the $\ln \eta$ curve when plotted as a function of the inverse temperature $\beta= \beta_{A}$,
we have for the coefficient at $T=T_{A}$, the amplitude 
\begin{eqnarray}
\label{Af}
A = {\sf n} he^{- \Delta S_{A}/k_{B}}.
\end{eqnarray}
As, by its definition,at any temperature $T$, the entropy difference $\Delta S \le 0$, the amplitude of the local Arrhenius form
in Eq. (\ref{Arr}) must always satisfy 
\begin{eqnarray}
\label{Ac}
A  \ge   {\sf n} h,
\end{eqnarray}
at any equilibrium temperature $T_{A}$; this relation is more experimentally pertinent than Eq. (\ref{nh}). As explained in Section \ref{measurements}, 
at the lowest equilibrium temperature $T_{A}^{\min}$, one may anticipate $\Delta S_{A} =0$ leading to a precise quantization of the prefactor $A$. 
Due to the availability of excited transition system states (vis a vis the minimal energy escape states) at high energies/temperatures, one may anticipate the effective gap $\Delta$ in Eq. (\ref{Arr}), i.e., the enthalpy difference $\Delta H_{A}$ increases with $1/T_{A}$.
Furthermore, if the entropy difference $\Delta S_{A}$ is monotonically decreasing in temperature $T_{A}$
at which it is evaluated (or, stated equivalently, the difference in entropies between the full system $S_{transition}$ and that comprised of escape level subsystem $S_{escape}$
is monotonically increasing in temperature $T_{A}$) then, as Eq. (\ref{Af}) attests, the prefactor $A$ becomes larger. Thus, when $\ln \eta$ is plotted as a function of $(1/T)$
the local slope at a temperature $T_{A}$ is the energy difference $\Delta H_{A}$ which may increase with the temperature $T_{A}$ concomitantly with an increase of the adduced  
prefactor of Eq. (\ref{Af}). Such a prediction may be empirically tested (this trend seems to be in accord with numerical data, e.g., \cite{numerics}).

Most generally, of course, the local Arrhenius form at $T=T_{A}$ is {\it exactly} given by
\begin{eqnarray}
\label{wgeq}
\eta &&= W({\sf n}, T_{A}) \times \eta(0) \nonumber
\\ &&=  W({\sf n}, T_{A}) \times {\sf n} he^{- \Delta S_{A}/k_{B}}   e^{\beta \Delta H_{A}},
\end{eqnarray}
with $W$ a density (and temperature) dependent function. One may trivially calculate this viscosity from Eqs. (\ref{ratio},\ref{enk}) for simple cases such as a particle in a box \cite{box2}.

In Section \ref{normal}, we will compare experimental results with what one would anticipate if the lower bounds of Eqs. (\ref{tmin},\ref{nh},\ref{Ac}) would be saturated-
that is, if these inequalities would become equalities in the extrapolated limit of dynamics and viscosity in the liquid phase to infinite temperatures. 
As it is illuminating, we first review Eyring's theory of the viscosity and then explain how our derived bounds and the exact relations of Eqs. (\ref{etax},\ref{wgeq})
lead to these results if an assumption identical to Eq. (\ref{EnT}) is invoked. 

\section{Possible strict bounds on viscosity/entropy values in general systems}
\label{strict_b}

Eqs. (\ref{tmq}, \ref{etaint'}) further suggest a restriction on the extrapolated viscosity of general (i.e., not necessarily semi-classical) liquids to high temperatures,
\begin{eqnarray}
\eta \ge \frac{{\sf n}h}{4 \pi^{2}}.
\end{eqnarray}
If the entropy of a high temperature liquid  is bounded from above by that of an ideal gas at the same temperature (Eq. (\ref{ST}))
then in the extrapolated high temperature limit (wherein the thermal de Broglie wavelength $\lambda_{T}^{nr} \to \infty$), the viscosity to entropy density ratio,
\begin{eqnarray}
\label{eh5}
\frac{\eta}{s} \ge \frac{\hbar}{5 \pi k_{B}},
\end{eqnarray}
where $s$ is the entropy per unit volume.
This bound is $(4/5)$ smaller than that of Eq.(\ref{Eq:visc-bound}) originally motivated by holography in string theory. 
(We find a lower bound that is twice as large as Eq. (\ref{eh5}) if we invoke $\tau_{quantum}(T) \gtrsim \frac{h}{2 \pi^{2} k_{B} T}$ of \cite{hod}.)  
In general spatial dimension $D$, our above derivation of Eq. (\ref{eh5}) will lead to $\frac{\eta}{s} \ge \frac{1}{D+2} (\frac{\hbar}{k_{B}})$.

\section{The Eyring theory for the viscosity (with additional constraints) from the single particle Boltzmann equation}
\label{s:Eyring}

We now will briefly review Eyring's theory for the viscosity (which gives rise to the first term in Eq. (\ref{etax})). In doing so, we will highlight a property that has been overlooked until now. Namely, we will explain how, by its mean single particle nature, this theory must give rise to the leading (single particle) contribution to the viscosity in the Boltzmann equation. Thus, contrary to what Eyring advanced, in thermodynamic equilibrium this theory {\it must lead to the aforementioned first term Eq. (\ref{etax}) with no undetermined prefactors} (as present in his original derivation). The absence of widely varying
additional prefactors is important; in Section \ref{normal} we will find that, on average, the prefactors in fitting the viscosity with Eq. (\ref{Arr}) at the lowest temperature at which the system remains in equilibrium are very close to those anticipated from our simple kinetic theory considerations. That is, contrary to earlier approaches, we suggest that at this lowest equilibrium temperature, the prefactor $A$ may be very nearly equal to unity. 

Eyring \cite{eyring-viscosity} (and later Cohen and Grest \cite{CG}) advanced theories for the viscosity of liquid (and supercooled liquids) based on considerations of available volume for minimal processes. In what follows, we will review these ideas following \cite{eyring-viscosity} and then posit a new extension from these that may lead to a sharp quantization that saturates our derived bound of Eq. (\ref{nh}). 
As in the Cohen-Grest free volume theory \cite{CG}, Eyring assumed that excess volume was cardinal in order to enable motion. Eyring posited the existence of ``holes'' in a liquid to which particles may move to (leaving a hole in their former location). 
Effectively, he viewed the combination of a particle and a ``hole'' as an ``activated molecular complex'' and tried to view motions as transitions via such molecular systems. He modeled the rate of particle motion to such holes as a reaction process and subtracted the rate of backward motion (namely, that in a direction opposite to an applied shear) from that of forward motion. The notion of a ``hole'' (or vancacy) is, e.g., well defined for a solid system and electronic Fermi liquids, yet not quite so for general gaseous and liquid systems.
We remark that although holes \cite{eyring-viscosity,tabor} in a liquid might be ill-defined, the notion of examining, in a broader and more rigorous context, pairs of time reversed states (any motion with or without imaginary holes) as a way to sum over
the evolution of all states is viable. Towards that end, we may revisit and amend Eyring's derivation and consider all processes $\Upsilon_{\mp}$ for the rate of forward and backwards motion relative to the applied external shear direction. These time reversed states have energy shifts relative to each another.

Following his logic and terminology, Eyring considered neighboring ``layers'' of particles (transverse to the $z$ direction in Section \ref{Boltzmann-sec}) each of width $\lambda_{1}$ drifting, in the notation of Section \ref{Boltzmann-sec}, along the $x$ direction with speed $\Delta u$ relative to the layer beneath them. The distance between viable low energy particle locations in the same layer is defined to be $\lambda$ (the distance between viable minima in the cartoon of Fig. (\ref{Potential}). The area per ion in the sheared layer is ${\sf{A}}$. Eyring equated the viscosity $\eta = f \lambda_1/\Delta u$ with $f$ the force per unit area leading to the displacement of one layer relative to another with relative speed $\Delta u$ (i.e., in more standard nomenclature, $f$ is the shear stress). The applied external shear force symmetrically lowers (when $f>0$) the energy barrier for forward motion (along the positive $x$ direction) while elevating it for backwards motion by amounts $(\mp( f \lambda {\sf{A}} /2))$ respectively. The rates of forward and backwards motion are (consistent with Eq. (\ref{ratio})) are
\begin{eqnarray}
r_{forward}= r e^{\beta f {\sf{A}} \lambda/2}, ~ r_{backwards} = r e^{-\beta f {\sf{A}}  \lambda/2},
\end{eqnarray}
with $r$ the rate in the absence of an applied external shear.  
Thus \cite{eyring-viscosity}, the difference in velocities between layers and consequent viscosity are given by
\begin{eqnarray}
\Delta u = \lambda r(e^{\beta f {\sf{A}} \lambda/2}- e^{-\beta f {\sf{A}} \lambda/2}), \nonumber
\\ \eta = \frac{f \lambda_1}{ \lambda r(e^{\beta f {\sf{A}} \lambda/2} - e^{-\beta f {\sf{A}} \lambda/2})} \nonumber
\\ \sim \frac{k_{B} T \lambda_1}{r \lambda^2 {\sf A}},
\end{eqnarray}
where in the last line the asymptotic form for $f \lambda {\sf{A}} \ll k_{B} T$ was invoked. 
This can be re-expressed as $\eta = {\sf n} k_{B} T/r \times (\lambda_{1}/\lambda) = {\sf n} k_{B} T \tau \lambda_{1}/\lambda$.
The ratio $(\lambda_{1}/\lambda)$ is that between the layer width- the inter-particle distance along the direction transverse to the applied shear- 
to the distance of the jump along the applied shear direction. Clearly, in a uniform isotropic fluid, if the jump distance ($\lambda$) is equal to the inter particle distance 
along any direction (including the transverse direction for which the inter particle distance is $\lambda_1$) then the viscosity will be given by the first term of Eq. (\ref{etax})-- the dilute fluid result of Eq. (\ref{enk}). As noted in the beginning of this section and as the reader can now verify the above derivation of Eyring invokes only a uniform single particle relaxation time. As it considers only single particle motion with uniform relaxation rate, whatever the details of derivation and underlying assumption, if it is consistent, the result {\it must} reduce to the uniform single particle contribution to the general Boltzmann equation described in Section \ref{Boltzmann-sec}. This is so as the single particle Boltzmann derivation assumes nothing but the existence of a uniform relaxation time and the existence of an equilibrium Boltzmann distribution. Thus, we see that as promised in the beginning of this section, within the picture of single particle motion with constant $\tau$ (regardless of its microscopic origins associated with volumes and ``holes''), effectively the ratio $(\lambda_{1}/\lambda)$ introduced by Eyring {\it cannot, contrary to \cite{eyring-viscosity}, be left arbitrary}; this ratio
{\it must be replaced by unity}. 

If one goes beyond the uniform single particle picture it is evident that unlike in a  periodic crystal, in the many disparate processes that may summed over (as in Eq. (\ref{ratio})), the local values of the ratio $(\lambda_{1}/\lambda)$ may vary. Spatially averaged over each individual liquid, this distribution may lead to an effective shift of $(\lambda_1/\lambda)$ from unity. 

\section{Comparison with measured viscosities of metallic fluids}
\label{normal}

In this section, we contrast the considerations of 
the previous sections with experimental measurements of complex metallic fluids exhibiting nontrivial atomic composition and interactions,
e.g., \cite{mendel}. We will perform our analysis by fitting the viscosity of Eq. (\ref{wgeq}) setting $We^{- \Delta S_{A}/k_{B}}$ to unity. As discussed in subsection \ref{measurements}, 
we may set at the lowest temperature at which the system is in equilibrium, $\Delta S_A =0$ (we will  return to this in Section \ref{ETH_section}
and further conjecture that (the random) corrections in the second term of Eq. (\ref{etax}) lead to a value of $W=1$). We now fit the 
viscosity in the vicinity of this lowest temperature at which the system is still in equilibrium to an Arrhenius form (Eq. (\ref{rA})),
\begin{eqnarray}
\label{AnH}
\boxed{\eta = A {\sf n} h \exp(-\beta \Delta {\sf H}_{A})}
\end{eqnarray}
and then extract
the prefactor $A$. An Arrhenius form is a prediction of several theories of liquid (and supercooled) dynamics, e.g., \cite{eyring-viscosity,tabor,avoided_critical,
non-abelian_PRB,Review_LPS,egami,itamar_also_change_at_TA}. The lowest temperature at which the liquid is still in equilibrium was gauged by (i) its smooth viscosity form as the liquid  was rapidly cooled from high temperatures (see Appendix \ref{experiment-appendix} and  \cite{metallic-glass})
as well as (ii) observing, numerically, what is the lowest temperatures at which the Stokes-Einstein relation still holds \cite{numerics}. 

If the hypothesis concerning the saturation of the exact inequalities of Eqs. (\ref{tmin}, \ref{nh}, \ref{Ac})
in the extrapolated infinite temperature limit and/or the assumption of the ensemble average of Eq. (\ref{EnT}) is correct
then, on average, ${\mathbb{E}} \{A_{a}\} =1$. We first focus on the assumption of an exact saturation of Eqs. (\ref{tmin}, \ref{nh}, \ref{Ac})
with forms more complicated than Eq. (\ref{AnH}) and explain why this will canonically give rise to a Gaussian 
distribution of the prefactors $A$ about an average value of unity. 

 We remark that following the spirit of Jaynes \cite{jaynes}, given a large ensemble liquids $\{a\}$, we can examine the probability distribution $Prob(A)$ for having a particular prefactor in Eq. (\ref{AnH}).
If we define a function $Q(A)$ by setting
\begin{eqnarray}
Prob(A) \equiv {\cal{N}} e^{-Q(A)},
\end{eqnarray}
with ${\cal{N}}$ a normalization constant and require that the average value 
$\int dA ~Prob(A)~ A  = 1$ 
(that may be mandated by a choice of the Lagrange multiplier as in canonical ensembles for the average anergy instead of an average value of $A$)
and that  
$Prob(A)$
has its maximum at $A=1$ (requiring that $Q$ have a quadratic minimum at $A=1$),
we find a Gaussian form for the probability distribution of $A$ values. 
Stated alternatively,
the exact relation of Eq. (\ref{rTL}) will become inexact when we use the approximate form of Eq. (\ref{AnH}). This unbiased approximate form will satisfy 
this relation only on average as information is lost when we use the less detailed approximate form of Eq. (\ref{AnH}). In general data sets, one may work backwards from the distribution of measured values such as those of $A_{a}$ to find a corresponding ``effective energy'' $Q$.
That is, in general problems, given a distribution value for fit parameter values, exact constraints such as those concerning a saturation of Eq. (\ref{nh}) will become
embedded in the logarithm ($Q$) of the probability function. If no other constraints or biases appear then $Q$ will have its minimum about the 
value that it should have attained as an exact equality (viz., a Gaussian distribution appears). On a more rudimentary level, if the higher order corrections in Eq. (\ref{etax}) (or those form higher order multi-particle correlation functions) from different liquids in an ensemble are 
uncorrelated with each other and effectively randomly distributed in a Gaussian fashion about the universal leading order term of ${\sf n} k_{B}T \tau$ then one would anticipate a corresponding scatter in the values of $A_{a}$. In Section \ref{ETH_section}, we will further explain that on supercooling a liquid, a distribution of effective temperatures may result;
this may broaden the spectrum of prefactors $A_{a}$ in Eq. (\ref{AnH}).

\begin{figure}[htp] 
\centering
\includegraphics[width=6cm, angle=-90]{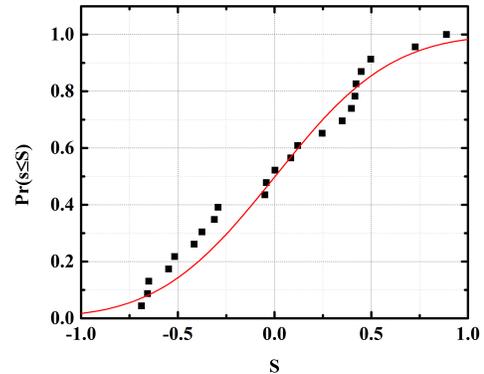}
\caption{(Color online.) Contrast with experiment. This graph illustrates how close the average of the prefactor in the Arrhenius form of Eq. (\ref{AnH}) is to unity (Eq. (\ref{eah})).
Here it is possible to gauge the distribution of extrapolated high temperature viscosity of supercooled metallic liquids about the 
quantum limit set by ${\sf n} h$ with ${\sf n}$ the particle density and $h$ Planck's constant. Specifically, in this figure, we plot along the vertical axis is the fraction of liquids in the ensemble of 23 liquids experimentally studied (see the Appendix for details) and in \cite{metallic-glass}, i.e., the probability  $Prob(s  \le S)$, as a function of the extrapolated high temperature viscosity assuming the single exponent form of Eq. (\ref{AnH}) with $s = (A- 1)$ (black square symbols associated with the 23 data points). 
The red curve corresponds to the cumulative function 
associated with the normal distribution, 
$\frac{1}{2}[ 1+Erf(S - \langle s \rangle)/(\sigma_{s} \sqrt{2})]$.}
\label{Gauss}
\end{figure}

In Figure \ref{Gauss}, we compare these expectations with the data reported in \cite{metallic-glass}  (in particular, table S2 therein). 
The values of  $\{A_{a}\}$ reported in \cite{metallic-glass} were found by fitting the viscosity of supercooled fluids at temperatures just slightly above the temperature (labelled $T_{coop}$ in Figures (\ref{universal_collapse}, \ref{standard_dev})) at which each liquid adheres to Eq. (\ref{AnH}) (as described in more detail in the Appendix). In the context of Section \ref{measurements},
$T_{coop} \equiv T_{A}^{\min}$.

At lower temperatures, $T<T_{coop}$, the system falls out of thermal equilibrium (as, e.g., seen by the violation of the Stokes-Einstein relations) and cooperative effects set in \cite{metallic-glass,numerics}. At such low temperatures, departure from equilibrium becomes more noticeable and the viscosity grows faster than Eq. (\ref{AnH}) with a constant energy barrier $\Delta {\sf H}_{a}$.
That is, in Figure \ref{Gauss} and Eq. (\ref{AnH}), we choose the temperature $T_A$ of Section \ref{measurements} to be $T_{coop}$- the lowest temperature in which thermal equilibrium still holds. In Section \ref{ETH_section}, we briefly discuss a particular interpretation that may be associated with this lowest equilibrium temperature, $T_{coop} \equiv T_{A}^{\min}$.

As explained and invoked in subection \ref{measurements} and in Section \ref{lower_b} (in particular, Eq. (\ref{wgeq}) therein), this relates to our form of Eqs. (\ref{particle_derivation},\ref{ratio}) with an effective $\mu(T)$ set to its limiting value in the lower temperature end of
the equilibrated liquid phase (where equilibration still holds yet will break down at lower temperatures, $T<T_{coop}$). 
Figure \ref{Gauss} depicts the probability of the deviation of the extrapolated high temperature viscosity from the theoretical value of ${\sf n}h$.  The 
$s \equiv (A-1)$ values are contrasted with those anticipated from a Gaussian distribution.  The comparison suggests that the measured values of $A$ 
are indeed normally distributed around the theoretical value of unity. Over the ensemble of measured 23 liquids, 
the mean value, when averaged over all liquids ($1 \le a \le 23$), of 
\begin{eqnarray}
\boxed{{\mathbb{E}}(\{A_{a}\}) = 0.99353}, 
\label{eah}
\end{eqnarray}
and
the width of the distribution $\sigma = 0.453$. For the average value ${\mathbb{E}}(\{A_{a}\})= \frac{1}{23} \sum_{a=1}^{23} A_{a}$
the corresponding standard deviation about the mean value is $\sigma/\sqrt{22} \sim 0.09$. 

 The activation barriers $\{\Delta {\sf H}_{a}\}$ that we found when fitting the data to Eq. (\ref{AnH}) are reasonably close to the $k_{B} T_{\sf evaporation}$ as is consistent with the discussion following Eq. (\ref{particle_derivation}). 
 
 \section{The lowest equilibrium temperature and possible links to quantum thermalization}
 \label{ETH_section}
 It may appear to be little connection between the lowest temperature at which the liquid remains in equilibrium $T_{coop} \equiv T_{A}^{\min}$  to the melting or, more precisely, the liquidus temperature. The liquidus temperature is, by its definition, the lowest temperature at which a material may remain completely liquid.
 In the current section, we would like to suggest that a relation between the two might appear quite naturally.
 This connection may, yet again, be a consequence of the underlying quantum nature of the liquid. As has been argued for by numerous researchers \cite{eth1,eth2,eth3,von_neumann}, thermalization may be an inherently quantum phenomenon of many body systems. More specifically, a quantum system with an energy density (or temperature) lying above the mobility edge \cite{Mobility_Edge} may become thermalized at long times: the long time average in quantum state may equal to a static average
 computed via statistical mechanics. Any quantum state may be expanded in eigenstates. Ergodicity then suggests that 
 such a thermalization might be manifest even at the level of a single eigenstate (as advanced by the ``Eigenstate Thermalization Hypothesis'' \cite{eth1,eth2,eth3}). 
 In Appendix (\ref{ETH-A}), we discuss the essentials of this hypothesis as it pertains to our work.
 According to the simplest rendition of this hypothesis, for pertinent operators ${\cal{O}}$, the average
 \begin{eqnarray}
 \label{ETH-h}
 \langle n| {\cal{O}}| n \rangle = Tr (\rho {\cal{O}}).
 \end{eqnarray}
On the lefthand side of Eq. (\ref{ETH-h}), the expectation value of ${\cal{O}}$ is computed within an eigenstate $|n \rangle$ of energy $E$;
on the righthand side a statistical mechanical average is performed with a density matrix $\rho$ (e.g., micro-canonical or canonical) associated
 with the same energy $E$. Within such a picture, eigenstates with an energy above the mobility edge may become thermalized.  By contrast, eigenstates of sufficiently low energies (below the mobility edge), need not thermalize and instead the system may become ``many body localized''. In random systems, it has indeed been established that many body localization may even persist to energy densities associated with the system at infinite temperatures \cite{MBL}. When eigenstate thermalization holds, the microcanonical average is equal to that evaluated within a single stationary eigenstate (related extensions even appear for systems with time dependent Hamiltonians \cite{von_neumann}). In such thermalized quantum states, the system is ergodic. If the Eigenstate Thermalization Hypothesis is taken seriously then this might imply that (i) the loss of full ergodicity of the liquid as it slowly cooled to form a solid breaking translational, rotational, or other symmetries and (ii) the non thermalized states breaking the Stokes Einstein relations at low enough temperatures \cite{numerics} might need to share a common origin. That is, at sufficiently low energy densities, the system eigenstates lie below the mobility edge and are largely localized; at these temperatures or energy densities, the eigenstates no longer satisfy the Eigenstate Thermalization Hypothesis (as indeed occurs in many body localized systems).  Accordingly, a consequence of this viewpoint is that associated energy scales below which the eigenstates no longer satisfy the Eigenstate Thermalization Hypothesis need to be the same for both (a) the melting or liquidus temperature $T_{l}$ and (b) the lowest equilibration temperature of the liquid ($T_{coop} \equiv T_{A}^{\min}$) at which, e.g., thermodynamic considerations leading to decay rates of the form of Eq. (\ref{ratio}) as well the Stokes Einstein relations  still hold may be the same. Taken together, (a) and (b) imply that 
 \begin{eqnarray}
 \label{TLTC}
 T_{l} \approx T_{coop}.
 \end{eqnarray} 
 This viewpoint further suggests that at the lowest equilibration temperature $T_{A} = T_{coop} \approx T_{l}$, as no fully ergodic thermal states lie below the mobility edge, the entropy difference may indeed precisely saturate, i.e., $\Delta S_{A} =0$ in  Eq. (\ref{wgeq}) (as invoked in Section \ref{normal} and Eq. (\ref{AnH})). For hydrodynamic motion restoring ergodicity (and full thermalization) there are available states at corresponding energies (or associated temperatures).  Below that, the states are non-ergodic and the full thermalization over all of phase space is not restored (even though there are, of course, numerous low energy states that are partially localized and non-ergodic). At the energy associated with $T=T_{l}$ there may be only one ergodic state. At lower energy, no fully ergodic states remain. Semi-classically, for a single ergdoic state at the energy scale associated with $T \approx T_{l}$  (leading to $\Delta S_{A} =0$ at such a lowest equilibration temperature), one may anticipate that the average $\langle \tau U_{1z}^{2} \rangle
= \langle \tau \rangle \langle U_{1z}^{2} \rangle$ in Eq. (\ref{etaint}) leading to vanishing corrections in Eq. (\ref{etax}).
Invoking Eqs. (\ref{ratio}, \ref{etax}) and that $\tau = 1/r$, all of these considerations further motivate that the prefactor in 
Eq. (\ref{AnH}) for the viscosity at the lowest equilibration temperature $T_{A}^{\min} = T_{coop}$ satisfies $A \approx 1$. 
  
\begin{figure}[htp] 
\centering
\includegraphics[width=6cm, angle=-90]{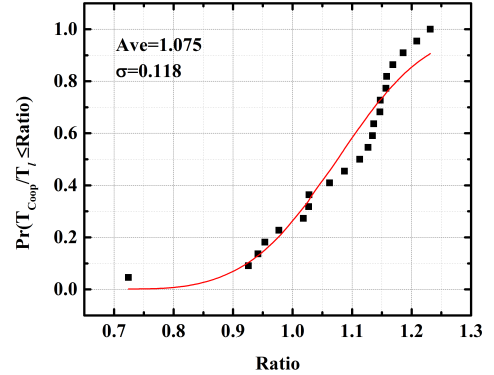}
\caption{(Color online.) An experimental comparison between the lowest temperature $T_{coop} \equiv T_{A}^{\min}$ at which Arrhenius dynamics appear and the system may be equilibrated
(and can, e.g., satisfy, the Stokes Einstein relation \cite{numerics})
to the liquidus temperature $T_{l}$ . Specifically, we plot, based on the data of  \cite{metallic-glass},  along the vertical axis is the fraction of liquids in the ensemble of 23 liquids studied here (see the Appendix for details) 
for which the ratio $(T_{coop}/T_{l})$ is smaller than a given value. The black square symbols associated with the 23 data points); the continuous red curve corresponds to the cumulative function  associated with the normal distribution, 
$\frac{1}{2}[ 1+Erf({\sf Ratio} - {\sf Ave})/(\sigma \sqrt{2})]$
where {\sf Ave} $=1.075$ denotes the average value of $(T_{coop}/T_{l})$ for the 23 examined supercooled metallic liquids.
A similar average appeared for the $11$ non metallic glass formers studied in \cite{non-abelian_PRB}.}
\label{Gauss'}
\end{figure}

In what will follow, we will elaborate on the consequences of possible widening of the energy distribution function upon supercooling of liquids. These may render some of our suggested exact equalities in earlier sections only correct, on average.  
 In Figure \ref{Gauss'}, we contrast $T_{l}$ and $T_{coop}$ for 23 experimentally examined supercooled metallic fluids and indeed find them to be close to one another. Quite naturally,  in a supercooled liquid, the system might not have ``enough time'' to thermalize at the lowest equilibration temperature as it is rapidly cooled below that temperature before becoming ``frozen'' in a quenched state. On very short time scales, only local readjustments might be more readily possible. Consequently, the effectively measured lowest such equilibration temperature may be higher than the true ideal minimum amongst the system eigenstates. Below $T_{coop}$ there may be an effective crossover coexistence of the many body localized and ergodic states; eigenstates that do not fully thermalize and become ergodic (i.e., do not satisfy the Eigenstate Thermalization Hypothesis) may be occupied. Only a fraction of these eigenstates might become delocalized and equilibrate. 
 In \cite{non-abelian_PRB}, a near equality between the two temperatures (for other fluids) 
 was found to be consistent with simple expectations for ideal tessellations for a simple liquid favoring icosahedral structures (as found in metallic glasses, e.g., \cite{numerics,icosahedral}). 
 For instance, for a simple Lennard Jones liquid that favors local low energy icosahedral structures (possibly associated with an ``avoided'' transition at $T_{coop}$ \cite{avoided_critical,non-abelian_PRB, Review_LPS}), the local energy density is 8.4\% lower
 than that of a regular lowest energy crystal associated with the global energy minimum (connected with $T_{l}$). Such a deviation is not far off the mark from the average ${\mathbb{E}} (T_{coop}/T_{l}) = 1.075$ found for the 23 examined metallic fluids in Figure \ref{Gauss'}.  For the 11 non-metallic supercooled liquids studied in \cite{non-abelian_PRB},
 this average was $1.096$. Thus, for the total ensemble of 34 supercooled liquids studied here and in \cite{non-abelian_PRB}, the average value of ${\mathbb{E}}
 (T_{coop}/T_{l}) = 1.082$ 
 (a value that happens to be proximate to the 8.4\% difference in energy densities between the minimal energy global structures (a face-centered cubic (FCC) structure)
  and lowest local energy density icoashedral configuration in the quintessential simple Lennard Jones liquid \cite{non-abelian_PRB}). 

We now reiterate and formalize our considerations. This will enable us to argue for more general stringent inequalities augmenting the approximate equality of Eq. (\ref{TLTC}). 
Starting from any high temperature equilibrium state of the liquid (for which the Eigenstate Thermalization Hypothesis of Eq. (\ref{ETH-h}) applies), we can cool via 
the unitary operator 
\begin{eqnarray}
\label{uti}
U(t_{final}, t_{initial}) = {\cal{T}}  e^{-\frac{i}{\hbar} \int_{t_{initial}}^{t_{final}} dt' H(t')},
\end{eqnarray}
where ${\cal{T}}$ is the time ordering operator between the initial and final cooling times and $H$ the system Hamiltonian with time dependent parameters that 
capture the cooling starting from the high energy equilibrated system (associated with a temperature $T_{initial}>T_{coop}$) at an initial time $t_{initial}$ to lower energies at a time $t_{final}$ (characterized by low measured temperature $T_{final} = T$)
For supercooled fluids, the evolution
operator $U$ may, e.g., encode locally preferred structures in glassy systems,  \cite{avoided_critical,non-abelian_PRB, Review_LPS}. In slowly cooled annealed systems,
$U$ may give rise to an ordered crystalline state. Generally, 
the state 
\begin{eqnarray}
\label{pfi}
\psi(t_{final} )= U(t_{final}, t_{initial}) \psi(t_{initial}).
\end{eqnarray}
This state may be expanded in the eigenbasis
of $H(t_{final})$- the time independent Hamiltonian with which it evolves
for all subsequent times $t \ge t_{final}$. That is, we may, of course, write
\begin{eqnarray}
\label{dec}
\psi(t_{final}) = \sum_{j} c_{j} \phi_{j},
\end{eqnarray} 
where $\{\phi_{j}\}$ are the eigenstates of $H(t_{final})$. 
The full system Hamiltonian $H(t_{final})$, that of the many atom system with all kinetic and interaction terms in tow, may admit crystalline or other symmetry breaking eigenstates
at low energy densities (or temperatures); these states may be experimentally observed below
the melting/liquidus temperature in slowly cooled liquids with an essentially time independent $H$. 
By contrast, at sufficiently high energies $E_{j}$, the associated eigenstates $\phi_{j}$ correspond to 
the uniform liquid. (At yet higher energies, gaseous and other phases may appear.) 
If the system is held below the liquidus temperature $T_{l}$ then, clearly, the eigenstates $\{\phi_{j_{low}}\}$
having an energy density lower than that of melting
enjoy a sizable total probability, $\sum_{j_{low}} |c_{j_{low}}|^{2} ={\cal{O}}(1)$. (For clarity, we emphasize that these low energy eigenstates constitute a subset of all eigenstates of $H(t_{final})$,
i.e., $\{\phi_{j_{low}}\} \subset \{\phi_{j}\}$.)
The eigenstates $\{\phi_{j_{low}}\}$ no longer satisfy the Eigenstate Thermalization Hypothesis
as the crystal (or other solid state with which they are associated) is no longer ergodic. 
Instead, such low energy states $\{\phi_{j_{low}}\}$ may lie below the mobility edge and be many body localized \cite{MBL}. 
Thus, at these energy densities, the state $\psi(t_{final}) $ is not completely ergodic as it has, in Eq. (\ref{dec}), components $\phi_{j_{low}}$ lying below the mobility edge. Consequently, the long time evolution $(t>t_{final})$ generated by the time independent $H(t_{final})$ is not ergodic- the low energy states will remain 
localized at long times. Putting all of the pieces together, augmenting the approximate equality of Eq. (\ref{TLTC}), we may have the following as a corollary:

$\bullet$ (i)  {\it If a system has a melting temperature (or more general liquidus temperature) $T_{l}$ at which ergodicity is broken 
for an annealed liquid, then regardless of how rapidly this liquid is cooled from high temperatures, it must become non-ergodic at a temperature} 
\begin{eqnarray}
\label{tctl}
\boxed{T_{coop} \ge T_{l}.}
\end{eqnarray} 

Both Eqs. (\ref{TLTC}, \ref{tctl}) seem to be largely in accord with the experimental data as displayed in Figure \ref{Gauss'}. 

For, e.g., a crystal formed below the melting temperature, the system states are not ergodic as translational and other symmetries are broken (but are ergodic only within a subspace of
 phase space). In such a case, we may make the system effectively ergodic by augmenting the system Hamiltonian (such as that appearing in the density matrix in Eq. (\ref{ETH-h}))
by external fields and/or boundary terms. If
such fields are added then the system will remain ergodic at all positive temperatures. By the Eigenstate Thermalization Hypothesis, if the slowly cooled liquid becomes a crystal or
another solid that remains thermal then even if such a system is supercooled below its liquidus temperature then

$\bullet$ (ii)  {\it No finite temperature
``ideal glass'' transitions (at which a supercooled liquid falls out of equilibrium on {\bf all} times scales) exist}.  

This possible corollary is at odds with the most prominent fit for glassy dynamics. Indeed, according to the celebrated Vogel-Fulcher-Tammann-Hesse (VFTH) fit \cite{vft1}, the viscosity
of a supercooled liquid is given by
\begin{eqnarray}
\label{VFT}
\eta_{VFTH} = \eta_{0} e^{DT_{0}/(T-T_{0})},
\end{eqnarray}
where $\eta_{0}, D,$ and $T_{0}$ are liquid dependent constants.
As Eq. (\ref{VFT}) makes clear, the VFTH function 
posits an ideal glass transition temperature $T_{0}$
at which the relaxation times and viscosity diverge. 
Other expressions, e.g., \cite{metallic-glass, avoided_critical,non-abelian_PRB, Review_LPS} for the viscosity are consistent with both (i) and (ii). In what follows, we derive similar  forms consistent with both corollaries. 

We briefly further underscore and reiterate the possible physical content of item (ii). Starting from a high temperature state, a slowly cooled liquid may become
a solid. The resulting solid breaks ergodicity via, e.g., translational and rotational symmetry breaking. Nevertheless, such an annealed solid formed by gradual cooling
generally remains thermal and is ergodic within a subvolume of phase space. According to (ii) above if, starting from the initial high temperature state, 
this very same liquid (that forms a thermal solid upon slow cooling) is rapidly supercooled then this system may, 
correspondingly, still remain thermal at all positive temperatures irrespective of the cooling rate.

We now turn to an examination of the possible more specific {\it predictive quantitative consequences} of the Eigenstate Thermalization Hypothesis-many body localization paradigm at low temperatures. Our earlier analysis led to Eq. (\ref{wgeq}) at sufficiently high temperatures ($T \ge T_{coop}$) such that the system remains in equilibrium; in the vicinity of the lowest temperature  ($T=T_{coop}$) at which the system still remains in equilibrium, the viscosity was of the form of 
Eq. (\ref{AnH}) with $A =1$ (as was tested in Section \ref{normal}). 
We now turn to the situation of supercooled fluids at temperatures $T <T_{coop}$. As emphasized earlier and touched on in corollary (i), at these temperatures the system is no longer ergodic. In the eigenstate decomposition of Eq. (\ref{dec}), the sum $\sum_{j_{low}} |c_{j_{low}}|^{2} ={\cal{O}}(1)$- a large component of the system is many body localized. 
A complementary high energy piece (that associated with energy densities above melting or liquidus temperature $T_{l}$) of the wavefunction lies above the mobility edge.
In the simplest analysis, only this higher energy part of the wavefunction of Eq. (\ref{dec}) can contribute to transport. If the Eigenstate Thermalization Hypothesis holds for this constituent of the wavefunction
then its contribution to the system properties will be identical to that anticipated from equilibrium thermodynamic considerations. We may therefore anticipate that the relaxation rate
for transport will be given as a weighted superposition of Eq. (\ref{ratio})) at different effective temperatures (corresponding to the ``stretching'' of the energies values appear
in the decomposition of Eq. (\ref{dec})). For the states formed by supercooling (Eqs. (\ref{uti},\ref{pfi},\ref{dec})), the energy distribution function is given by ${\it{p}}(E) = \sum_{j} \delta(E_{j} - E) |c_{j}|^{2}$. 
As noted above, this broad energy spectrum may rationalize the inequality of Eq. (\ref{tctl}). In the vicinity of $T_{coop}$, it is expected that
energy densities corresponding to effective temperatures in its proximity are superposed in the rate of Eq. (\ref{ratio}).
These may lead to a Gaussian broadening for the effective free energy barrier in the likes of Eq. (\ref{wgeq}) and Eq. (\ref{AnH}) with the associated prefactor $A$.
At far lower temperatures, the contribution to transport from 
energies corresponding to effective temperatures near $T_{l}$ may become smaller. If one ``blocks out'' the localized low energy components 
$\{\phi_{j_{low}}\}$ and focuses only on the high energy part of the wavefunction (that corresponding to energies above the melting temperature)
then at final temperatures $T<T_{coop} \sim T_{l} $, we will anticipate the viscosity to the of the form of Eq. (\ref{AnH}) yet with a prefactor $A<1$.
Specifically, at low temperatures/energies, the equilibrium calculation relaxation rate contribution of Eqs. (\ref{ratio},\ref{rA}) will be multiplied by the probability of the system to be in a thermalized eigenstate. The viscosity is calculated from the transverse component of the retarded current correlation function,  Eqs. (\ref{Eq:eta-C},\ref{Eq:Cij}). 
When evaluated in the state $|\psi_{final} \rangle$, the current operator commutator 
\begin{eqnarray}
\label{Eq:Cij_long}
\langle \psi_{final} | [ j_i(t,{\bf r}),j_j(t',{\bf r}')] | \psi_{final} \rangle = \nonumber 
\\  \sum_{\zeta,\zeta' = high,low}  \langle \psi_{\zeta} | [ j_i(t,{\bf r}),j_j(t',{\bf r}')] | \psi_{\zeta^{'}}\rangle,
\end{eqnarray}
where $|\psi_{high} \rangle$ and $| \psi_{low} \rangle$ are the un-normalized projected high- and low-energy components of $|\psi_{high} \rangle$.
That is, $| \psi_{high} \rangle$ includes only eigenstates (such as those in Eq. (\ref{dec})) with an energy above that of melting while $| \psi_{low} \rangle$
only has components lying in $\{\phi_{j_{low}}\}$. The state $| \psi_{final} \rangle$ is equal to the sum of these two vectors, 
$|\psi_{final} \rangle = |\psi_{high} \rangle + | \psi_{low} \rangle$. As highlighted in Section \ref{Sect:correlation}, caution is required when taking the $\omega\to 0, q\to 0$ limits of the correlator in Eq. (\ref{Eq:Cij_long}). Now, here is a trivial yet important point: the cross-term $\langle \psi_{high} | [ j_i(t,{\bf r}),j_j(t',{\bf r}')] | \psi_{low}\rangle$ is vanishingly small as the {\it local} current operators $j$ cannot link global states of different macroscopic energy. We next consider what occurs if no hydrodynamic transport is present at low energies when the system solidifies such that $\langle \psi_{low} | [ j_i(t,{\bf r}),j_j(t',{\bf r}')] | \psi_{low}\rangle$ may be further omitted from the sum in Eq (\ref{Eq:Cij_long}). 
The resultant simple equality $\langle \psi_{final} | [ j_i(t,{\bf r}),j_j(t',{\bf r}')] | \psi_{final} \rangle 
=   \langle \psi_{high} | [ j_i(t,{\bf r}),j_j(t',{\bf r}')] | \psi_{high}\rangle$ simplifies our analysis as, by the Eigenstate Thermalization Hypothesis,
this latter expectation value may be calculable by thermodynamic considerations. In other words, we may invoke our results from
the previous sections when focusing on $|\psi_{high} \rangle$. With this simplification, in the simplest approximation, our previous calculations for the relaxation rate $r(T)$ and the viscosity $\eta$ will then be trivially amended as we that may need to do is to find the contribution from the high energy eigenstates in $| \psi_{high} \rangle$. Their contribution is associated with the ``high energy tail'' of the energy probability function distribution ${\it{p}}_{T}(E)$ associated with the final measured temperature $T$ characterizing the final 
state of Eq. (\ref{dec}). (In a related vein, e.g., \cite{two_fluid}, the relaxation rate associated with an elastic response function in a
system, comprised of a nearly non-dissipative solid land a dissipative liquid type dissipative parts is proportional to the strength of the liquid component response \cite{triv_example}.) Thus, invoking the Eigenstate Thermalization Hypothesis, and an equivalence of ensembles, the transition rate in the supercooled ({\sf{s.c.}}) system,
\begin{eqnarray}
\label{rt++}
r_{\sf{s.c.}} (T) = \int_{T_{l}}^{\infty} dT' r(T') ~{\it{p}}_{T} \big(E(T')\big) ~ \Big( \frac{d E}{d T'} \Big),
\end{eqnarray}
where $E(T')$ is the internal energy of an equilibrated system at a temperature $T'>T_{l}$ and $r(T')$ is given by Eq. (\ref{ratio})
(i.e., the equilibrium specific heat of the liquid/solid at temperature $T'$ is $C_{v}(T')=\frac{d E}{d T'}$). 
In a simple and very crude approximation, the viscosity set by $1/r_{\sf{s.c.}}$ is, approximately, 
\begin{eqnarray}
\label{sumhigh}
\eta \sim 
\frac{\eta_{coop}}
{\sum_{j_{high}} |c_{j_{high}}|^{2}}.
\end{eqnarray}
If the rates associated with transitions from different energy eigenstates are independent of each other, a possible and more refined expression
for the viscosity in which 
the variation of the equilibrium thermalization rate (i.e., with a temperature dependent $r(T')$ in Eq. (\ref{rt++}) as given by (Eq. (\ref{ratio}))) is incorporated is, for temperatures $T<T_{coop}$, given by
\begin{eqnarray}
\label{sumexact} 
\eta = \eta_{coop} \Big( \frac{r_{\sf{s.c.}}(T_{coop})}{r_{\sf{s.c.}}(T)} \Big).
\end{eqnarray} 
In Eq. (\ref{sumhigh}), the high energy sum ${\it{p}}_{high}= {\sum_{j_{high}} |c_{j_{high}}|^{2}}$ is performed over those eigenstates in Eq. (\ref{dec}) for which the energy density is larger than that associated with $T_{coop} \sim T_{l}$, (i.e., ${\sum_{j_{high}} |c_{j_{high}}|^{2}} = \int_{E_{coop}}^{\infty} dE~ {\it{p}}(E)$). In Eqs. (\ref{sumhigh},\ref{sumexact}), $\eta_{coop}$ is the viscosity at $T=T_{coop}$. The amplitudes $\{c_{j}\}$ (and thus the cumulative probability sum ${\it{p}}_{high}$) depend on the details of cooling protocol embodied in the evolution operator $U(t_{final}, t_{initial})$, see Eqs. (\ref{uti},\ref{pfi}).
As such, nothing precise can be stated about these. If we assume a generic ``random'' state for which the average energy density $\langle E \rangle$ is that associated with a temperature $T$ for which, similar to the considerations in Section \ref{normal}, a Gaussian distribution of width $\sigma$ will appear then the probability to be in a state of energy exceeding $E_{coop}$ (i.e., an energy above that associated with the lowest equilibration temperature $T_{coop}$ or proximate melting temperature) is
\begin{eqnarray}
\label{ph}
{\it{p}}_{high} = \frac{1}{2} erfc \Big( \frac{E_{coop}-\langle E \rangle)}{\sigma_{E} \sqrt{2}} \Big).
\end{eqnarray}
The width $\sigma_{E}$ of the energy distribution ${\it{p}}_{T}(E)$
will clearly be a function of the temperature $T$ to which we supercooled below the lowest equilibration temperature $T_{coop}$.
That is, if an initially equilibrated annealed system at $T_{initial}>T_{coop}$ is supercooled to a final temperature $T= T_{initial}^{-}$, only barely below the initial temperature, then the probability
${\it{p}}(E)$ will be nearly a Dirac delta function centered about an average $\langle E \rangle$ set by the initial temperature $T_{initial}$.
Thus, we anticipate $\sigma$ to monotonically increase in the size of the temperature interval $(T_{coop}-T)$; starting from a given initial state 
the larger the supercooling interval, the more mixing of different energies is anticipated in Eq. (\ref{dec}).
If, in Eq. (\ref{sumhigh}), we replace the energies $E_{coop}$ and $E$ by the lowest equilibrium and final temperatures $T_{coop}$ and $T$ to which they respectively correspond (and, by equipartition, to which they are equal to up to a constant prefactor in harmonic systems) or, equivalently, assume a simple Gaussian distribution of width $\overline{\sigma}$
 for the effective temperatures corresponding to the energies in the distribution ${\it{p}}(E)$) and employ the asymptotic relation $erfc(z \gg 1) \sim \frac{e^{-z^{2}}} {z \sqrt{\pi}}$ then, by Eqs. (\ref{sumhigh}, \ref{ph}), for $(T_{coop} - T) \gg \overline{\sigma}$, the viscosity
\begin{eqnarray}
\label{geoni}
\eta(T) \sim \frac{\eta_{coop} \overline{\sigma} e^{(T_{coop}-T)^{2}/(2 \overline{\sigma}^{2})}}
{{\sqrt{2 \pi}(T_{coop}-T)}}.
\end{eqnarray}

Thus putting all of our simple considerations together, we predict that (a) at temperatures above $T_{coop}$, the viscosity may be given by the likes of Eq. (\ref{wgeq}),
(b) for temperatures in the vicinity of the lowest equilibration temperature, $T \gtrsim T_{coop}$, the viscosity is set by Eq. (\ref{AnH}) with an average prefactor $A=1$,
while (c) at temperatures far below $T_{coop}$, it can be generally given by Eqs. (\ref{sumhigh},\ref{ph}) that may asymptotically simplify to Eq. (\ref{geoni}). 
Furthermore, as we argued for in the beginning of this section, the lowest equilibration temperature is proximate to the melting or liquidus temperature $T_{coop} \sim T_{l}$
(Eqs. (\ref{TLTC},\ref{tctl})). For a fixed $\overline{\sigma}$, the low temperature expression of Eq. (\ref{geoni}) is somewhat similar to that the ``BENK'' fit in \cite{metallic-glass} and the modified parabolic fit of \cite{ejcg} (the latter forms would be arrived at if one were to assume a Gaussian distribution in the effective inverse temperature instead of the temperature)
as well as the avoided critical fit \cite{metallic-glass, avoided_critical,non-abelian_PRB, Review_LPS}. In general, of course,
neither the distribution ${\it{p}}(E)$ nor its corresponding variant for the effective temperatures nor inverse temperatures
need to be Gaussian. More detailed forms of ${\it{p}}_{T}(E)$ beyond a minimal Gaussian and invoked all too simple lrelation between the internal energy and temperature yield more intricate expressions for dynamical attributes such as relaxation rates and the viscosity, as well as  long time expectation values of single particle observables and thermodynamic quantities. As a hypothetical illustrative example, if as the system is supercooled and evolved according to Eqs. (\ref{uti},\ref{pfi}), certain components of 
$\psi(t_{final})$ may not have sufficient time to change greatly relative to their initial value while other low energy eigenstates $\phi_{j}$ in Eq. (\ref{dec}) are populated then, instead of a single Gaussian, the distribution  ${\it{p}}_{T}(E)$ may be the sum of two 
Gaussians- (i) one at energies near those of the initial equilibrated system at $T_{coop}$ of total weight $x(T)$ the and (ii)  another Gaussian at lower energies, i.e., explicitly,
\begin{eqnarray}
\label{2G}
 {\it{p}}_{T}(E) =  \Big[ \frac{x(T)}{\sqrt{2 \pi \sigma^{2}_{1}(T)}}  e^{-(E-E_{coop})^{2}/(2 \sigma^{2}_{1}(T))} \nonumber
 \\ + \frac{1-x(T)}{\sqrt{ 2 \pi \sigma^{2}_{2}(T)}}  e^{-(E-E_{2}(T))^{2}/(2 \sigma^{2}_{2}(T))} \Big],
 \end{eqnarray}
 with $E_{2}$ far smaller than $E_{coop}$,  
such that the average of $E$ computed with this distribution is affiliated with the internal energy at the final temperature $T$. As the system may be evolved for longer times when cooling to lower final temperatures, the overlap between ${\it{p}}_{T}(E)$ with that of higher energy system (set by $x$ for narrow Gaussians) is expected to diminish. Similarly, as we more broadly discussed earlier, the width of the lower energy distribution may increase as the system is evolved with an external Hamiltonian for a longer time. 
For this hypothetic example, in the extreme limit of $x=0$ at low temperatures, a dramatic ``fragile'' \cite{fragile} increase of the viscosity with temperature (such as that present in Eq. (\ref{geoni})) is expected while diametrically opposite limit ($x=1$), just below $T_{coop} \sim T_{l}$ with a narrow $\sigma_{1}$, a near Arrhenius (or ``strong'') behavior arises from the distribution of Eq. (\ref{2G}).
Along different lines, in examining the distribution of effective barriers in a simulated glass former,
the authors of \cite{itamar_also_change_at_TA} found the latter to be a sum of two Gaussians (with temperature dependent amplitudes and widths)
with precisely such trends. Regardless of physical motivations which are prone to inaccuracies, Eq. (\ref{2G}) is simply a rather trivial example of distributions beyond the minimal single Gaussian one that we invoked earlier. Generally, an approximation such as that of Eq. (\ref{sumhigh}), or its trivial rewriting, 
\begin{eqnarray}
\label{solvep}
\frac{\eta_{coop}}{\eta(T)} = \int_{E_{l}}^{\infty} dE ~{\it{p}}_{T}(E),
\end{eqnarray}
may enable the practical determination of ${\it{p}}_{T}(E)$ at different temperatures $T$ 
given disparate forms of the viscosity $\eta(T)$. Similarly, the correspondence of Eq. (\ref{solvep}) and its counterparts may, in principle, be tested numerically in simulated (classical) model systems. 

In numerous liquids, structure may be similar to that of (the low energy) crystal states on short and intermediate length scales, e.g., \cite{makov}. Thus far, apart from the variation of the barrier $\Delta G$ was ignored for energies above $E_{coop}$. Likewise, we neglected relaxation rates associated with the low energy solid like eigenstates and summed, in Eq. (\ref{rt++}), only over contributions from the ergodic higher energy states. Often, in solids, relaxations (including those of vacancy, interstitial, dislocation, or other defects) are of an Arrhenius type, with a rate $e^{-\beta \Delta_{d}}$, where $\Delta_{d}$ is the barrier for defect motion. If the low energy solid like states do not greatly differ from the higher energy equilibrated liquid states then, correspondingly, the dynamics (including the viscosity) is not expected to change dramatically. Such low energy contributions may be of greater pertinence to the so-called ``strong'' glass formers \cite{fragile} while the analysis of Eqs. (\ref{rt++},\ref{sumhigh},\ref{ph},\ref{geoni},\ref{solvep}) in which the low-energy contributions have been ignored can be of more direct relevance to ``fragile'' glass-formers \cite{fragile}. These effects may augment those attributed to Eqs. (\ref{2G},\ref{solvep}) alone. A temperature variation of the distribution ${\it{p}}_{T}(E)$ may not only trigger changes in the dynamics but also in thermodynamic measurements or general long time averages of various local quantities. Albeit not being ergodic at temperature $T<T_{coop}$ and having a non-equilibrium $p_{T}(E)$, if similar to phase cancellations at long times, standard long time averages ($l.t.a.$) such at that of Eq. (\ref{llong}) still hold (as we underscored earlier, the disorder free final Hamiltonian $H(t_{final}) = H$  is ergodic in a subspace of phase space (see corollary (ii))), then $\overline{{\cal{O}}}_{l.t.a} = \int dE~ {\it{p}}_{T}(E) ~ {\cal{O}}(E)$. In such a case, a change in the form of the distribution function $p_{T}(E)$ (e.g., in the hypothetical example of Eq. (\ref{2G}), a near vanishing $x$ (effectively leading to a single Gaussian giving rise to the ``fragile'' form of Eq. (\ref{geoni})) beyond a certain $T=T_{1}$) may give the impression of a ``phase transition'' at $T\sim T_{1}$. Similarly, if, hypothetically, ${\it{p}}_{T}(E)$ becomes finite for the ground state energy $E=E_{g.s.}$ below another temperature, $T<T_{2}$, then measurements associated with such a non-analytic change of $p_{T}(E)$ may suggest a thermodynamic phase transition at $T=T_{2}$ although, as our corollaries (i) and (ii) demonstrate, none might exist. Lastly, we remark that the breaking of translational and rotational symmetries below the melting temperatures (or, for eigenstates 
below the corresponding melting energy) may naturally lead to spatially non-uniform dynamics. Indeed, liquids below their melting temperatures rather ubiquitously 
exhibit spatially non-uniform dynamics- so called {\it dynamical heterogeneity} \cite{DH}. At temperatures only slightly below $T_{l}$, a significant component of
$p_{T}(E)$ is still associated with the ergodic states. At far lower temperatures, however, what largely remains
is the low energy non-ergodic component.  Dynamical heterogeneities may naturally to the breaking of 
ergodicity and translational and rotational symmetries within the low energy crystalline or other solid eigenstates of $H$ of different energies. In particular, at low energies, in the sum of Eq. (\ref{dec}), certain low energy states
$|\phi_{j} \rangle$  in Eq. (\ref{dec}) may exhibit disparate phonon and other excitations in different solids that are rotated and translated relative to one another (all of which have energies $E$
with the aforementioned distribution). At higher energies (yet below melting), anharmonic plastic and other effects may appear in pertinent high weight states $|\phi_{j} \rangle$
(as they do in solids). 

 An analog of Eq. (\ref{rt++}) holds for general energy dependent observables 
${\cal{O}}(E)$ and other quantities that may be effectively thermal in the low energy eigenstates of $H$. 
If the system is localized and non-ergodic below the meting temperature yet its distribution in a phase space subvolume is such that still exhibits a thermal distribution of ${\cal{O}}$ at positive temperatures, then
the mean value of ${\cal{O}}$ in the supercooled liquid, 
\begin{eqnarray}
\label{thermog}
{\cal{O}}_{\sf{s.c.}} (T) =\int_{E_{g.s.}}^{\infty} dE~  {\it{p_{T}}} (E) ~ {\cal{O}} (E) \nonumber
\\ =  \int_{0}^{\infty} dT' ~{\cal{O}}(T') ~{\it{p}}_{T} \big(E(T')\big) ~ C_{V}(T'),
\end{eqnarray}
with the integrand evaluated for an equilibrated system at $T'$. If the interval of energies in which ${\it{p}}_{T}(E)$ has its relevant support
does not correspond to an energy (or associated temperature) at which the low energy solid exhibits a phase transition then, similarly, ${\cal{O}}_{\sf{s.c.}}$ may not
 display a phase transition (as in corollary (ii)). Approximate equalities of the type of Eqs. (\ref{rt++}, \ref{solvep}) and Eq. (\ref{thermog}) may, respectively, relate
 (via the distribution ${\it{p_{T}}} (E)$)
 {\it dynamics to thermodynamics} and may rationalize the ``Kauzmann paradox'' \cite{walter} sans an assumption of an ideal glass transition. Kauzmann observed that
 the entropy of supercooled liquids appears to veer below an extrapolated temperatures $T_K$ to values below that of the solid (an impossibility- hence the name ``paradox'') at nearly the same temperature
at which the extrapolated viscosity seems to diverge (as the temperature $T_{0}$ of Eq. (\ref{VFT}) empirically satisfies $T_{0} \sim T_{K}$). Eqs. (\ref{solvep}, \ref{thermog}) are consistent with such a trend. That is, if there exists a temperature $T_{K}$ below which, upon extrapolation from high temperatures, ${\it{p_{T}}} (E)$ appears to have all of its support from the low energy states of the solid then the viscosity of Eq. (\ref{solvep}) will diverge in unison with the tendency of static thermodynamic quantities to approach those of the equilibrated low energy solid.  This is so as the latter thermodynamic quantities are functions of the internal energy, ${\cal{O}}(E)$ of Eq. (\ref{thermog})
and their derivatives relative to $T$ and will thus yield values that may coincide with those of the low temperature solid. 
In principle, measurements of general thermodynamic quantities ${\cal{O}}$ over a wide range of temperatures $T$ (such that Eq. (\ref{thermog}) may
be solved to obtain ${\it{p_{T}}}(E)$), may leads to the thermalization rate of Eq. (\ref{2G})) or approximations to the viscosity 
(Eq. (\ref{solvep}) or Eq. (\ref{sumexact})) having {\it no adjustable parameters}. In future work, we hope to test these predictions. In practice, however, current data might not allow for
such a close scrutiny of our approach. While there is no shortage of theories of the glass transition, e.g., \cite{CG,avoided_critical,non-abelian_PRB, Review_LPS,ejcg,
glass-theory} including, in particular, those specifically associating sluggish dynamics to, local in space, solid like clusters and defects, e.g., 
\cite{avoided_critical,non-abelian_PRB, Review_LPS,glass-theory} none have directly derived dynamics from a direct correspondence 
with the eigenstates of the equilibrium solid and liquid states at different temperatures. An Occam's razor type appeal of the strategy outlined in this section is the theory is free of {\it any specific structural or other details or assumptions}. In our approach, based on quantum dynamics, both the dynamics and expectation values of thermodynamic observables of supercooled liquids are dictated by the evolution of the supercooled liquid state into a superposition of different energy (or effective equilibrated temperature) eigenstates whose properties are known (i.e., the equilibrium high temperature liquid and low temperature solid states). 
 
 \section{Viscosity of quantum critical systems}
 \label{vis:qc}

We conclude our discussion of the viscosity fit with a brief remark and additional prediction concerning {\it quantum critical} systems. The derivation of Eq. (\ref{etaint}) does not assume 
Eq. (\ref{rTL}). Thus, if similar to the analysis in subsection \ref{qcp:sec}, the rate is given by Eq. (\ref{rt:qcp}) then we will
find that instead of Eq. (\ref{nh}), the extrapolated high temperature viscosity
\begin{eqnarray}
\label{nh+}
\lim_{T \to \infty} \eta^{\sf quantum~critical} = e^{a_{+}} {\sf n} h.
\end{eqnarray}

Since quantum critical systems with dynamic exponent $z=1$ resemble conformal invariant relativistic systems, it is useful 
to explore these similarities in the context of the viscosity. For example, 
the free energy volume density, ${\cal F}$,  of a scale invariant relativistic system is typically given by \cite{Subir-polylog,Appelquist} 
\begin{equation}
{\cal F}={\cal F}_0-f(k_BT)^{D+1}/(\hbar c)^D
\end{equation}
where ${\cal F}_0$ is 
the ground state (free) energy density, $c$ is the speed of light, and $f$ is a universal constant. This yields an entropy density, 
\begin{equation}
s=\frac{f(D+1)k_B}{(\hbar c)^D}(k_BT)^D.
\end{equation}
Scale invariance also implies that the viscosity can be determined by dimensional analysis, so we can introduce another universal number, 
$c_\eta$, to write $\eta/s=c_\eta\hbar/k_B$. Holographic arguments lead to $c_\eta=1/(4\pi)$.  In Ref. \cite{Appelquist} it was 
conjectured that the number $f$ satisfies the inequality $f_{IR}\leq f_{UV}$ for theories having infrared (IR) and ultraviolet (UV) fixed 
points, with the UV fixed point being asymptotically free. Assuming the validity of this inequality, we obtain, 
\begin{equation}
\left(\frac{\eta}{c_\eta}\right)_{IR}\leq\left(\frac{\eta}{c_\eta}\right)_{UV}.
\end{equation} 

\section{The Navier Stokes equation and possible minimal time scale for discrete dynamics}
\label{ns}

Liquid dynamics adheres to the Navier Stokes equation,
\begin{eqnarray}
\rho\left[\frac{\partial {\bf v}}{\partial t} + ({\bf v} \cdot \nablab) {\bf v}\right]
= -{\nablab} P + {\nablab} \cdot {\cal{T}} + {\bf f}.
\end{eqnarray}
Extensions of this equation appear in Korteweg hydrodynamics \cite{kdv}.
Here, $\rho$ the mass density, ${\bf v}$ the velocity, $P$ the pressure, 
${\bf f}$ external body forces (per unit volume), and ${\cal{T}}$ is the stress tensor. In the canonical setting, 
${\nablab} \cdot {\cal{T}} = \eta \nabla^{2} {\bf v}$. Albeit its simplicity, the regularity
of the solutions of the Navier Stokes equation in $D=3$ dimensions is an open problem. Given the results 
in the current work, $\eta$ may be bounded by its limiting
high temperature value in Eq. (\ref{nh}). This implies that the Reynolds number, 
$Re = \frac{\rho v \ell}{\eta}$,
with $\ell$ the typical mean free path and $v$ the average speed,
may saturate to $\rho/({\sf n} h) \times (vL) = p \ell/h$ (with $p = mv$ the linear momentum scale) 
at high temperatures. In this limit, $Re \sim L/\lambda_T^{nr}$ with $\lambda_T^{nr} = h/p$ the de-Broglie wavelength. For particles to be well defined, $\ell$
cannot be significantly smaller than the de-Broglie wavelength.

While the Navier Stokes equations and their viable extensions are those of a continuous velocity field, the results that we discussed highlight that transitions and motion on a microscopic scale are not continuous but rather set by {\it minimal time increments} of order $\tau_{\min}$ (see Eq. (\ref{tmin})) for basic 
processes to occur.  Taken literally, this smallest time step set by the reciprocals of 
Eq. (\ref{rTL}) suggests that in fluids motion cannot be considered as continuous 
and an effective time lag may be microscopically present between molecular transitions/motions. Such a requisite time separation is reminiscent to that introduced in recent analysis of the Navier-Stokes equation \cite{tao} in studying (ir)regularity properties of viable solutions.

\section{Electrical response- possible exact bounds and asymptotic saturation values}
\label{ec}

The approach that we invoked in discussing viscosity may, in principle, be replicated 
for other transport measurements for general multi-particle systems with dynamic degrees of freedom (e.g., point like charges, topological defects or large clusters, etc.). 
To conceptually illustrate the basic premise
(with no pretense of rigor), we consider a dirty insulator with charges ($e$) 
of volume density ${\sf{n}}$ in the presence of an applied electric field ${\bf E}$.
In what follows, we extend and apply the considerations of Section \ref{s:Eyring} to this system.
If the mean free path of between the ions is $\ell = {\sf n}^{-1/D}$ in $D$ dimensions then
the difference between the forward and backwards flow along and opposite the applied
field will be
\begin{eqnarray}
\label{dr}
\Delta r(T) =r_{forward}(T) - r_{backwards}(T) = r(T)  \beta e |{\bf E}| \ell,
\end{eqnarray}
where $r(T)$ is the rate of Eqs. (\ref{particle_derivation},\ref{ratio}) in the absence of an applied field,
and we Taylor expand the exponential of Eqs. (\ref{particle_derivation},\ref{ratio})  
for both forward and backward
motion ($r_{forward}, r_{backwards}$) to obtain the factor of $\beta e |{\bf E}| \ell$. The lowering/increase in the energy barrier for backwards and forward motion is linear in the distance $\ell$ between charges on adjacent ions and the applied field. 
Pictorially, the applied field will tilt the potential barrier in Figure \ref{Potential}.
In the high temperature limit, we may invoke Eq. (\ref{rTL}), and Eq. (\ref{dr}) becomes
$\Delta r(T) \sim e |{\bf E}| \ell/h$. The charge velocity between ions is $v= \ell \Delta r$
and the total current $j = {\sf n} e v$ becomes, at high temperatures, $j \sim {\sf n} e^{2} \ell^{2}  |{\bf E}|/h$
and the conductivity is thus 
\begin{eqnarray}
\sigma \sim {\sf n}^{1-2/D} \Big( \frac{e^{2}}{h} \Big) \equiv \sigma_{\infty}.
\end{eqnarray} 
Within this simple approach, in two dimensions, the high temperature resistance will decrease with temperature and tend, in the high temperature limit, to $h/e^{2}$- a value akin to the quantum of resistance. Resistivities (both longitudinal as above as well as transverse) in $D=2$ dimensional systems might extrapolate and saturate to $e^{2}/h$ at high temperatures, e.g., \cite{QHR,amir} in accord with the 
the Ioffe-Regel criterion \cite{Ioffe}.

We next consider a classical metal with a mean free path $\ell$ that is a function
of temperature and is larger than the inter particle separation ${\sf n}^{-1/D}$.
We now examine what occurs as temperature is increased within our framework. 
Towards this end, we invoke the Drude formula, $\sigma = {\sf n} e^{2} \tau/m$ with $m$
the (quasi) particle mass and $\tau = 1/r(T)$ of Eqs. (\ref{particle_derivation},\ref{ratio}). 
In this case, at asymptotically high temperatures, the resistivity is linear in temperature and at finite temperatures, this is multiplied by an effective factor of $\exp(-\beta \Delta {\sf H}_{A})$
akin to Eq. (\ref{particle_derivation}),
 \begin{eqnarray}
 \label{rHA}
 \rho \sim \Big( \frac{mk_{B}T}{{\sf n}e^{2}h} \Big) e^{-\beta \Delta {\sf{H}}_{A}}. 
 \end{eqnarray}
 The considerations that we have invoked throughout this work apply when the system is at sufficiently high temperatures so that it is at thermal equilibrium (so that
 the kinetic theory Einstein relations (including, e.g., the Stokes-Einstein relation \cite{numerics}) holds) \cite{MBL'}.  Kinetic theory considerations (including the Einstein diffusivity relations) do not hold in non-equilibrium systems
 such as supercooled liquids and many body localized systems. When it holds, as seen from Eq. (\ref{rHA}), at temperatures $T \gg \Delta {\sf H}_{A} /k_{B}$, the resistivity may increase linearly with temperature until the mean free
 path saturates to the inter particle distance ${\sf n}^{-1/D}$. When this saturation occurs, 
 $\sigma \sim \sigma_{\infty}$ as in the insulating case above. Some time after our work 
 initially appeared \cite{qv}, we learned \cite{jan_private} that related bounds for the diffusion in strongly correlated electronic systems sans the exponential 
 factor in Eq. (\ref{rHA}) were also very recently suggested by \cite{hartnoll} assuming that the relaxation time is of order ${\cal{O}}(\hbar/(k_{B} T))$. In strongly correlated electronic systems, the notion of quasiparticles may be ill-defined. The diffusivity bounds of \cite{hartnoll} might be violated in the strong correlation regimes as evinced by a numerical study of a Hubbard model on a Bethe lattice \cite{pakhira}. We reiterate that our considerations above may hold only for {\it non-localized equilibrated thermal} systems continuously connected to the high temperature liquid sans intervening transitions or crossovers.

\begin{figure}[htp] 
\centering
\includegraphics[width=6cm]{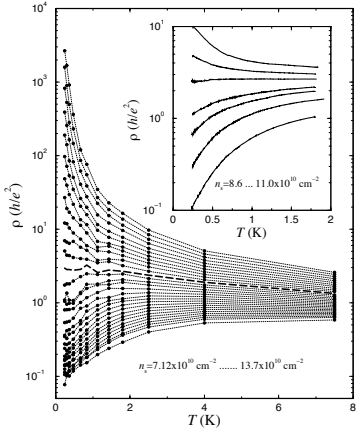}
\caption{(Color online.) {Reproduced from \cite{kravchenko}.
The resistivity of a two dimensional electronic system
in  a disordered MOSFET for different electronic densities.
The electronic densities are $8.6, ~8.8,~9.0, ~9.3, ~9.5, ~9.9,$ and $11 \times 10^{10} cm^{-2}$.
At higher temperatures, both insulators and conductors veer towards the dashed separatrix for which
$\rho \sim h/e^{2}$.}}
\label{mit}
\end{figure}

 In the current work, we demonstrated that, in many instances, the semiclassical relaxation in general multi-particle systems time may generally be exactly equal to $\tau_{\min}$.  More generally, 
 from our Eq. (\ref{tmq}) for arbitrary quantum systems (also those with no well-defined quasi-particles), we propose that for {\it all} electronic systems (whether having semi-classical relaxation times or not) that adhere to an effective Drude form, 
 \begin{eqnarray}
 \label{resist_bound}
 \rho \le \frac{2 \pi m k_{B} T}{{\sf n} e^{2} \hbar}.
 \end{eqnarray}
 Naturally, of course, the basic premise of the Drude form holds for quasi-particle systems. 
 When the Drude form is valid, we arrive at a yet more stringent bound (a factor of two smaller than the righthand side of Eq. (\ref{resist_bound})) if we invoke the decay rate limits suggested by
  \cite{hod}, viz., $\tau_{quantum}(T) \gtrsim \frac{h}{2 \pi^{2} k_{B} T}$. 
 These possible rigorous bounds (when a Drude form applies) seem to be experimentally realized and nearly saturated. Indeed, looking at Fig. (2) of \cite{bruin}, the resistivities of all materials with linear in temperature increase (``bad metals'') compiled therein (CeCoIn$_{5}$, UPt$_{3}$, CeRu$_{2}$Si$_{2}$, ~Sr$_{3}$Ru$_{2}$O$_{7}$, ~BaFe$_{2}$(P$_{0.3}$As$_{0.7}$)$_{2}$, ~Bi$_{2}$Sr$_{2}$Ca$_{0.92}$Y$_{0.08}$Cu$_{2}$O$_{8+ \delta}$, ~(TMTSF)$_{2}$PF$_{6}$, 
 ~Pb, ~Nb,~Pd, ~Al, ~Au, ~Cu, and ~Ag) seem to saturate this bound. The highest reported value of the linear in temperature coefficient corresponds to 
 \begin{eqnarray}
 \rho^{empirical}_{\max} \simeq 2.7 \frac{m k_{B} T}{{\sf n} e^{2} \hbar},
 \end{eqnarray} 
 associated with elemental metals such as lead. Thus, our bounds seem to indeed hold empirically. 
 We remark that in invoking the Drude formula for the conductivity and employing Eqs. (\ref{etaint}, \ref{etac}) for the high temperature viscosity, 
 we derive a simple result in which the relaxation time and particle density drop
 out. That is, at high temperatures where the the conductivity is semi-classical, 
 \begin{eqnarray}
 \frac{\sigma}{\eta} =  \frac{e^{2}}{mk_{B} T}.
 \end{eqnarray}
 As seen, Planck's constant drops out in this ratio. 
 
Thus, our simple considerations above suggest that the conductivities of both insulators and metals may veer to similar limiting high temperature behavior, $\sigma \sim \sigma_{\infty}$. For insulators, the conductivity tends to this value from below (with the conductivity scaling as $r(T)$ of Eqs. (\ref{particle_derivation}, \ref{ratio}))
whereas for metals, the conductivity approaches this value from above (scaling as $1/r(T)$).
Interestingly, in $D=2$ dimensional systems similar behavior is found as in Figure \ref{mit}.
In bad metals \cite{bad}, a linear $T$ increase of the resistivity persists to high temperatures. 
\section{Thermal Conductivity}
\label{tc}

The thermal conductivity  $\kappa$ is set by the size of the heat current generated by a thermal gradient,
$\vec{j}_{Q} = - \kappa \vec{\nabla} T$. 
In a classical $D=3$ dimensional metal with free particles, the ratio between the thermal conductivity and the Drude conductivity, i.e., the Lorentz number $L = \kappa/(\sigma T) =  (3/2) \times (k_{B}/e)^{2}$. More generally, $V \kappa = \langle {\bf{v}}^2  \rangle \tau C_{v}/D$ with $V$ the system volume and $C_{v}$ the specific heat. By Eq. (\ref{resist_bound}), an analogous relation implies
\begin{eqnarray}
\label{kp}
\kappa \ge \frac{3 {\sf n} \hbar k_{B}}{4 \pi m}.
\end{eqnarray}
Inserting Eqs. (\ref{particle_derivation}, \ref{rTL}) for $\tau$ and employing the equipartition theorem, 
we trivially find that, in semi-classical metals, prior to the saturation of the mean free path (when the mean free path is larger than inter particle distance), 
\begin{eqnarray}
\kappa \sim  \frac{hC_{v}}{mV} e^{\beta \Delta {\sf H}_{A}}.
\end{eqnarray} 

\section{Conclusions and outlook.}

While certain prevailing wisdom argues for a dichotomy of low temperature quantum and ``classical'' high temperature effects, our analysis illustrates that the two may be more ``entangled'' than is widely appreciated and that quantum phenomena may emerge rather sharply in the high temperature limit. We showed how this may occur in computing the viscosity of liquid and extrapolating the result to high temperatures. We sketched how similar results may appear for other transport functions. Our calculations and suggesting principles lead to results that are summarized below. Principally, we

{\bf (1)} Established that at high temperatures, WKB type quantization leads to the exact well known
substitution of the form of Eq. (\ref{sums}). This celebrated correspondence has earlier been 
motivated by heuristic considerations for which we now propose an underlying rigorous basis. 

{\bf (2)} Derived new general results connecting semi-classical dynamics to thermodynamic quantities. In particular, in subsection \ref{new_thermo}, we found a relation between entropy, dynamics, and Planck's constant in non-chaotic semi-classical systems. 

{\bf (3)} Illustrated that the transition state theory type result of Eq. (\ref{ratio}) may be derived
for general Hamiltonians by invoking the WKB results and deformation of the potentials.
In particular, we found that in the high temperature limit, irrespective of the complexity 
of the possible system trajectories, the relaxation time $\tau$ in semi-classical systems universally scales as Eq. (\ref{rTL})
and more generally a lower bound on the relaxation times (Eq. (\ref{tmin})) was found.
We demonstrated (Section \ref{measurements}) how to relate measured dynamics
to thermodynamics.

{\bf{(4)}} Derived lower bounds on the relaxation times in general equilibrated quantum systems (Eq. (\ref{tmq}))
and quantum critical theories therein (Eq. (\ref{rt:qcp})). We connected chaos to relaxation rates,
(further suggesting that the bounds on chaos of \cite{juan} might be derived from those invoking 
bounds on information transfer in quantum systems \cite{hod}) and
suggested new bounds on Lyapunov exponents in semiclassical systems (Eq. (\ref{chaosus})).
Albeit short compared to everyday experience, apart from their possible appearance in the viscosity and other transport functions listed below,
these thermalization time bounds (both in general quantum and semiclassical systems) 
may be of empirical relevance and tested by femtosecond spectroscopy, e.g., \cite{photo-exp,nadav} and 
other probes. 

{\bf (5)} Showed that a current correlation function calculation for general systems with relaxation time $\tau$
yields, at high temperatures and/or low densities, a viscosity given by Eq. (\ref{enk}). 

{\bf (6)} Invoking the Boltzmann equation while keeping arbitrarily high order correlations, we found that the viscosity attains a similar form
with a properly averaged relaxation time. More precisely, Eq. (\ref{etaint'}) was obtained with $\tau'$ a weighted relaxation time. The general result of Eq. (\ref{etax}) suggested 
how corrections will appear to the leading order result that coincides with that of item (4) above.

{\bf{(7)}} Demonstrated how a bound similar to the AdS-CFT bound of Eq. (\ref{Eq:visc-bound}) need not be arrived at by string theory considerations but rather by far more standard thermofield, Boltzmann equation, and ideal gas entropy bounds (Eq. (\ref{eh5})).

{\bf{(8)}} Obtained lower bounds on the viscosity and  explained how non-trivial information may be extracted from a linear tangent to the plot of the log of the viscosity as function of inverse temperature. In particular, we observed that the viscosity may be generally given by Eq. (\ref{wgeq}). In Section \ref{lower_b}, we further explained why in an Arrhenius fit to the viscosity of the liquid at a temperature $T_{A}$ (with this temperature adduced from a linear fit to the graph of $\ln \eta$ as a function of inverse temperature at $T_{A}$) leads to Eq. (\ref{Arr}) with a prefactor (Eq. (\ref{Af})) that may saturate to the quantized value ${\sf n} h$ at the lowest equilibrium temperature below which cooperative many-body effects may render the system non-ergodic $T_{coop} \equiv T_{A}^{\min}$. From Eqs. (\ref{Arr},\ref{Af}), the Arrhenius form prefactor
may monotonically increase when $T_{A}$ is made larger.

{\bf (9)} Found that when fitting the experimentally measured viscosity of 23 metallic glass formers,
the {\it ensemble average} of the extrapolated high temperature viscosity is exceedingly close to
that arrived by a simple implementation of the Boltzmann equation (with an error of $0.6 \%$ and standard deviation 
of order $9 \%$).

{\bf (10)} Suggested that the lowest temperature at which the ``classical'' liquid remains in equilibrium may be naturally related to the melting/liquidus temperature of
the system via the Eigenstate Thermalization Hypothesis (Section \ref{ETH_section}). The data suggest that these two temperatures (i.e., the lowest equilibration temperature $T_{coop}$ of the classical liquid and the melting or liquidus temperature) are indeed very close to one another. An extension of these ideas led to possible forms for the viscosity of supercooled liquids with a dependence on $T_{coop}$. To illustrate, we sketched how to  argue for possible crude approximate functions such as that of Eq. (\ref{geoni}) for temperatures below $T_{coop}$. More generally, we suggested that relations such as that of Eq.  (\ref{sumexact} or Eq. (\ref{solvep}) 
might determine, with no adjustable parameters, the viscosity of supercooled liquids, given an energy distribution adduced from thermodynamic measurements (Eq. (\ref{thermog}).
The latter possibility needs to be tested against  experimental data.

{\bf (11)} Noted that a similar possible corollary of the Eigenstate Thermalization Hypothesis is that if disorder free slowly cooled liquids evolve into crystalline or other thermalized solids 
at all positive temperatures then even if supercooled such liquids might not exhibit a finite temperature ``ideal glass'' transition (Section \ref{ETH_section}). That is, at all positive temperatures, liquids will equilibrate at long enough times.  

{\bf (12)} Proposed how to extend how own analysis to other transport properties including conductivities.
Using very simple considerations we almost immediately found that possibility of a resistivity saturation
values set by the quantum of resistance $e^{2}/h$ with $e$ the electronic charge. We furthermore
found that linear in temperature resistance (such as that found in ``bad metals'') 
is quite natural within this framework. 

{\bf{(13)}} Suggested, in the context of the last item, that apart from near semi-classical inequalities for fluids and possible 
saturation values, a possible general bound on the resistivity of bad metals (see Section \ref{ec}, Eq. (\ref{resist_bound}) in particular).  
This bound seems to be satisfied by empirical data on numerous materials. Similarly, a universal bound on the thermal conductivity of metals may be proposed (Eq. (\ref{kp})).

{\bf (14)} Remarked that our results suggest bounds on the extrapolated values of the Reynolds numbers
at high temperatures and that there is a minimal time scale for discrete molecular dynamics that may
amend the continuous fluid dynamics descried by the Navier-Stokes equation.

{\bf (15)} In item {\bf (9)} above, we suggested the specter of a near quantization regarding {\it average value} measurements. The viscosity (or other transport function), may still provide a relatively good estimate of the number of particles in the system. Thus, in systems in which the particle density is not known or effectively {\it ``dark''}, the viscosity may provide an approximate value of the particle density.  

We caution that additional effects in the analysis of the experimental data
(apart from the inexact single exponential approximation found in Eq. (\ref{AnH}))  
may lead to a distribution about the computed theoretical values. The same may hold true for other transport 
functions such as the conductivities that we briefly discussed. 
It would be interesting to carefully analyze the viscosity of liquids other than the 23 metallic fluids examined here and in \cite{metallic-glass}, to see the distribution of their high temperature values to see if our results to do not change.
Another possible extension is the analysis of the transverse (or Hall) viscosity complementing the standard longitudinal viscosity studied in the current work and \cite{metallic-glass}. Hall viscosity has recently been examined in Quantum Hall systems \cite{qhv}. In the context of our current work and that of \cite{metallic-glass}, this may be experimentally achieved by measuring the viscosity of a rotating levitated droplet 
(the levitation scheme is briefly described in the Appendix); such a rotating motion may engender synthetic magnetic fields \cite{Gil} as in cold atom systems \cite{synthetic}.

\acknowledgments 

We thank Oded Agam, Amnon Aharony, Mark Alford, Ehud Altman, Ariel Amir, Carl Bender, Gil Young Cho, Doron Cohen, Takeshi Egami, Eduardo Fradkin, Anup Gangopadhyay, Hans Hansson,  Eyal Heifetz, Joe Imry, Jainendra Jain, Nadav Katz, Steven Kivelson, Mo Li, Richard Mabbs, Gerald Mahan, Mark Newman, Cristiano Nisoli, Shmuel Nussinov, Dror Orgad, Gerardo Ortiz, Itamar Procaccia, Louk Rademaker, Filip Ronning, Joost Slingerland, Lee Sobotka, Ryan Soklaski, Boris Spivak,  Ady Stern, Vy Tran, Stuart Trugman, Li Yang, and Jan Zaanen for discussions and comments. We are grateful for support by the US National Science Foundation (DMR-11-06293, DMR-12-06707, and DMR-14-11229), the Deutsche Forschungsgemeinschaft (DFG) via collaborative research center SFB TR 12, and the US National Aeronautics and Space Administration (NNX10AU19G).

\appendix

\section{Experimental details}
\label{experiment-appendix}

\begin{figure}
\centering
\includegraphics[width=9cm]{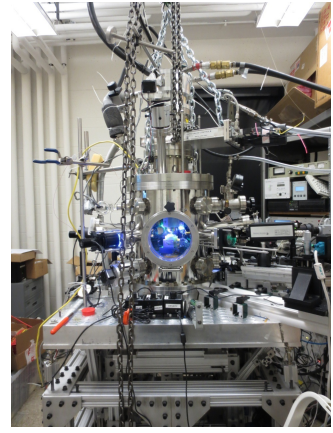}
\caption{(Color online.) {
The Washington University Beamline ElectroStatic Levitation Facility (WU-BESL) in which
fluid droplets are electrostatically levitated  \cite{cite:besl}. }}
\label{BESL}
\end{figure}

\begin{figure} 
\centering
\includegraphics[width=4cm]{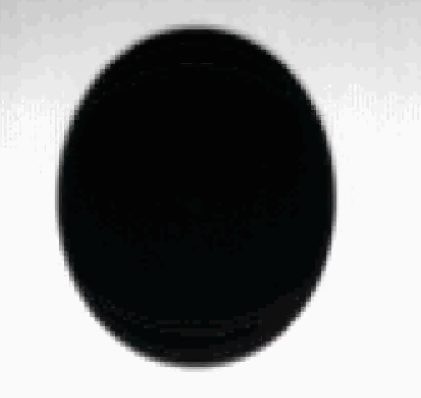}
\caption{(Color online.) {
A time frame in the oscillations of the droplet once an $l=2$ mode
is excited.}}
\label{p1}
\end{figure}

\begin{figure} 
\centering
\includegraphics[width=4cm]{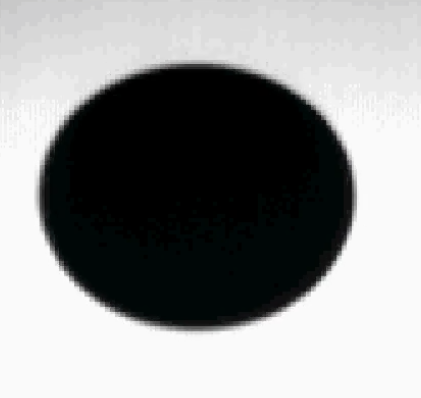}
\caption{(Color online.) {A later snapshot oscillation of the shadow of the $l=2$ mode oscillations.}}
\label{p2}
\end{figure}


In sections \ref{normal},\ref{ETH_section} (and in Figures \ref{Gauss},\ref{Gauss'} therein, in particular), 
we compared theoretical results to those recently reported for numerous metallic glass forming liquids \cite{metallic-glass}. 
In what follows, for self-completeness, we review important aspects of these experiments as they pertain to our
analysis of the viscosity of high temperature fluids. In \cite{metallic-glass}, samples were levitated and melted in the high-vacuum containerless environment of the Washington University Beamline ElectroStatic Levitation Facility (WU-BESL), see Figure \ref{BESL}. \cite{cite:besl}.
The viscosity was subsequently measured via the oscillating drop method \cite{bendert}. Specifically, given the $l$-th multipole mode, 
the viscosity
\begin{eqnarray}
\eta = \frac{\rho_{m} R_{0}^{2}}{(l-1)(2l+1) P},
\label{P}
\end{eqnarray}
where $P$ attenuation time, $R_{0}$ unperturbed radius of the sample, and $\rho_{m}$ the mass density (i.e., mass per unit volume) \cite{bendert}.
The droplet size oscillated as 
\begin{eqnarray}
\label{R}
R = R_{0} \Big( 1+ \delta (\cos \Omega t) e^{-t/P} \Big),  \nonumber
\\ \Omega^2 =  \frac{l(l-1)}{\rho  R^{3}}\Big( (l+2) \sigma - \frac{\overline{Q}^{2}}{16 \pi^{2} \epsilon  R^{3}} \Big),
\end{eqnarray} 
where $\delta$ is a constant, $\sigma$ is the surface tension, $\overline{Q}$ is the charge, and $\epsilon$ the dielectric constant.
The voltage on the vertical electrode was modulated at a frequency that was close to the $l = 2$ spherical harmonic mode resonant frequency (typically 120-140 Hz) of the
liquid to induce surface vibrations. A high-speed camera (1560 frames per second) recorded the shadow of the oscillating sample, as depicted in Figs. \ref{p1}, \ref{p2}. After a sufficiently long time during which spurious transient oscillations faded, the perturbative voltage was removed and the time-dependent amplitude of the decaying surface harmonic oscillations was probed. The viscosity was determined from the decay time for the oscillation, $P$ 
via Eqs. (\ref{P}, \ref{R}). 
 
\begin{figure}
\centering
\includegraphics[width=6.3cm]{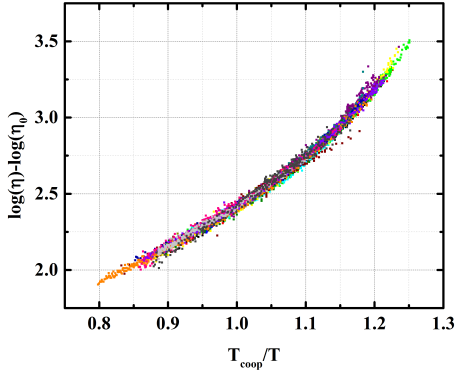}
\caption{(Color online.) {
The universal collapse of the viscosity of metallic glass forming fluids.
The collapse is obtained by scaling the viscosity by a system dependent constant $\eta_{0}$
(which close to ${\sf n} h$) and by scaling the temperature by $T_{coop}$- a temperature above which,
as seen from the straight line at high temperatures, the viscosity is well described by a single exponential form of Eq. (\ref{AnH}).
Numerically, $T_{coop}$ is identified with the lowest temperature at which the liquid is still in equilibrium \cite{numerics}.
The values of $\eta_{0}$ in \cite{metallic-glass} were obtained by the assuming the form of Eq. (\ref{AnH}) for the viscosity at temperatures
just above the limiting temperature $T_{coop}$.}}
\label{universal_collapse}
\end{figure}

\bigskip

\begin{figure} 
\centering
\includegraphics[width=6.3cm]{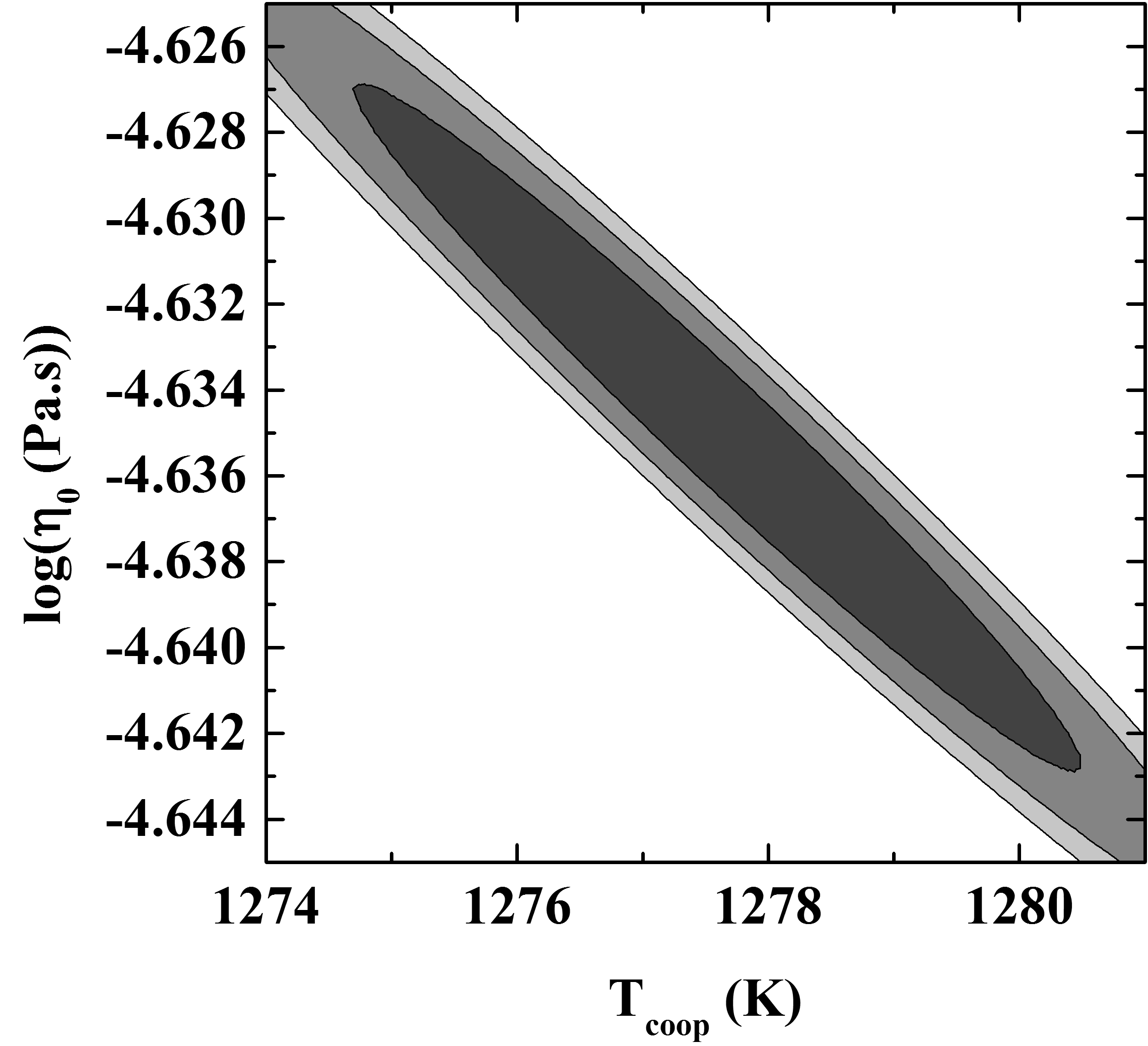}
\caption{(Color online.) {Fitting parameters and associated standard deviations in fitting the experimental data
of a metallic fluid (Vit1) \cite{metallic-glass}.
A standard deviation of 1 sigma separates consecutive shown ovals.
The data severely constrain the extrapolated high temperature viscosity $\eta_{0}$ and the scaling temperature $T_{coop}$
above which the dynamics are well described by Eq. (\ref{AnH}).}}
\label{standard_dev}
\end{figure}

The measured viscosity data (including that reported by works other than Ref. \cite{metallic-glass}) of numerous metallic fluids collapses onto a single master curve, 
as evinced in Figure \ref{universal_collapse}
with $\eta_{0}$ the extrapolated high temperature viscosity. This value of $\eta_{0}$ was compared with theory in Figure \ref{Gauss}. The value
of $\eta_{0}$ as well as the temperature $T_{coop}$ above which Arrhenius type dynamics is manifest is severely constrained by the data, 
see Figure \ref{standard_dev}. Thus, the displayed points in Figure \ref{Gauss} have small error bars if good fit of the data is to be achieved.

\section{The Eigenstate Thermalization Hypothesis}
\label{ETH-A}

In this brief appendix, we review the rudiments of the Eigenstate Thermalization Hypothesis \cite{eth1,eth2,eth3,von_neumann} employed in Section \ref{ETH_section}.
Given a state $|\psi(t_{final}) \rangle$ at time $t=t_{final}$ of the form of Eq. (\ref{dec}) we evolve it for times $t> t_{final}$ with the time independent Hamiltonian $H(t_{final}) = H$ that has eigenstates 
$\{|\psi_{n} \rangle\}$ with
corresponding energies $\{E_{n}\}$. With these, the long time average (l.t.a.) of the expectation value of a general operator ${\cal{O}}$ then reads longhand
\begin{eqnarray}
\label{llong}
{\overline{{\cal{O}}}}_{l.t.a.} = 
\lim_{ {\mathcal{T}} \to \infty}
\frac{1}{\mathcal{T}}
 \int_{t_{final}}^{t_{final} + {\mathcal{T}}} dt'~  \langle \psi_{final} | {\cal{O}}(t') | \psi_{final} \rangle \nonumber
 \\ = \lim_{ {\mathcal{T}} \to \infty}
\frac{1}{\mathcal{T}}  \int_{t_{final}}^{t_{final} + {\mathcal{T}}} dt'~ \sum_{n,m} c_{n}^{*} c_{m} \langle n| {\cal{O}} | m \rangle e^{i(E_{n}-E_{m})t'/\hbar}.
\end{eqnarray}
As $\int_{0}^{\infty} dt' e^{i (E_{n}- E_{m}) t'} = (1+i) \pi~\delta (E_{n} - E_{m}) + \frac{1}{E_{n} - E_{m}}$,
if the spectrum is non-degenerate ($E_{n} \neq E_{m}$) then the off-diagonal contributions in the last line of Eq. (\ref{llong}) will vanish in the long time (${\cal{T}} \to \infty$) limit.
Assuming the lack of sufficient phase coherence between these different off-diagonal contributions, 
only diagonal contributions will remain in Eq. (\ref{llong}). (We remark that in the case of a degenerate spectrum, block diagonal contributions
will arise from the degenerate subspaces.) Thus, the long-time average will be given by the weighted sum of expectation values 
in the eigenstates of $H$. Within the micro-canonical (m-c) ensemble, the average of ${\cal{O}}$ is 
\begin{eqnarray}
\label{m.c.}
{{\mathbb{E}}}_{m.c.} ( {\cal{O}}) = \frac{1}{{\cal{N}}(E,\Delta E)} \sum_{n; ~ E \le E_{n} \le E+ \Delta E} \langle n | {\cal{O}}| n \rangle,
\end{eqnarray}
where 
${\cal{N}}(E,\Delta E) \equiv (\sum_{n; ~ E \le E_{n} \le E+ \Delta E} 1)$
is the number of micro-states of an energy $E$ that lies in the interval $[E,E+\Delta E]$.
If the diagonal expectation values of the ${\cal{O}}$ do not vary for states $|\psi \rangle$ that have components only in this interval of fixed energy then, by virtue of normalization
$\sum_{n} |c_{n}|^{2} =1$, the expectation values in Eqs. (\ref{llong},\ref{m.c.}) are equal to each other, 
\begin{eqnarray}
\label{ethe}
{\overline{{\cal{O}}}}_{l.t.a.} = {{\mathbb{E}}}_{m.c.} ( {\cal{O}}).
\end{eqnarray}
Eqs. (\ref{llong},\ref{m.c.},\ref{ethe}) capture the guiding principle of the Eigenstate Thermalization Hypothesis.
If, for any real system (ultimately specified by some microscopic quantum state), the expectation values of a general ${\cal{O}}$ may be found by thermodynamic ensemble averaging (over all states in the energy shell $[E,E+\Delta E]$), then Eq. (\ref{ethe}) will hold. In Section \ref{ETH_section}, we have repeatedly invoked the equality of an diagonal matrix element expectation value in a given state to that found by thermodynamic averaging in an equilibrated ergodic system (for which the thermal expectation value is
performed using Eq. (\ref{m.c.})). We wish to remark that similar to the considerations that we invoked for the local current operators in Section \ref{ETH_section}, if the matrix elements of local operators ${\cal{O}}$ between orthogonal degenerate states vanish then, even in the degenerate case, the remaining sum in Eq. (\ref{llong}) will be purely over diagonal states. Such a vanishingly small matrix elements between degenerate states is found for many local operators in numerous systems. Typically, such off-diagonal matrix elements of local operators are either exactly zero or vanish in the large system size limit.


\begin{thebibliography}{99}
\bibitem{remark1} In line with dimensional arguments and a main premise of the article to be detailed later,
the prefactor multiplying $e^{\Delta/(k_{B} T)}$ is ${\sf n}_{eff}(T) h$ with  ${\sf{n}}_{eff}(T)$ an effective temperature dependent density
and $h$ Planck's constant. This functional form is akin to that derived in this article for normal high temperature fluids.
\bibitem{explain_J} Below, we trivially illustrate that Eq. (\ref{Eq:tau}) indeed holds in all dimensions. An alternate, more standard, demonstration of this assertion proceeds by using contact transformations. The approach below is more mundane. With $p$ and $x$ set along the arc length of the system evolution (so that $dx = (p/m) dt$), the action variable
$J = \int_{0}^{\tau(E)} \frac{p^{2}}{m} dt$ and thus, by immediate differentiation, $\frac{\partial J}{\partial E} = \frac{1}{m} \int_{0}^{\tau(E)} \frac{\partial (p^{2})}{\partial E} dt = \frac{1}{m} \int_{0}^{\tau(E)}  \frac{p}{\dot{x}}dt = \tau(E)$ where, in the second equality, we invoked Hamilton's equation $\dot{x} = \frac{\partial H}{\partial p}$ with $H=H(x,p)=E$ along the integrated time. Clearly in any approach once the phase space coordinates along the arc (viz., $x$ and $p$) are employed, the one-dimensional result of Eq. (\ref{Eq:tau}) must follow for trajectories in any number of spatial dimensions.
\bibitem{evaporation_footnote} That evaporation and other temperatures may be equal to activation energies has long been observed. We are not examining a phase transition between different states of matter. Rather, we are investigating motion in a many body system and propose that the activation barrier $E_{\rm escape}$ may be of the scale of the transition temperature from the liquid into a gas. This proposal concerning gas like momentum eigenstates at high enough energy scales is similar yet a bit different from Eyring's \cite{eyring-viscosity}; Eyring literally thought of particles and ``holes'' in a liquid as constituents forming molecular type structures (``activated complex'') and of the evaporation of such constructs.
\bibitem{identical_n}
A semi-classical system initially evolving along some direction $\hat{n}$ at time $t=0$ in configuration space may obtain both longitudinal (i.e., along the direction $\hat{n}$) as well as transverse momenta from the heat bath. Longitudinal momentum changes are those that render the problem one-dimensional along $\hat{n}$ (and more generally the plane defined by this direction as we will explain momentarily). By contrast, transverse momenta changes link a state initially evolving along $\hat{n}$ to an evolution along another direction. 
For rapidly moving systems at high temperatures, external transverse momenta changes (small relative to the large initial momenta) will leave motion in a narrow cone about the initial momentum direction.
Even if general transverse changes occur, the resultant excitation may still be expressed as an excited state $|\psi_{\hat{n'}} \rangle$ associated with another initial state along another direction $\hat{n'}$. Thus averaging over an ensemble of initial possible states renders the semi-classical evolution of the system to that of decoupled one-dimensional rays,
We now regress and further discuss the motion along each direction ${\hat{n}}$. As explained in Eq. (\ref{MBarc}), along one initial direction $\hat{n}$ of the momentum in configuration
space (or that for a single particle), the shape of the one-dimensional
semi-classical particle or system trajectory in higher dimensional space
configuration space (or three-dimensional shape for a single particle) will
generally change with increasing energy $E_{n}$. Similarly, the
corresponding potential energy evaluated along the particle trajectory, the
function $V_{E,\hat{n}}(x)$, will change its form. Nevertheless, what is of pertinence 
in evaluating the rate is the Boltzmann factor and the existence of a cutoff energy $E_{escape}$
for the integral of Eq. (\ref{particle_derivation}) for any
given path set by the direction ${\hat{n}}$ in configuration space; in the integrand
of Eq. (\ref{particle_derivation}) we may employ the partial differentiation in Eq. (\ref{Eq:tau}) when the 
initial system coordinates and initial velocity directions are fixed in any number 
of dimensions \cite{explain_J}. The uniform sign of the derivative in Eq. (\ref{Eq:tau}) implies
that, as in the one-dimensional case, for any fixed initial coordinate and velocity direction,
the change of integration from that relative to $J$ to an integration over energy values $E$ may
performed as in Eq. (\ref{particle_derivation}). Thus, our calculation will be
identical along any ray ${\hat{n}}$ regardless of the change in the explicit
functional form of $V_{\hat{n},E}$ which is immaterial in Eq. (\ref{particle_derivation}).
\bibitem{caging} In Eq. (\ref{r(T)}), we examine
transport within the energetic region (of energies $E>E_{escape}$).
At temperatures far below $E_{escape}/k_{B}$, the system is most of the time at far lower energies ($E \ll E_{escape}$),
particles oscillate in wells far depressed beneath $E_{escape}$.
That time spent is, trivially, accounted for in our calculation. This is so because as long
as the particle is trapped in the well, it does not contribute to global transport.
The fraction of the time that the particle is at high energies (enough to overcome the energy barrier), for a state of energy $E_{n} > E_{escape}$ 
is the probability of occupying such states $\langle {\sf n}_{n}(T) \rangle$.
It is only motion during those times that contributes to the rate in Eq. (\ref{r(T)}).
\bibitem{explain-TST} In the original version of transition state theory \cite{eyring-tst}, the rate linked with the translational degrees of freedom, was arrived at by multiplying ``the number of available states'' associated with motion in the forward direction (i.e., half the partition function which, for free particles which can move forwards and backwards in a one dimensional system of size $\overline{L}$) given by $\overline{L}/(2 \lambda^{nr}_{T}) = \overline{L} \sqrt{2 \pi m k_{B} T}/(2h)$ with the average transition rate per individual state (in other words., the average velocity in the forward direction divided by the distance $L$) namely $\langle v \rangle/\overline{L} = \int_{0}^{\infty} dv~ v~e^{-\beta mv^{2}/2}/(\overline{L}\int_{0}^{\infty} dv~ e^{-\beta mv^{2}/2}) = \sqrt{(2 k_{B} T)/(\pi m \overline{L}^{2})}$). The product of these two quantities is $\langle v \rangle/(2  \lambda^{nr}_{T}) =(k_B  T)/h$. The ratio $\langle v \rangle/(2  \lambda^{nr}_{T})$ is reminiscent of Eq. (\ref{TRP}) with the length $L$ therein set equal to thermal de Broglie wavelength $\lambda^{nr}$. The notion of transition state theory was pioneering since its inception \cite{eyring-tst}. It is, nevertheless, not obvious that one can assume that all available states contribute equally (with this uniform average velocity $\langle v \rangle$) to the rate. Explicit normalization does not appear in these considerations. In equilibrium statistical mechanics, at very high temperatures, the number of available states (the partition function) diverges and thus, following the logic above, one
would seem to arrive at the paradoxical situation that an initial single particle can seemingly ``flow'' to a final location with unit probability through each of the channels provided by the different eigenstates $|n \rangle$. In our equilibrium based derivation of Eqs. (\ref{particle_derivation},\ref{ratio}), unitarity and ensuing normalization are enforced via a chemical potential (we divide by a partition function to normalize the probability distribution).
\bibitem{particle_in_a_box} 
At a temperature $k_{B} T \ll (E_{escape}- E_{0})$, for, e.g., a particle with a minimal energy $E_{0}$ in a deep double-dimensional potential well of total length $L$ (set by the boundaries
at $x_{initial}$ and $x_{final}$) that extends up to an energy $E_{escape}$ as a rendition of Figure \ref{Potential},
the exponential $\exp(-\beta \Delta G) \sim \lambda_{nr}/L \exp(-\beta \Delta)$
where $\Delta = E_{escape} - E_{0}$.
Consequently, the relaxation time $\tau \sim 1/r(T) \propto \frac{1}{\sqrt{T}} \exp(\beta \Delta)$,
with $\Delta$ the energy barrier. Such a form is in accord with Eqs. (2,4) in \cite{itamar_also_change_at_TA}.
More complicated effects including those of the ``caging'' type effects in deep interconnected wells lead to a similar result.
\bibitem{box2} For the case of a deep well (see., e.g., \cite{particle_in_a_box}),  Eqs. (\ref{ratio},\ref{enk}) imply that the viscosity
$\eta \propto {\sf n}{\sqrt{T}} \exp(\beta \Delta)$ with ${\sf n}$ the particle density. 
If the temperature $T_{A}$ is far below the box excitation energy (an effective bonding energy)
then the system will remain in its ground state and the 
entropy difference associated with Eq. (\ref{dg})
will, essentially, be nil, $\Delta S_{A} \sim 0$.
\bibitem{explain_fd} Within linear response theory, the change $\delta W$ in an observable $W$ to a perturbation $V$, is given by a convolution of the perturbation with the susceptibility $\delta W(t) = \chi * V \equiv \int_{- \infty}^{t} \chi(t-t') V(t') dt'$. By the fluctuation-dissipation theorem, $\chi$ is set by the connected correlators (i.e., those involving the initial time of the perturbation $V$ and the later time of measurement of $W$). For an equilibration time $\tau$, the response function decays as $e^{-t/\tau}$. For computational convenience, the correlator of Eq. (\ref{fte}) was examined in \cite{juan}. instead of the bare two operator correlator yet it encodes the same time. Thus, we see that it is this thermalization time of Eq. (\ref{fte}) that is of relevance to our calculations of the viscosity.
\bibitem{maxwell} We remind the reader of an approximate derivation of the Maxwell relation. Replacing $\langle U_{z}^{2} \rangle$ by $\frac{1}{3} (\ell/\tau)^{2}$ in the last line of Eq. (\ref{etaint}) leads to $\eta \approx \rho \ell v/3$.
\bibitem{triv_example} 
The relaxation rate associated with an arbitrary response function $\chi$ is given by the absolute value of the imaginary part of the poles $| \Im{\omega}^*|$ of such a response function. For $\chi$ having a low energy solid contribution that exceeds that of the high energy fluid component at low temperatures, the Dyson series for consecutive solid-like and fluid-like temporal evolutions leads to $\chi(\omega) = [(\chi^{solid}(\omega))^{-1}- \chi^{liquid}(\omega)]^{-1}$ with $\chi^{solid}$ and $\chi^{liquid}$ the low and high energy response functions. For caricatures of a simple dissipative liquid response of $\chi^{liquid} =  g (1 - i \omega \tau)^{-1}$ and  harmonic $\chi^{solid} = (-M \omega^{2}  + K)^{-1}$, the imaginary part $| \Im{\omega}^*|= \frac{g \tau}{2 M [1+ (\omega_{0} \tau)^{2}]}$ (with $\omega_{0} \equiv \sqrt{K/M}$ denoting the resonant (Einstein phonon type) frequency of the solid) is linear in the strength of the the liquid component contribution.
\bibitem{MBL'} Certain ``many body localized'' systems \cite{MBL} are exceptions to this rule .
\end{thebibliography}
\end{document}